\documentclass[12pt]{amsart}
\usepackage{amsfonts,amssymb,amscd,amsmath,epsfig}
\usepackage{latexsym}
\usepackage{mathrsfs}
\usepackage{tocvsec2}

\usepackage[%dvipdfm,
hyperindex=true,pagebackref=true,bookmarks=true,
colorlinks=true,linkcolor=blue,citecolor=red]{hyperref}
\def\~{{\rm --}}
\begin{document}

%MACOS FOR LECTURES ON DAHA
\renewcommand{\tilde}{\widetilde}
\renewcommand{\hat}{\widehat}

\newcommand{\BR}{{\mathbb R}}
\newcommand{\BQ}{{\mathbb Q}}
\newcommand{\BC}{{\mathbb C}}
\newcommand{\BP}{{\mathbb P}}
\newcommand{\BZ}{{\mathbb Z}}
\newcommand{\BN}{{\mathbb N}}
\newcommand{\BS}{{\mathbb S}}

\newcommand{\cH}{{\mathcal H}}
\newcommand{\cA}{{\mathcal A}}
\newcommand{\cB}{{\mathcal B}}
\newcommand{\ccF}{{\mathfrak F}}
\newcommand{\cD}{{\mathcal D}}
\newcommand{\cL}{{\mathcal L}}
\newcommand{\cF}{{\mathcal F}}
\newcommand{\cP}{{\mathcal P}}
\newcommand{\cX}{{\mathcal X}}
\newcommand{\cY}{{\mathcal Y}}
\newcommand{\cS}{{\mathcal S}}
\newcommand{\cSol}{\hbox{$\mathcal Sol$}}
\newcommand{\cT}{\hbox{$\mathcal T$}}

\newcommand{\Z}{{\mathbb Z}}
\newcommand{\Q}{{\mathbb Q}}
\newcommand{\N}{{\mathbb N}}
\newcommand{\C}{{\mathbb C}}
\newcommand{\R}{{\mathbb R}}
\newcommand{\X}{{\mathbb X}}
\newcommand{\Y}{{\mathbb Y}}

\newcommand{\CH}{{\mathcal H}}
\newcommand{\CA}{{\mathcal A}}

\def\HH{\mbox{${\mathcal H}$\kern-5.2pt${\mathcal H}$}}

\newcommand{\binomial}[2]{\genfrac{(}{)}{0pt}{}{ #1 }{ #2 }}
\newcommand{\qbinomial}[2]{\genfrac{[}{]}{0pt}{}{ #1 }{ #2 }_q }
\newcommand{\qbinom}[3]{\genfrac{[}{]}{0pt}{}{ #1 }{ #2 }_{ #3 } }

%%SPECIAL SEC 1.0

\def\der{\partial}
\def\tensor{\otimes}
\def\gam{\gamma} \def\Gam{\Gamma}
\def\del{\delta} \def\Del{\Delta}
\def\kap{\kappa}
\def\lam{\lambda} \def\Lam{\Lambda}
\def\Comp{{\mathbb C}}
\def\sM{{\mathcal M}}

\newtheorem{theorem}{Theorem}[section]
\newtheorem{maintheorem}[theorem]{Main Theorem}
\newtheorem{proposition}[theorem]{Proposition}
\newtheorem{definition}[theorem]{Definition}
\newtheorem{lemma}[theorem]{Lemma}
\newtheorem{corollary}[theorem]{Corollary}
\newtheorem{notation}[theorem]{Notation}
\newtheorem{remark}[theorem]{Remark}
\newtheorem{example}[theorem]{Example}

\newtheorem{theorem }{Theorem}[section]
\newtheorem{maintheorem }[theorem]{Main Theorem}
\newtheorem{proposition }[theorem]{Proposition}
\newtheorem{definition }[theorem]{Definition}
\newtheorem{lemma }[theorem]{Lemma}
\newtheorem{corollary }[theorem]{Corollary}
\newtheorem{notation }[theorem]{Notation}
\newtheorem{remark }[theorem]{Remark}
\newtheorem{example }[theorem]{Example}

\newtheorem{ maintheorem }[theorem]{Main Theorem}
\newtheorem{ theorem}{Theorem}[section]
\newtheorem{ proposition}[theorem]{Proposition}
\newtheorem{ definition}[theorem]{Definition}
\newtheorem{ lemma}[theorem]{Lemma}
\newtheorem{ corollary}[theorem]{Corollary}
\newtheorem{ notation}[theorem]{Notation}
\newtheorem{ remark}[theorem]{Remark}
\newtheorem{ example}[theorem]{Example}

\newtheorem{thm}{Theorem}[section]
\newtheorem{prop}[thm]{Proposition}
\newtheorem{lem}[thm]{Lemma}
\newtheorem{cor}[thm]{Corollary}
\newtheorem{conj}[thm]{Conjecture}
\newtheorem{con}[thm]{Conjecture}
\newtheorem{dfn}[thm]{Definition}
\newtheorem{df}[thm]{Definition}
 \newcommand{\rem}{{\bf Comment.\ }}
 \newcommand{\rmk}{{\bf Comment.\ }}
 \newcommand{\exmp}{{\bf Example.\ }}
 \newcommand{\ex}{{\bf Example.\ }}
 \newcommand{\prob}{{\bf Problem.\ }}

\newtheorem{note}{Note} 
\renewcommand{\thenote}{}
\newtheorem*{acka}{Acknowledgments}
\newtheorem{ack}{Acknowledgments}
\renewcommand{\theack}{}
\renewcommand{\appendixname}{\bf Appendix}
\renewcommand{\proof}{{\em Proof.\ }}

\hyphenation{
ap-pen-dix as-ymp-tot-ic at-trib-uted at-trib-ut-able
Bry-li-n-sky com-mu-ta-tion de-ge-ne-rate
de-riv-a-tive dis-trib-ute equi-vari-ant ex-tra-or-di-nary  
geo-met-ric griev-ance griev-ous grad-ed ho-lo-no-my ho-mo-thetic
in-fin-ite-ly in-fin-i-tes-i-mal Ha-rish Cha-n-dra mul-ti-plic-able 
non-euclid-ean non-iso-mor-phic non-smooth par-a-digm 
par-a-bol-ic pa-rab-o-loid pa-ram-e-trize phe-nom-e-non 
post-script pseu-do-dif-fer-en-tial pseu-do-fi-nite 
qua-drat-ics quad-ra-ture Han-kel rec-tan-gle semi-def-i-nite 
set-up wide-spread Euler-ian Feb-ru-ary Gauss-ian Grothen-dieck 
Hamil-ton-ian Her-mi-t-ian her-mi-t-ian Jan-u-ary 
Japan-ese Ka-shi-wa-ra Kor-te-weg Le-gendre No-vem-ber Rie-mann-ian 
Sep-tem-ber Za-mo-lo-d-chi-kov Kni-zh-nik quan-tum Op-dam
Mac-do-nald Ca-lo-ge-ro Su-ther-land Mo-ser 
Ol-sha-net-sky  Pe-re-lo-mov in-de-pen-dent ope-ra-tors 
cy-clo-to-mic ra-tio-nal de-gen-er-a-tion 
in-ter-est-ing de-for-ma-tions de-for-ma-tion pro-ce-dure 
fol-lows ope-ra-tors  pre-serve suf-fices ap-proach 
for-mu-las con-sider its com-ple-tion cor-re-spond-ing 
au-to-mor-phism be-cause pro-por-tional fi-nal-ly let-ting 
equi-v-a-lence ge-n-er-al-ized Mac-do-nald iden-ti-ties 
cor-re-s-pond sub-dia-grams par-ti-tion na-t-u-ral-ly 
or-dered stan-dard de-for-ma-tion ar-gu-ment com-bined 
sphe-r-i-cal rep-re-sen-ta-tions tri-go-no-me-t-ric
ge-n-er-al-ly speak-ing pri-m-it-ive ir-re-du-cible 
sum-ma-tion  rep-re-sen-ta-tives pro-por-ti-o-na-li-ty
ultra-sphe-ri-cal Ro-gers}

\def\ffor{\quad\hbox{ for }\quad}
\def\wwhen{\quad\hbox{ when }\quad}
\def\wwhere{\quad\hbox{ where }\quad}
\def\aand{\quad\hbox{ and }\quad}
\def\for{\  \hbox{ for } \ }
\def\iif{ \ \hbox{ if } \ }
\def\when{ \ \hbox{ when } \ }
\def\where{\  \hbox{ where } \ }
\def\and{\  \hbox{ and } \ }
\def\and{\  \hbox{ and } \ }
\def\oor{\  \hbox{ or } \ }
\def\proof{{\em Proof. \  }}

\def\equal{\stackrel{\,\mathbf{def}}{= \kern-3pt =}}

\def\la{\lambda}
\def\La{\Lambda}
\def\om{\omega}
\def\Om{\Omega}
\def\Th{\Theta}
\def\th{\theta}
\def\al{\alpha}
\def\be{\beta}
\def\ga{\gamma}
\def\ep{\epsilon}
\def\up{\upsilon}
\def\Up{\Upsilon}
\def\de{\delta}
\def\De{\Delta}
\def\ka{\kappa}
\def\kapp{\hbox{\bf \ae}}
\def\si{\sigma}
\def\Si{\Sigma}
\def\Ga{\Gamma}
\def\ze{\zeta}
\def\io{\iota}
\def\bio{b^\iota}
\def\aio{a^\iota}
\def\twio{\tilde{w}^\iota}
\def\hwio{\hat{w}^\iota}
\def\gio{\g^\iota}
\def\Bio{B^\iota}

\def\del{\delta}
\def\pa{\partial}
\def\vp{\varphi}
\def\ve{\varepsilon}
\def\inf{\infty}

\def\vph{\varphi}
\def\vps{\varpsi}
\def\vPh{\varPhi}
\def\vep{\varepsilon}
\def\vpi{{\varpi}}
\def\vth{{\vartheta}}
\def\vsi{{\varsigma}}
\def\vrh{{\varrho}}

\def\bph{\bar{\phi}}
\def\bsi{\bar{\si}}
\def\bvp{\bar{\varphi}}

\newcommand{\bS}{{\mathbf S}}
\newcommand{\bH}{{\mathbf H}}
\newcommand{\bF}{{\mathbf F}}
\newcommand{\bE}{{\mathbf E}}

\def\tal{\tilde{\alpha}}
\def\tbe{\tilde{\beta}}
\def\tde{\tilde{\delta}}
\def\tpi{\tilde{\pi}}
\def\txi{\tilde{\xi}}
\def\tPi{\tilde{\Pi}}
\def\tPhi{\tilde{\Phi}}
\def\tV{\tilde{V}}
\def\tJ{\tilde{J}}
\def\tla{\tilde{\lambda}}
\def\tga{\tilde{\gamma}}
\def\tGa{\tilde{\Gamma}}
\def\tvs{\tilde{{\varsigma}}}
\def\tu{\tilde{u}}
\def\tU{\tilde{U}}
\def\tw{\widetilde w}
\def\tW{\widetilde W}
\def\tB{\tilde B}
\def\tv{\tilde v}
\def\tV{\tilde V}
\def\tz{\tilde z}
\def\tb{\tilde b}
\def\ta{\tilde a}
\def\tih{\tilde h}
\def\trh{\tilde {\rho}}
\def\tx{\tilde x}
\def\tf{\tilde f}
\def\tg{\tilde g}
\def\tG{\tilde G}
\def\tk{\tilde k}
\def\tl{\tilde l}
\def\tL{\tilde L}
\def\tD{\tilde D}
\def\tR{\tilde R}
\def\tP{\tilde P}
\def\tH{\tilde H}
\def\tp{\tilde p}

\def\hH{\hat{H}}
\def\hh{\hat{h}}
\def\hR{\hat{R}}
\def\hY{\hat{Y}}
\def\hX{\hat{X}}
\def\hP{\hat{P}}
\def\hT{\hat{T}}
\def\hV{\hat{V}}
\def\hG{\hat{G}}
\def\hF{\hat{F}}
\def\hw{\widehat{w}}
\def\hW{\widehat{W}}
\def\hu{\hat{u}}
\def\hs{\hat{s}}
\def\hv{\hat{v}}
\def\hb{\hat{b}}
\def\hB{\widehat{B}}
\def\hze{\hat{\zeta}}
\def\hsi{\hat{\sigma}}
\def\hrh{\hat{\rho}}
\def\hth{\hat{\theta}}
\def\hy{\hat{y}}
\def\hx{\hat{x}}
\def\hz{\hat{z}}
\def\hg{\hat{g}}
\def\he{\hat{e}}
\def\hE{\widehat{E}}

\def\B{\mathbf{B}}
\def\I{\mathbf{I}}
\def\P{\mathbf{P}}
\def\G{\mathbf{G}}
\def\S{\mathbf{S}}
\def\F{\mathbf{F}}
\def\one{\mathbf{1}}
\def\Sn{\mathbf{S}_n}
\def\0{\mathbf{0}}
\def\H{\mathbf{H}}
\def\V{\mathbf{V}}

\def\f{\mathcal{F}}
\def\çF{\mathcal{F}}
\def\o{\mathcal{O}}
\def\t{\mathcal{T}}
\def\r{\mathcal{R}}
\def\l{\mathcal{L}}
\def\m{\mathcal{M}}
\def\k{\mathcal{K}}
\def\n{\mathcal{N}}
\def\d{\mathcal{D}}
\def\p{\mathcal{P}}
\def\cP{\mathcal{P}}
\def\a{\mathcal{A}}
\def\h{\mathcal{H}}
\def\c{\mathcal{C}}
\def\y{\mathcal{Y}}
\def\e{\mathcal{E}}
\def\v{\mathcal{V}}
\def\z{\mathcal{Z}}
\def\x{\mathcal{X}}
\def\s{\mathcal{S}}
\def\g{\mathcal{G}}
\def\u{\mathcal{U}}
\def\w{\mathcal{W}}
\def\i{\mathcal{I}}
\def\j{\mathcal{J}}
\def\b{\mathcal{B}}

\def\lan{\langle}
\def\llb{(\!(}
\def\ran{\rangle}
\def\rrb{)\!)}
 \def\dim{{\hbox{\rm dim}}_{\mathbb C}\,}
\def\lng{\hbox{\rm{\tiny lng}}}
\def\sht{\hbox{\rm{\tiny sht}}}
\def\sph{\hbox{\rm{\tiny sph}}}
\def\inv{\hbox{\rm{\tiny inv}}}

\def\br#1{\langle #1 \rangle}

\def\rank{\hbox{rank}}
\def\gl{\mathfrak{gl}_N}
%\def\sgn{\hbox{sgn}}
%\font\germ=eufb10 %at 12pt 
%\def\mathfrak#1{\hbox{\germ #1}}

\newcommand{\Aut}{\operatorname{Aut}}
\newcommand{\Hom}{\operatorname{Hom}}
\newcommand{\End}{\operatorname{End}}
\newcommand{\Ind}{\operatorname{Ind}}
\newcommand{\ad}{\operatorname{ad}}
\newcommand{\pr}{\operatorname{pr}}
\newcommand{\aweyl}{\tilde{\mathbb S}_n}
\newcommand{\hec}{{\mathcal H}^t_n}
\newcommand{\Func}{{\mathcal F}({\mathbb C}^n,{\mathcal H}^t_n)}
\newcommand{\tr}{\operatorname{tr}}
\newcommand{\Out}{\operatorname{Out}}
\newcommand{\Rad}{\operatorname{Rad}}
\newcommand{\Spec}{\operatorname{Spec}}
\newcommand{\id}{\operatorname{id}}
\newcommand{\Int}{\operatorname{Int}}
\newcommand{\ct} {\operatorname{ct}}

\newcommand{\rat}{{\mathbb Q}}
\newcommand{\real}{{\mathbb R}}
\newcommand{\cplx}{{\mathbb C}}
\newcommand{\zint}{{\mathbb Z}}

\newcommand{\sq}{\phantom{1}\hfill$\qed$}
\newcommand{\Rea}{\Re}
\newcommand{\Ima}{\Im}

\newcommand{\st}{\bowtie}
\newcommand{\modd}{\mbox{\,mod\,}}
\newcommand{\lr}{\langle}
\newcommand{\rr}{\rangle}
\newcommand{\eps}{\varepsilon}
\newcommand{\phk}{\phi^{(k)}}
\newcommand{\psk}{\psi^{(k)}}
\newcommand{\Res}{\mbox{Res}\;}
\newcommand{\sgn}{\mbox{sgn}}
\newcommand{\mn} {\left\{ \begin{array}{c}m\\
n\end{array}\right\}}

\def\sX{\mathscr{X}}
\def\sH{\mathscr{H}}
\def\sY{\mathscr{Y}}
\def\TT{\mathfrak{T}}
\def\JJ{\mathfrak{J}}
\def\HH{\mathfrak{H}}
\def\FF{\mathfrak{F}}
\def\GG{\mathfrak{G}}
\def\CC{\mathfrak{C}}
\def\LL{\mathfrak{L}}

\def\BB{\mathfrak{B}}
\def\AA{\mathfrak{A}}
\def\ZZ{\mathfrak{Z}}
\def\HH{\hbox{${\mathcal H}$\kern-5.2pt${\mathcal H}$}}
\def\HHH{\hbox{${\mathbb H}$\kern-4.2pt${\mathbb H}$}}
\def\tHH{\widetilde{\HH\ }}

\font\smm=msbm10 at 12pt 
\def\symbol#1{\hbox{\smm #1}}
\def\lsmash{{\symbol n}}
\def\rsmash{{\symbol o}}
\def\#{\sharp}

\font\tenbf=cmbx10
\font\tenrm=cmr10
\font\tenit=cmti10
\font\ninebf=cmbx9
\font\ninerm=cmr9
\font\nineit=cmti9
\font\eightbf=cmbx8
\font\eightrm=cmr8
\font\eightit=cmti8
\font\sevenrm=cmr7
\font\sevenbf=cmbx7

%END MACROS

\title [AI approach to momentum risk-taking]
{Artificial intelligence approach to momentum risk-taking}
\author[Ivan Cherednik]{Ivan Cherednik, UNC Chapel Hill $^\dag$}
%\date{February 2, 2014}

\begin{abstract}
We propose a mathematical model of momentum risk-taking, which 
is essentially real-time risk management focused on short-term 
volatility of stock markets. Its implementation, our fully automated 
momentum equity trading system presented systematically, 
proved to be successful in extensive historical and real-time 
experiments.  Momentum risk-taking is one of the key components of 
general decision-making, a challenge for artificial intelligence 
and machine learning with deep roots in cognitive science; its
variants beyond stock markets are discussed. We begin with a 
new algebraic-type theory of news impact on share-prices, which 
describes well their power growth, periodicity, 
and the market phenomena like price targets and 
profit-taking. This theory generally requires
Bessel and hypergeometric functions. 
Its discretization results in some tables of bids, which are 
basically expected returns for main investment horizons,
the key in our trading system. The ML procedures we use are 
similar to 
those in neural networking. A preimage of our approach is the new 
contract card game provided at the end, a combination of bridge 
and poker. Relations to random processes and the fractional 
Brownian motion are outlined.   

%Stock markets are quite a test for any risk-management theories.

\end{abstract}

%\thanks{$^\dag$ \today}

\address[I. Cherednik]{Department of Mathematics, UNC
Chapel Hill, North Carolina 27599, USA\\
chered@email.unc.edu}

\thanks{$^\dag$ \today.
\ \ \ Partially supported by NSF grant
DMS--1901796}
% and Simons Foundation}

 \def\sht{\raisebox{0.4ex}{\hbox{\rm{\tiny sht}}}}
 \def\bysame{{\bf --- }}
 \def\~{{\bf --}}
 \def\rr{{\mathsf r}}
 \def\cc{{\mathsf c}}
 \def\mm{{\mathsf m}}
 \def\pp{{\mathsf p}}
 \def\ll{{\mathsf l}}
 \def\aa{{\mathsf a}}
 \def\bb{{\mathsf b}}
 \def\NS{\hbox{\tiny\sf ns}}
 \def\ssum{\hbox{\small$\sum$}}
\newcommand{\comment}[1]{}
\renewcommand{\tilde}{\widetilde}
\renewcommand{\hat}{\widehat}
\renewcommand{\V}{\mathbb{V}}
\renewcommand{\F}{\mathbb{F}}
\newcommand{\dagx}{\hbox{\tiny\mathversion{bold}$\dag$}}
\newcommand{\ddagx}{\hbox{\tiny\mathversion{bold}$\ddag$}}
\newtheorem{conjecture}[theorem]{Conjecture}
\newcommand*\toeq{
\raisebox{-0.15 em}{\,\ensuremath{
\xrightarrow{\raisebox{-0.3 em}{\ensuremath{\sim}}}}\,}
}

\vskip -0.0cm
%\par
%{\centering
%\medskip
%\par}
%\vskip -0.0cm
\maketitle
\smallskip


{\em\small \ \ \ \ \ \ \ \ \ \ \  \ \ 
92C30, 93E35, 33D90, 91E10, 91E45, 68T27, 68T37   }

\vskip -0.5cm
\renewcommand{\baselinestretch}{1.2}
{\vbadness=10000\textmd
{\small \tableofcontents}
}
\renewcommand{\baselinestretch}{1.0}
%$\mathfrak{a,b,c,d,e,f,g,h,i,j,k,l,m,n,o,p,q,r,s,t,e^u,e^v,e^w}$
\vfill\eject

\renewcommand{\natural}{\wr}

\setcounter{section}{0}
\setcounter{equation}{0}
\section{\sc Introduction}

\vbadness=3000
\hbadness=3000

%\centerline{\bf  MOMENTUM APPROACH TO ARTIFICIAL RISK-TAKING}
%\centerline{Ivan Cherednik}
%\vskip 0.3cm 
\baselineskip 12pt
%\vfill

\subsection{\bf Objectives and tools}
We propose a new theory of momentum 
risk-taking, which is basically real-time risk management, one 
of the key components of general decision-making. We focus 
on momentum risk-taking, {\it MRT}, when our decisions must be fast 
and mostly short-term. This is a development of 
"thinking fast" from \cite{Ka}.  Stock markets are the key to us;
a new approach to short-term volatility and high 
frequency trading is the main theoretical result of this paper. 
Its implementation is a momentum trading system, which was
extensively tested in stock markets, including real-time trading.
The discussion of its performance is an important part of the
paper. Stock markets provide a unique opportunity to test
our theory, but the core mechanisms of {\it MRT} 
seem quite universal, well beyond investing. We will argue that 
{\it MRT\,} is a major purpose of any intelligence (not only 
with humans). Our results indicate
that modeling such mechanisms is within the reach of artificial 
intelligence systems; they can be natural "ends" 
and also indispensable research "means", as we try to demonstrate. 
\vfil

We propose a new continuous mathematical 
model of news impact on share prices, and then describe   
its "stratified discretization", necessary to deal with
discontinuous functions. The news impact of a single
event is basically  $\,t^r\,$ in terms of time $\,t\,$ 
with fractional powers (exponents) $r$
multiplied by proper functions in the form
$\cos(A\log(t)+Bt)$. The $\log(t)$-periodicity here resembles 
that of Elliot waves. The $t$-periodicity is related to profit
taking, which we associate with the asymptotic periodicity
of Bessel functions. Mathematical understanding profit-taking 
is of obvious importance in the theory of market volatility; 
see e.g. \cite{ACH,EN,FL,FPSS} concerning the latter.
Hypergeometric functions serve the case of two 
events, which is connected with certain types
of hedging. Generally basic (difference) hypergeometric 
functions appear here, but we stick to continuous models
in this part of the paper.

\vskip 0.2cm
As any theory, our one must be checked experimentally.
Stock charts are the main examples for us; stock markets are 
quite a test for any risk-management theories. An obvious problem
is that such charts are discontinuous,
so the differential equations must be replaced by difference
ones. Novel approaches to the discretization appeared
necessary; cf. \cite{ChS}.  We restrict ourselves with relatively 
short time periods after the event; high volatility right after 
the news is mostly avoided. Then the core of 
our approach is the usage of tables of bids, which is based on 
ranked collections of sample time-forecasts for different 
time-horizons. 
 
\vskip 0.1cm 
Given a chart, these tables provide a short-term  
prediction of its evolution, which 
involves the prior behavior in some "non-linear" way;
cf. \cite{Gu}.
Forecasting here is not on the basis of derivatives  
of charts or their difference counterparts,
though they are of course employed. The usage of 
"global factors" and prior experience is the key for us,
which is no different from the way our brain works.  
 
\vskip 0.2cm 
Our tables are actually similar to 
bidding tables 
in contract card games, though the role of time,
the non-linearity of our tables, and some other features
have no counterparts with cards. 
In the realm of stock markets, the tables determine
optimal horizons, expected durations of the investments,
and also provides the corresponding short-term forecasts 
for the share-prices. We think that our brain "employs" 
similar auction-type procedures when
risk-taking, so such {\it bidding tables\,} are universal
well beyond playing cards and trading stocks.  

%\vfil
A usual approach to understanding the ways our brain 
works is via carefully designed 
experiments, which are mostly focused on very specific, basic,
simplified and sometimes artificial tasks.
However, the simpler the challenges the more special and
primitive tools our brain invokes. So laboratory  experiments
can generally clarify only very basic features; 
they are games in a sense. With any game, our brain readily 
switches to the corresponding optimal thinking mode, at least 
upon some training; we are very good with this.  So the experiments
mostly measure our ways to play this particular game. 
The risks must be as real 
as possible to force our brain to use its full potential,
which is hardly possible in experiments. Real behavior of 
people is difficult to recreate in artificially 
designed situations, even well crafted.  
 
It seems that the most promising, if not the only,
rigorous  approach to understanding  risk-taking
and other processes of this kind is to do our
best with creating artificial intelligence systems and
then comparing their decisions in real situations with
those of people. Of course the "super-aim" here is 
to reach some "superhuman" levels, but even any
"simple" reproduction of our real behavior is a breakthrough.
\vfil
  
The automated momentum trading system based on our approach 
can be seen as a step in this direction; it is
discussed in the second part of this paper.
Its preimage 
is a new contract card game presented at the end. 
By design, this system uses only share-prices;
i.e. it operates only on the basis of the
technical analysis. So it is inevitably "late" 
with any decisions vs. professional traders and investors,
and is subject to the bid-ask spread and many other 
factors reducing profitability. In spite of
such disadvantages, the system proved to be
profitable, which is some justification of our approach. 
The success of any AI system
can be of course only an {\it indirect\,}
confirmation of its principles.

\vskip 0.2cm
We discuss the main features of our trading
system in the paper and provide some typical results of 
its performance.
Designing historic experiments was a serious consideration,
to prevent the usage of any kind of "future". 
At least, this is impossible
with real-time trading, where the system was tested systematically
(with about 1000 companies). The results we provide can supply
those who will implement our approach and the tables practically
with some benchmarks. We think that the {\it pont-tables\,}
from Section \ref{sect:PP} can significantly
help to understand and get used to the
{\it 2-bid tables\,} from Section \ref{sect:2bids}. 

%\vfil
Importantly, we can always "explain", interpret to be
exact,  the trades  
our system makes; our system is not a black box and
can shed light on the risk-taking preferences of the traders.
This provides
some {\it quantitative\,} model of  "thinking fast" from
\cite{Ka}. Traders must quickly react to many unknown factors, which
creates some special
{\it market intuition\,} of obvious interest to 
cognitive theory and behavioral finance. 

\subsection{\bf Organization of the paper}
In the current section, we describe our approach and 
discuss its general origins, including risk-management
and cognitive theory. {\it Momentum risk-taking\,}
is essentially short-term forecasting based on whatever 
information available (frequently incomplete). We demonstrate that
it can be modeled mathematically. Due to our focus on 
professional trading, we can disregard 
the {\it expected utility hypothesis\,} 
originated by Daniel Bernoulli, the asymmetry
between loss and gain from {\it
prospective theory}, and similar. The market agents are 
assumed to act {\it "rationally"}
on the basis of the current news impact, 
so the purpose of our AI system is to capture their preferences.  

\vskip 0.1cm
{\it Section} \ref{sec:2}
is on our mathematical model of market news
impact, based on systems of differential equations 
resulting in Bessel functions, hypergeometric functions, and their
degenerations. These systems are closely connected with 
new tools in harmonic analysis and random processes. 
As a  demonstration of the universality of our system, 
we show that it models well the tree growth in a 
{\it difference setting\,}. 

The connection with {\it fractional
Brownian motion\,} ({\it fBM}) is briefly
discussed at the end of Section \ref{sec:regimes}.
Our {\it price-functions\,} are related to the
standard deviations and
{\it transition probability densities\,}
 of the corresponding processes,
which provides a  statistical framework for our approach.
See  \cite{Che, GJR, GNR} and \cite{Bo} concerning 
{\it fBM} in  studying market volatility
and  {\it power-laws\,} for price functions. 

Using a single {\it fBM} with small {\it Hurst exponent\,} 
as a model for a price function creates a theoretical problem 
with the existence of {\it arbitrage} (some kind of
"free lunch") due to the negatively correlated increments. However, 
{\it mixed fBM} are arbitrage free, and anyway we are doing
only relatively short time-intervals, where this concern is not
quite relevant. The author thanks Patrick Cheridito for a discussion.

We mostly consider the impact of one-two events. 
Statistical ensembles of 
news are mathematically significantly more
challenging. The corresponding stochastic processes are similar to 
those in \cite{BC}. Also, our trading system provides
some experimental support for our approach to 
modeling the impacts of {\it isolated\,} consecutive events. 
The "multi-dimensional" theory of ensembles of events seems
difficult to check experimentally  and use practically. 

\vskip 0.1cm
{\it Section} \ref{sec:3} contains a reasonably complete 
description of the algorithms of our {\it trading
system}. The results of extensive experiments, including
real-time trading serve two purposes. {\it First,} we provide
evidence for power-laws for price functions with exponents
depending on the investment horizons (say, can be $0.137$
for day-trading).
{\it Second,} we provide some
performance {\it benchmarks} for those who will follow our 
approach in their own trading systems. Our system has many new
features, including the simultaneous running its  multiple 
variants (sometimes even with identical {\it opti-parameters}
but with different entry points), simultaneous
pro-trend and contra-trend trading, the usage of the results of 
optimization for creating  {\it weights\,} of companies, and so on. 
Potential followers must know what to expect, theoretically
and practically. We also explain how testing the system
was performed.

\vskip 0.2cm
We do not discuss much the {\it machine learning\,} procedures
we employ, namely the optimization of parameters and
creating the company weights. The discretization parameters,
counterparts of {\it action potentials for neurons}, are the
key for us, but there are also important {\it difference}
counterparts of first and second derivatives of the charts we use 
for
forecasting and trading. There is vast {\it ML\,}
literature on entropy, 
information theory, Bayesian predictive method and 
generative adversarial networks, {\it GAN}.
The latter approach is somewhat similar to our auction-type 
procedure, when different decision-making "bids" from  
different investment horizons contest with each other; 
cf. \cite{DB,HS}. See
also \cite{SC} about some general perspectives
of {\it deep learning}. 

\vskip 0.1cm
Since we deal with a limited number
of {\it opti-parameters\,} (all of them have theoretical meaning),
a relatively straightforward {\it gradient method\,}
is mostly used for the optimization. It is rare when our AI system 
cannot find the parameters providing a {\it solid} jump in performance 
for almost any "education periods", though their 
uniqueness is of course not granted. This is for individual
companies or portfolios. The weights of the companies we use
are based on the prior optimization. We
omit the discussion of the usage of correlations between equities in
this paper, which is common in automated investing systems.
  The impact of our optimization-based
weights can be significant, but using them generally
restricts the trading volumes, which is a consideration for us. 

\vfil
\vskip 0.1cm
{\it Section} \ref{sec:4} is exceptional. We motivate
our {\it 2-bids} by designing a contract card game, 
{\it pont}, 
combining the elements of {\it bridge} and {\it poker}.
It adds poker-style uncertainty to the bridge-type 
auction. The contract is declared on the basis of 6
cards, but the hand can consist up to 9, which is determined
by the declarer, the winer of the auction. Thus we add
poker risk-taking to bridge-type bids, which is 
similar to {\it 2-bids} in our trading system.
It appeared that the players can easily get used to such
"fractional bids"; the size of the hand is the denominator.
It is closely connected with our approach to discretization
(the key in any neural networking); our {\it 2-bids}
are discrete, though the threshold is subject to optimization.
The play (taking tricks) has nothing to do with
stock markets: {\it pont\,}  is just a game. However
the play makes the bidding sufficiently 
non-trivial to consider {\it pont\,} as a model of real risk-taking; 
this is missing in poker, where there is no actual "play".

%\vfil
\subsection{\bf AI and risk-taking}
The purpose of artificial intelligence (AI) systems is to
perform tasks that require human cognition. Actually, the aim  
here is to {\it exceed\,} human decision-making abilities using
computers and machine learning at full potential. 
Even if the quality of automated decisions is mediocre, the
cost efficiency, speed and the broad range of applications
can be "superhuman" and result in  great societal and economic 
benefits. There is a lot  of progress with {\it narrow AI\,},   
focusing at special tasks. 
However we are decades away from  {\it general purpose
AI\,} according to the conclusion from
"The National Artificial Intelligence Research and 
Development Strategic Plan (2019 update)" by the
National Science \& Technology Council (USA). 
Astonishing versality and flexibility of human intelligence
remains quite a challenge, and not only because our brain
contains about 100 billion of neurons.  

\vfil
{\it Decision-making\,} is the key test for any AI systems.
This is quite a complex process. {\it Risk management} is one 
of its important components, which is generally a system of 
protection measures aimed at reducing future risks. 
Here the focus is mostly on general adjustments, 
not on exact timing.  There is a direct analogy with 
predicting earthquakes; even if potential places are  
known, we do not know when, especially in advance. See e.g. 
\cite{FL} for various aspects of risk management, including
high-frequency trading, and \cite{EN} for mathematical aspects.
\vfil

{\it Momentum risk-taking\,} can be then broadly defined as 
{\it real-time\,} 
risk management, prompt responses to  
events and developments, short-term forecasting.
Here timing is the key, though a lot of 
prior knowledge and experience is needed; see e.g. \cite{EN}.
 The events we are reacting to are mostly not 
really new; almost always they have occurred before. 
The problem is to address quickly their 
strength and other factors involved. 
Real-time monitoring the developments before and
after our actions is an important part of risk-taking.
The response can be required immediately, so
it can be difficult to understand what really
affected our decision. 
{\it Thinking fast, intuition, subconscious processes\,} 
are certainly involved; this can be not too transcendental,
a special mode of our brain to quickly 
manage time-sensitive information. 
\vfil
 
If the subconscious processing the signals is essentially 
similar to the usual (rational) one, then risk-taking AI 
systems can be quite relevant. 
Moreover, if this is true, then
using AI can help a lot to understand which
kind of  "thinking fast" and "intuition" is involved here;
this alone is quite
a motivation of the present paper and our project, 
without any reference to stock markets. 
One of our main observations  (based on machine modeling) is 
that {\it core risk-taking} is actually controlled by very few
parameters. Moreover, these parameters seem to be of 
universal nature, 
though  they are obviously adjusted to 
serve concrete situations. 

Both findings can be seen in stock markets. As some
confirmation of broad nature of risk-taking, 
the results based on the optimization of individual companies 
are {\it only modestly\,} better
than the results based on the optimization performed for  the groups,
"portfolios", of companies. Generally, the greater variety 
of different risk-taking tasks you went through, games included, 
the greater your risk-taking skills. This sounds quite obvious, but
is very difficult to implement in any automated systems;
developing  {\it general purpose\,}
artificial intelligence systems is needed here,
not just those focused on specific tasks. 
%\vskip 0.2cm
%\vfil

\subsection{\bf Universality of MRT}
Let us try to outline which {\it  minimal tools\,} seem
necessary for any risk-taking. This will be not a 
biological (neuroscientific) or philosophical discussion. 
We mostly
rely on the fundamental mathematical nature of the corresponding
differential equations. Our brain does many things; 
the {\it decision-making\,} is one of the key for our (any)
intelligence. Many aspects of our intellectual activities has no 
or almost no clear connection with {\it risk-management}, however 
the origins of quite a few of them still have something to do with 
decision-making. Science is an obvious example. Even if some
research directions are focused on highly abstract objects, the
value and motivation of such research can be not
just our curiosity or sense of fulfillment; almost any research 
is somehow related to decision-making, at least historically. 
Also, many aspects of our intellectual life
that seem or are "artificial", say the games we design 
and play, serve a clear purpose of developing and training
our decision-making 
abilities (social skills included), including preparing 
us for real risk-taking.
\vskip 0.2cm 

With this pragmatic understanding of intelligence,
there must be no significant difference 
between humans, other creatures, and artificial systems. 
Here {\it long-term\,} risk-management is quite different. 
Unconditional instincts and reflexes are generally
the key for such management in nature. Our ability to 
"rationally" plan long-term is of course a special
and cultural phenomenon. This requires at least a highly sophisticated 
concept of time, among many other intellectual abilities unique 
for humans. Generally, anything about long-term forecasting
and planning seems beyond what we can hope to model 
mathematically, especially for the processes with
high uncertainty. 

\vskip 0.2cm
However, upon restricting ourselves with relatively 
short intervals
and with {\it momentum risk-taking,\,} for short {\it MRT\,}, 
mathematical modeling  
seems much more doable. Almost any creatures have 
some basic concepts of time in this range, sometimes at the 
level of chemical and physical processes. 
{\it Thinking\,} plays significantly greater role for short-term 
forecasting and performing the corresponding actions vs. 
instincts and reflexes; this is not only for humans.
Though some special {\it MRT}-type
intuition is of course heavily involved, humans included.

\vfil
It is not impossible that the core mechanisms
of  {\it MRT\,} 
can be observed in the "neural
architecture" of our brain. One of the main mechanisms 
is the well-studied
{\it action potential\,}; its counterparts are the key for the
discretization in this paper. The action-type procedures
between different options are present too. We claim
that there must be at least one more important component here.
Mathematically, any decision making requires some 
{\it price-functions}. Such "functions" must be present somehow, 
but it can be difficult to understand how such functions can
be practically formed  and "stored" in our brain. 

\vskip 0.1cm
Recall that {\it MRT} is news-driven short-term forecasting and
the corresponding risk-taking. Our brain does a lot of things.
For instance, about 50 percent of the cortex is doing vision.
Our model is naturally focused only on  
{\it newly emerged\,} events. 
After  {\it news\,}  is determined
(which is quite a process), the brain must do its initial 
classification and invoke the corresponding
{\it weight or rank\,} of this news within the current task, 
which is basically the expected significance.
The weight is of course based on the prior experience, 
learning included;
having it somehow is a must for any decision-making systems.

This is actually the key. We demonstrate that the 
differential (or difference) equations for the 
{\it news-function\,} alone 
is mathematically insufficient without adding the 
{\it price-function}.
Only with these two functions and their interaction, the
corresponding mathematical model becomes satisfactory.
The {\it news-function\,} basically measures the resources (the
number of neurons) currently involved in the analysis
of a particular event. The {\it price-function} measures
an expected importance of this particular event vs. other events and
the corresponding {\it expected\,} brain resources needed for
its analysis. The latter depends on the news-function: the
greater the number of neurons currently involved in the analysis of
the news, the greater its potential importance. The initial
"price of news" will increase when the news generates the
neural activities greater than it was expected. Vice-versa,
when the current number of neurons involved approaches some expected
levels (provided by the price-function),
the news "fades". Its impact can still continue
to grow, as well as the price-function, but the brain will
then attempt to reduce the resources used for its analysis.
\vskip 0.1cm

This sort of interaction is essentially the
system of differential equation we suggest, which can be
generally solved in terms of power-type functions
and Bessel functions.  
Accordingly, the prediction is that the {\it price-function}
must be somehow formed, constantly updated, and
stored in the brain to be used
again and again when the events of this kind emerge. 
We claim that such 2-function "interaction", {\it the actually
needed resources vs. those expected to be used\,}, is 
necessary for any kind of {\it MRT}. This is of course a
mathematical prediction, but the simplicity and
fundamental nature of the corresponding differential and
difference equations is a strong argument in its favor.
Actually, these equations are relatively new, though with strong
connection with classical special functions. 
{\it Nonsymmetric\,} Bessel functions, Dunkl operators and
other latest tools in 
harmonic analysis are involved here.
\vskip 0.1cm
\vfil

{\sf An example: driving}. Brain activities while driving a
car is a convenient example of "general" {\it MRT}. 
%It has little
%to do in common with trading, but the principles are actually
%very similar. 
Permanent basic visual information is a must here, and even
more such and similar information can be "requested"
to provide for {\it MRT}. This is
very resource consuming, much greater than {\it MRT} itself. 
%comparing with relatively small
%special division of our brain focused on momentum risk-taking.
Also, as with many tasks, driving can be combined with talking, 
listening radio, thinking about something etc.

\vskip 0.2cm
The actual beginning of {\it MRT\,} is
when our brain identifies {\it events}. 
They can be some turns or road curves, 
especially those requiring special attention, 
road signs, pedestrians, neighboring cars, navigation matters
and so on; see e.g. {\tt arXiv:1711.06976},
{\tt arXiv:1906.02939} on {\it self-driving cars}.
Importantly, such {\it categories\,} of potential events are 
supposed to be analyzed constantly and simultaneously, even if 
the next {\it news\,} is or will be in a particular 
category. There is an almost exact analogy with trading stocks,
especially when they are treated independently, the main
regime for our trading system:\, {\it all\,} stocks 
considered for potential investing and investment horizons
must be constantly monitored, regardless of the current
or expected positions.  

%This is the third level.
\vskip 0.1cm
Only some {\it events\,} will reach in our brain the level of
{\it signals}. The separation of the signals from {\it noise\,} 
is quite a problem, which requires a lot of prior experience. 
By {\it noise\,}, we mean "insignificant events", those that 
hardly require special consideration.  After the signals are
determined, our brain is supposed to estimate the resources 
needed for the analysis of the signal. 
This stage requires invoking from the memory
the latest {\it price\,} of the event,
which is essentially an estimate for the {\it expected\,} 
brain resources (neurons) needed for its analysis.
\vskip 0.2cm

Then a systematic {\it analysis\,} begins, which can trigger 
the number
of neurons significantly different from what was initially
"allocated". This
can be because of unexpected complexity of the event, 
due to changing its priority for driving and so on. 
The greater the neural activity,
the greater the growths of the corresponding 
{\it price-function}. However, when the activity becomes beyond
the projected {\it price-levels\,}, our brain will attempt
to reduce the number of neurons involved. 

The risk-taking is this {\it analysis\,} and the
corresponding driving maneuvers (when needed). 
The rank (category) of the news and the intensity of its impact  
must be high enough for this "action",
which is similar to the usage of our {\it 2-bids\,}. 
Our main assumption is that 
{\it short-term news impact grows polynomially with some 
fractional exponent}. This provides {\it predictions\,} for 
the intensity of the corresponding brain activities, which can be 
used to end or restrict our analysis when they drop below some 
{\it termination curves}. Similar to trading stocks,
this can help to optimize the usage of the brain resources
by switching to new events  when the importance of the current
ones begins to "fade".  

%\vskip 0.3cm

%\vfil
\subsection{\bf Games as concepts} 
AI systems not always follow the ways of our brain,
even if the problems are human-related. 
However  {\it nature\,}, our brain included,
is definitely the prime source of concepts for any AI.
Just to give an example, the airplanes are very  
different from birds, but the {\it concept} of 
flying is from nature. This is no different for AI. 
{\it Narrow AI systems\,}, in specialized well-defined domains,
can sometimes follow "non-human", ways. However general
AI systems are expected to borrow a lot from
human intelligence, though the final implementations can 
be quite different: "aircrafts vs. birds". 

\vskip 0.2cm
Importantly, many faces of decision-making are
reflected in the {\it games} we play. Some include timing,
some do not. For instance,  solving puzzles and playing chess 
are not focused on timing, unless in tournaments. 
On the other hand, poker and
contract card games are time sensitive. The interaction
and risks in card games are as close as possible to real life,
for models of course, which they are.   Investing 
is obviously closer to playing poker than to playing
chess or bridge. Poker's bidding is a great model of 
dealing with uncertainties, but the risks are too "mathematical"
and the actual "play" is missing. Solid rules and protocols make 
stock market some kind of a game, but here the risks are more 
than real. From this perspective, it provides a highly 
developed  and quite universal "model" of risk-taking,
which is of obvious interest.  

\vskip 0.2cm
Psychologically, games 
reflect life in various ways, 
potentially preparing us for real challenges; we
naturally discuss only "intellectual games". 
Some are designed to deal with real tasks; playing
them can be more dangerous than life itself.
Using game theory, especially mean field games, is quite common
in financial mathematics; see e.g. \cite{GLL}. 
From this perspective, one can look for the games reflecting
our {\it concept\,} of momentum risk-taking. We found none 
and invented
a new one, {\it pont\,}, which is essentially
a version of bridge with poker-style bidding.
%\vfil

Stock markets by design mean that their "agents" look only 
to their own interests and to market prices \cite{GLL},
though investing is a complex and very much 
interactive process with 
solid grounds in our psychology. Nevertheless investing
is a great confirmation of the universality of 
momentum risk-taking, {\it MRT\,}. It looks like there is some 
general purpose risk-taking {\it source
code\,} in our brain in charge of 
{\em all\,} kinds of {\it momentum\,} risks, which constantly
improves itself whatever the nature of the risks and 
uncertainty. 
If this is true, then we can try to 
use AI to understand this code and to model it! 

Philosophically, we test
here Kant's antinomy 2 ({\it atomism}), by considering 
risk-taking as composite substance, and his
antinomy 3 ({\it causal determinism})
concerning the variability of decision-making. 
We study stock markets
as "an end in themselves", disregarding their 
economic and societal purpose for the sake of 
mathematical modeling.  

%The philosopher Immanuel Kant said that rational human beings 
%should be treated as an end in themselves and not as a means 
%to something else. The fact that we are human has value in itself. 

%\vfil
\subsection{\bf Momentum investing}
{\ }
\vskip 0.1cm

\noindent
{\sf Marshmallow test.} 
A well-known test 
for children, "one marshmallow now or two in 15 minutes",
is actually one of the origins of our modeling risk-taking
in stock markets. The latest 
psychological experiments found limited support to the 
thesis that {\it delayed 
gratification\,} with children 
leads to better outcomes in their future  \cite{WDQ}.
Waiting for 15 minutes here can be expected for those who
have already learned that "patience is rewarded", but
not for all and not always. 
Also, the 15 minute interval can be not that short
for little ones. Their impatience can depend on the age, 
social and economic background; some can simply 
favor  short-term approaches.  If the 
interval increases (say, days),
some uncertainty is added, or the reward
is diminished, the "impatience" can be well justified;
the problem is "how much?". 

\vskip 0.2cm
Similar  to \cite{Ka}, we began with some analysis of  
psychological roots. 
A starting point of our research was  a postulate that we 
have quite
a rigid "table" of risk preferences in our brain. 
To give the simplest example, if the return was
1\% today and you count on extra 1\% tomorrow, then 
do not sell. However, if only a fraction
is expected tomorrow and even this is not granted, 
then sell now and avoid extra risks.

\vskip 0.2cm
This is very basic; and
obviously the interval matters here. What 
justifies risks for a day-trader can be
not acceptable to the traders with greater investment horizons.
Of course, the better someone's trading experience
the more realistic expectations, and many other
factors are involved. However 
the mechanism in our brain that determines acceptable
risks  seems quite universal to us. To give a model
example: 
1.5 marshmallow tomorrow vs. 1 now "sounds reasonable". 
Indeed, the offer "2 tomorrow" is quite acceptable (for adults), 
but maybe just too good; however, receiving tomorrow only 1 makes no 
sense, since we can have it right now. So our brain 
possibly takes the average here,
which makes 1.5 tomorrow a reasonable compensation for the
delay.

\vskip 0.2cm
The auction in {\it pont\,}  does almost
exactly this. For instance, the smallest bid during the auction
is $3/6$ (3 from 6), which means that you are obliged to take 
3 tricks with 6 card (your initial hand), or,
upon the "increases", 4 from 7-8, or 5 from 9  
(requesting up to 3
additional cards). So the {\it pont-bids are actually
fractional\,}; the next bid is $4/7$, where your contract (if you
win the auction) must be $4/6, 4/7, 5/8$ or $6/9$. The number of
cards in your hand and the number of taken tricks reflect 
respectively the duration of the investment and the return.

%\vfil
The play itself is of course {\it not} market-related; this is 
simply a way to validate your bid.  In real 
investing, the "contract" means opening a position and the 
"play" is finding the moment of its termination. The resulting
return is similar to the value of the contract.
There are many successful
ways to invest; picking one of them 
resembles very much bidding in card games, but
real "timing" is  not well reflected in card games.
{\it Pont\,} somehow addresses this;
it is a model of our approach to
comparing returns for different durations of positions. 
Interesting mathematics is involved here, including 
Bessel and hypergeometric functions. This is
not like "comparing apples and oranges", though 
our brain routinely compares everything with everything.

There are of course stock market 
features well beyond {\it pont\,}. 
For instance, the execution risks are 
connected with the investment risks
\cite{EF}. Also, opening short positions and 
terminations of long ones are based on the same {\it sell 
signals\,}, so they are related. Mathematically, 
{\it pont-bids\,}
are linear (our tables are nonlinear), but it captures our 
method well.  
\vskip 0.3cm

{\sf Market implementation.}
The termination rules we use are based on {\it termination
curves\,}. These curves are directly linked to forecasting
share prices. The hierarchy of basic {\it pont-bids\,} 
discussed above
is such a curve (with 4 points):
3 tricks from 6 cards,  4 tricks from 7 or 8 cards,
and 5 from 9. 
As with cards, the discretization of 
stock market bids is necessary for
our trading system to work. The separation of
the signals from noise,
which we do successfully, absolutely requires such a discretization.
Stock market charts are 
discontinuous functions by their nature, especially short-term.
Also, the discretization of bids is 
closely related to the
discretization of time, which is inevitable
for finding the optimal time range of investments 
(in hours, days, weeks). 

\vfil
We will argue below that the prediction and termination
curves are of type $\,const\cdot t^r$, 
where $t$ is time and $r$ is
some fraction (generally, below $1/2$). This assumption
matches well {\it momentum investing\,}, which can be
defined as "investing on news". It works well
for individual companies,
portfolios of companies, market indexes, including\, SPY,
the spider, and for commodities; it seems to us of quite
general nature.
\vfil

We note that the optimization 
becomes significantly more 
involved for our trading system
when {\it strict hedging\,} was imposed, i.e. when for
any open position, an equal
amount is invested in the
opposite direction in \, SPY\, or similar; see \cite{BLSZ}.
Theoretically, hypergeometric functions are needed here;
using $\{t^r\}$ becomes too approximate.
The system works, but the returns are 
less impressive. More generally, the correlations between companies
are important; this is beyond the system we present, 
though we do
the group optimization.

In our approach, we do not even try
to evaluate the news itself. Its impact is measured through
the response of the markets via stock prices and trading volumes.
Thus, the parameters we find and use actually reflect 
{\it investor risk-taking preferences\,}, 
which can be expected sufficiently stable.  The 
trading frequency is one of the main 
factors here; see for instance \cite{Al,ChS,CS}. The risk 
preferences of day-traders are quite 
different from those  of mutual funds. The challenge is that
a stock
can be involved in trading with different frequencies and
horizons, which was addressed in our {\it bidding tables}.
 This is especially
applicable to  trading indices; see e.g. \cite{FPSS,GTW};
all kinds of trading patterns 
can be present here, but our system mostly manages them well.
See \cite{Bo,DB} on using typical time scales.

The design of our trading systems included many
special market twists. 
For instance, the counter-trend (contrarian) 
variants of our trading system frequently outperform pro-trend 
ones. We actually 
used both variants simultaneously, which is some kind of hedging.
Contra-trend trading can be successful because of several
reasons. Our system needs time to measure the impact of 
the news to be sure that this is  not "noise";
large trade sizes are a consideration too (see \cite{GRS}).
Counter-trend trading is not unusual in stock markets \cite{CK}.
Let us mention here that the initial version of our trading
system mostly relied 
on the intersections of termination curves with actual charts,
changing the directions of positions correspondingly. 
It worked reasonably, but reacted slowly to fast
market moves, which was improved via "start 2-bids",
where we used the same curves to produce signals for opening
new positions; this complimented well using the intersections.
%\vskip 0.1cm

%Quite a few of the features of our system 
%are actually of general nature (i.e. beyond stock markets). 
To trade real-time, our system was 
designed completely automated, a must for any AI even if they
are used interactively. See \cite{CJP}
concerning various aspects of automated
high-frequency trading. We note
that the trades of our system  
are  fully {\it explainable}; it is not a "black box".
Only such AI can be
really {\it trustworthy}; see e.g. \cite{HG}.

\vskip 0.3cm
{\bf Acknowledgements.} I am very thankful to David
Kazhdan, who greatly contributed to the success of this
project at many levels. The trading system discussed in the paper
was tested (improved
many times) thanks to support and supervision by Alexander
Sidorenko; I am very grateful to him. The author
thanks Mikhail Khovanov for various suggestions and
Jean-Pierre Fouque for his kind interest.

\vskip 0.3cm

\setcounter{equation}{0}
\section{\sc Modeling news impact} \label{sec:2}

In this section, a simple
mathematical model of short-term impact of
news is suggested. News-driven fluctuations of share prices
are the core examples. We come to certain
linear differential equations, which can be generally
solved in terms of
hypergeometric functions.
We focus on elementary solutions only. 
They have market applications, which 
were extensively tested in various stock markets,
including real-time experiments. There is another way to
obtain essentially the same equations via random processes, but
it will not be touched upon in this paper.  See \cite{EN}; e.g.
compare their News Impact Curves with our ones.

\subsection{\bf Hierarchy of news} Let us briefly describe 
the types of company or industry news, which can be {\it primary}
and {\it secondary}.  
 The primary ones are basically
{\it core} events and announcements. For instance, 
(a) new products or
acquisitions, (b) significant changes of earning
estimates by the company,
(c) upgrades or downgrades by leading market analysts. 
Major sector, industry or
economy news are of this kind too. 

Almost any
core news generates a flow of secondary news in the form of
(highly correlated) reports, reviews, and commentaries.
They mostly present the same core news, but sometimes
can  impact our behavior even greater than the original
event. In our model, commentaries will be generally treated 
on equal grounds with the core announcements.

By {\it reports}, we mean analysts'  
reports on the core event including perspectives and
predictions. 
Then {\it reviews} collect and present  the 
main findings in reports, mostly aiming
at professional investors.
Finally, the news itself and the findings above
reach all consumers via mass media
mostly in the form of  {\it commentaries}. 

Importantly, consumers will be influenced by all 
primary and secondary news  more or less 
regardless of the level, the "distance" from the actual event.
The actual originality is not the point here.
So the impact of the commentaries can be significant and quite 
comparable
with the impact of the event itself.

\subsubsection{\sf The basic equation}
We assume that 
{\it the impact of an
event at the moment $t$ is proportional to the $t$-derivative
of the total number of pieces
of news reflecting the event after it and before $t$.} 
The coefficient  of proportionality $0<c\le 1$ will be called the 
{\it reduction coefficient\,}; 
it depends on time, but mostly it will be treated as a constant. 

The value $c=1$ can be reached right after
the news, and then $c$ tends to $0$ with time, depending
on the "investment horizon" (hours, days, months); 
cf. \cite{DB}. Let us comment on this. Generally, 
 
{\bf (i)} analytic reports and all secondary news tend
to soften the expected implications of the core news,

{\bf (ii)} commentaries of all kinds disperse the 
original core news and diminish the expectations  even further, 

{\bf (iii)} the longer time passes after the core event and
the core news, the smaller their impact becomes.

\smallskip
All three mathematically mean that the coefficient
$c$ approaches zero as 
$t\to \infty.$ Indeed, putting news into perspective is the 
purpose of analysis and commentaries, but this almost always 
reduces the original expectations.
In the momentum investing, the impact of news
fades faster for short-term investing vs. long-term.
Approximately, if the trading positions are in days or weeks than 
$c\sim 1/2$ can be expected vs. $c=1$ for months; it can be
significantly smaller for high-frequency trading. Our tables provide
some "natural" $c$-coefficients for different trading
frequencies, investment {\it categories}. 
 
\vskip 0.2cm
From now on, {\it news will be represented by a 
positive or negative real number}, i.e. we assign a 
numerical value to it. Also, we assume that the 
time  distribution of
news is essentially uniform in the following sense.
\vskip 0.1cm

Let $N(t)$ be the total sum of news values (positive or
negative numbers) released from $0$ to moment $t.$
Then the number of pieces of news (their total value, to be exact) 
arriving from  $t$ to $t+\delta$ for some $\delta$, i.e.
$N(t+\delta)-N(t),$ equals approximately 
$c\cdot\delta\cdot N(t)/t,$ 
which is $\delta$ times the reduced average of {\it all\,} 
previous news 
from $0$ till $t.$ The greater the intensity (time-density) of
commentaries etc. triggered by an event, which is $N(t)/t$,  
the greater the number of {\em new\,} commentaries.
We come to the following differential equation:
\begin{align}\label{0}
&\frac{dN(t)}{ dt}\ =\ \frac{c}{ t}\, N(t).
\end{align}
It can be solved immediately if $c$ is a constant:
$N(t)=A\,t^c$ for a constant $A>0.$ When $c=1$ the growth
of $N(t)$ is linear, i.e. the event does not "fade" with time
and continues to attract {\it constant\,} attention.
We disregard that
$N(t)$ can be bounded; adding the "saturation" 
will be addressed later.
A physics-style
argument in favor of this equation is its {\it self-similarity}:
the solutions are
multiplied by some constants when the time units change, and
$c$ does not depend on the choice of units. 
%The value
%$c=1/2$ is connected with Brownian motion (random walk);
%we will discuss this below.

%{\comment
\vskip 0.2cm
{\sf Tree growth.} Equations 
of this type can be expected to have many applications.
Let us give one example. We will switch to a difference 
counterpart of (\ref{0}), naturally adding minimal "maturity":
\begin{align}\label{00}
&f_n-f_{n-1} =\ \frac{c}{n-2}\,f_{n-2} -\la f_{n-1}
\for  n>2, \ \la\ge 0.
\end{align}
When $\la=0$, this is a variant of
the famous {\it Fibonacci recurrence} 
with the birth rate $\frac{c}{n-2}$, i.e. when it is  
inversely proportional to "time". The term $-\la f_{n-1}$
restricts here
over-population by allowing "emigration".

Setting $\la=0, c\!=\!1,$ the fundamental solutions are:
$f^0_n=n,\  f^1_n=D_n/(n-1)!$, where $D_n$ is the number of
{\it $n$-derangements}, permutations of $n$ elements without
fixed elements. The second solution approaches $n/e$ as
$n\to \infty$, so both have linear growth at infinity.
We argue that $f_n$ in (\ref{00}) basically describes the size of 
a tree at its $n${\tiny th} year.  
\vskip 0.2cm

In contrast to the "Fibonacci rabbits", trees grow
linearly at most. The corresponding  $f_n-f_{n-1}$ is 
proportional to the corresponding $f_{n-2}$, where
the coefficient of proportionality, "the birth rate",
is roughly the surface area of the root
system divided by the volume of the tree. I.e. it is
approximately
$r^2/r^3=1/r$, where the "radius" of the tree $r$ can be assumed 
linearly depending on $n$. The radius is qualitatively about
the same for the tree and its root system. It is proportional 
to the number of {\it tree rings}; we obtain $\frac{c}{n-2}$.
This describes the "middle stage" of tree growth. 
In the beginning, the
volume of the tree is rather $r^2$ than $r^3$, so the tree can
grow exponentially for a short period of time. Also, only
the "active part" of its root system contributes to the growth,
which eventually diminishes $r^2$ to $r$ or so. Mathematically,
this gives the term $\frac{c}{(n-2)^2}f_{n-2}$ in (\ref{00})
and results in the saturation of the tree size at the late stage 
of its life cycle, which matches real tree growth.

This is somewhat parallel to the claim above 
that  the {\it reduction coefficient\,}
$\,c\,$ for $N(t)$ tends to $0$ when $t\to \infty$. We note here
that the saturation here can
be of course simply due to the upper bounds for
$N(t)$ or $f_n$, i.e. not because of the "geometric" argument 
we suggested above (via the root area divided by the
volume); see Section \ref{sec:log}.

There are obvious differences between {\it news impact}
and {\it tree growth}. For instance, adding $-\la f_{n-1}$
is secondary for trees (due to their aging or similar
growth reductions), but this 
term is of fundamental importance for the news. It reflects 
"pricing news"; see Section \ref{sec:prtarget}. Surprisingly
such different processes are quite similar mathematically, 
which clearly indicates that  (\ref{0}) and
(\ref{00}) are of  universal nature. 

Without going into detail, let us mention 
that solving (\ref{00}) and similar  difference Dunkl-type
equations 
generally requires {\em basic}
(difference) hypergeometric functions and their variants.
This is actually a relatively recent direction; 
see e.g. \cite{ChB}.
%} %%end of comment

%\vskip 0.3cm
\subsection{\bf Adding price targets} \label{sec:prtarget}
So far we have not considered the following
market-style response to news: when the news is already 
{\it priced in}, i.e. the current share-price already
includes it,  
the effect of further (secondary)  news goes down.
Similarly, when the stock is considered underpriced, 
positive commentaries have greater impact. There is a specific 
market way to address
this: {\it upgrades and downgrades}. They generally set
new {\it share-price targets}.
The main difference here from general
news is the dependence  on the 
current share-price. 
Generally,  {\it upgrades\,}, 
are {\it all} market, company or equity 
news of any levels  {\it addressing (depending on)
 the share-prices\,}.

Similar to $N(t)$, we represent {\it upgrades\,}
by positive
or negative numbers, using the notation  $U(t)$ instead of $N(t).$
Thus $U(t)$, the sum of values of upgrades, depends on the 
share-price. The following normalization
 $u(t)=U(t)/U(0)-1$ will be convenient below. 

Let $P_t$ be the share price and $p(t)=(P_t-P_0)/P_0$ be
the rate of return from the price-level $P_0$. The equation
above must be corrected for $u(t)$, since $U(t)$
goes down if the share-price "sufficiently" went up
 after the event, i.e. 
the news is already {\it priced in}. Similarly,
it goes up if the stock
is considered undervalued. This correction can be assumed
proportional to $p(t)/t,$ the average rate of change 
of $p(t)$ from $0$, which is more
"balanced" vs. taking $dp(t)/dt$ here.  Thus we 
arrive at the differential equation:

\begin{align}\label{1}
&\frac{du(t)}{ dt}\ =\ \frac{c}{ t}\, u(t)-\frac{1}{\si t}\ p(t). 
\end{align}

We note that the term $p(t)/t$ can be replaced
by "more aggressive" $p(t)t^{\nu -1}$ for $0<\nu\le 1$;
see system (\ref{16}\&\ref{17}) below. For instance 
it means for $c=0$:  
the longer $p(t)$ grows as 
$t^{\nu-1}$, the greater the number of downgrades. I.e.
$p(t)\approx \hbox{Const\,} t^{1-\nu}$ is considered 
non-sustainable.

\vskip 0.1cm
We will switch from now on from $N(t)$ to $u(t)$. 
Here $\si$ is qualitatively
proportional
to the $P/E$ or $P/S.$ More generally, it  reflects 
the expected growth  of the company. 
Mathematically, $\si$ is essentially as follows.

Let us assume that $p(t)$ is basically
linear in terms of $t$ and "shift" $t=0$ to the moment when 
the company is rated  "strong buy". 
For sufficiently large $t$, we can assume
that $u(t)\sim U(t)/U(0)$ and ignore $u(t)/t$; so
$u(t)\sim 1-p(t)/\si$ and 
$p(t_{max})=\si$ at $t=t_{max}$ such that $u(t)$, the 
current rating of the company, 
becomes $0$. This moment of time, $t_{max}$, 
is when analysts change their stock
ratings from "buy" to "neutral" 
on the basis of its price valuation.
So $\si$ is essentially the
{\it relative price-target,} i.e. $\si\sim 
p(t_{max})=(P_{targ}-P_0)/P_0,$ where $t_{max}$
is the moment of time when the news is fully priced in. 
We will make this analysis somewhat more rigorous 
in Section \ref{sec:log}.
\vskip 0.2cm

Now let us involve the 
differential equation for the share-price.
Almost no company event or news influences the 
share-price directly; this depends on the way the market reads
the news. The simplest  news-driven equation for $p(t)$ is as 
follows:
\begin{align}\label{2}
&\frac{dp(t)}{ \si dt}\ =\ \frac{a}{ t}\, u(t)+b\, 
\frac{du(t)}{dt}.
\end{align}
As with $N(t)$, here $u(t)/t$ is the {\it average upgrade\,} from 
the zero moment of time, which measures the {\it global\,}
news impact from $0$,
essentially the commonly used consensus rating  of the 
company shares. 
The term with $\frac{du(t)}{dt}$ is {\it local\,}:
the response to the
rate of change of $u(t)$ at $t$. 
 
\vskip 0.2cm
\subsection{\bf Logistic modification} \label{sec:log}
Before further analysis, let us touch upon 
the modification of equation (\ref{1}) under the assumption
that the number of upgrades or downgrades is limited.
Let $\widetilde U(t)$ be the sum of $\pm 1$ for
upgrades and downgrades, an integer. 
The relation with $U(t)$ is basically as follows:
$\widetilde U(t)=[U(t)]$ for the integer part $[x]$ of real $x.$
%Equation (\ref{1}) for  
%$\widetilde U(t)$ and
%$\widetilde u(t)=\widetilde U (t)/\widetilde U_{max},$
%becomes now {\it discrete\,}.

Since $\widetilde U(t)$ is bounded, let
$u(t)=U(t)/U_{top}<1$ for some bound $U_{top}$. Then 
(\ref{1})  must be modified 
if we want to use it for sufficiently
large $t$.  Namely, we must multiply
the right-hand side of (\ref{1}) by 
$(1-u(t)),$ which 
reflects the "number of  remaining commentators". One has:
\begin{align}\label{3}
&d\,u(t)/dt\ =\ (1-u(t))(
\frac{c}{ t}\,u(t)-
\frac{1}{ \si t}\ p(t)).
\end{align}
 
In the absence of the price-term, it is 
a well-known
{\it logistic equation}, with the following
modification:  the interaction coefficient
is proportional here to $1/t.$
When $p(t)\equiv 0$, it can be readily integrated.

Equation  (\ref{2}) remains unchanged:
\begin{align}\label{4}
&\frac{dp(t)}{ \si dt}\ =\ \frac{a}{ t}\, u(t)+
b\, \frac{d\, u(t)}{ dt}.
\end{align}

System (\ref{3})\&(\ref{4}) has no elementary solutions
for $a\neq 0$. 
Let us solve it when $a=0$, for {\it $b$-investing\,}
in the terminology below. One has:
\begin{align}\label{8}
&u(t)= (\be+Bt^{r-\be})/ (r+Bt^{r-\be}),\ r=c-b,\\
\label{9}
&p(t)=\si (b u(t)+\be),\ \ 0\le \be < r,\ \ B\ge 0. 
\end{align}
If  $B>0,$ then
$u(0)= \be/ r,\ p(0)=\si c\be/r,\ \
u(\infty)= 1,\ p(\infty) = \si (b+\be). 
$

Let us assume that $u(0)=0$, 
i.e. the rating of the company  is "neutral" at $t=0.$  
Then $\be=0,$ 
$\ p(0)=0$ and $p(\infty)=(P_{targ}-P_0)/P_0=\si b$.
%We come to the following re-interpretation of the
%coefficient $g$ under {\it logistic $b$-investing}.
So $\si$ is  $(P_{targ}-P_0)/(bP_0)$ for the 
price-target $P_{targ},$ 
which matches the interpretation
of $\si$ from Section \ref{sec:prtarget} for $b\sim 1$.

When $a\neq 0,$ the system can be solved {\it numerically},
but it is not clear whether the corresponding
solutions are more relevant than those obtained from
the original system (\ref{1})\&(\ref{2}). This is especially
true if we do not focus on large $t$,
and the simpler the better! Also, 
the stochastic and discontinuous nature of price fluctuations
restricts using differential equations here.  
Furthermore, $a,b,c,\si$  can depend
on time and do depend on the basic time-intervals,
which is another reason to stick to the simplest assumptions. 

Thus, we will continue with  system 
(\ref{1})\&(\ref{2}). Furthermore, 
to address the discontinuous and discrete nature of 
share-prices, we will later switch from this system   
to "tables" of its "basic solutions". The main conclusion
we will need from the analysis performed above is that 
$(P_t-P_{t_0})/P_{t_0}$ after the  news at $t_0$ can be 
assumed \,Const$\,(t-t_0)^r$ for some $r$ for short, but not
too short, time intervals $[t_0,t]$.

\vskip 0.3cm

\subsection{\bf Investing regimes}\label{sec:regimes}
Let us solve system (\ref{1})\& (\ref{2}). Recall that it
describes  fluctuations of share-prices 
under 
{\it news-driven investing}.
Both $a,b$ there are non-negative. 
The term $b\,du/dt$ in (\ref{2}) or
(\ref{4})  is typical for "local"
{\it pure momentum investing}, when only the
{\it latest\,} upgrades are taken into account.
The term $a\,u(t)/t$ reflects a more "global", balanced and
less "aggressive" approach, when the average of
{\it all\,} news values after the event is 
considered.  

We call the case $b=0$ pure 
{\it $a$-investing}, and the case $a=0$ 
pure {\it $b$-investing}.
If both terms are non-zero, it is naturally
 {\it mixed investing}.
The greater $t-t_0$ after the major event at $t_0$, 
the greater chances that $a$-investing dominates.

Equations (\ref{1}) and (\ref{2}) can be readily integrated.
Substituting $u(t)=t^r$,
the roots of the characteristic equation are:
\begin{align}\label{5}
&r_{1,2}=d\pm \sqrt{D},\ d=(c-b)/2,\  D=d^2-a.
\end{align}
Accordingly, unless $D=0$,  the formula for $p(t)$ is as follows:
\begin{align}\label{6}
&p(t)=C_1 t^{r_1}+C_2 t^{r_2}\, \hbox{\ if\ } D>0 
\ \, \hbox{for some constants\,}\ \, C_1, C_2,\\
&p(t)=t^{d}(C_1 \sin(\sqrt{-\!D}\log(t))\!+\!
C_2 \cos(\sqrt{-\!D}\log(t)))\, \hbox{\ if\ } D<0
\label{7}
\end{align}  

We will consider only $d>0$. For negative $d$,
$p(t)$ approaches zero for large $t$ and therefore this
is focused on  "the final stage" of the impact of an event;
our model and trading system are designed to serve mainly the 
beginning of this period. We also assume
that $c$ is a constant and that  $0\le c\le 1,$ so $d\le 1/2$.
In fact, $c$ slowly goes to zero as $t$ increases and the 
impact of the event gradually diminishes, but we will not
do large $t$.  Similarly, $c$ may be greater than $1$ right
after the major event, but this stage is disregarded
too; this is addressed in our trading systems 
by proper "discretization". 

\vskip 0.1cm
Let us briefly discuss the oscillatory regime 
in (\ref{7}).
It can happen only for $a$-investing or for the mixed one.
According to (\ref{7}), the {\it quasi-period} in terms of  
$\log(t)$ is $2\pi/\sqrt{-D}.$ 
So the durations of the oscillations 
form a geometric sequence. The magnitude  will grow 
in time as a power function of degree $0\le d\le 1/2.$
If $b=c$, then $d=0$ and  the function $p(t)$ is bounded. 

If the news is important for the share-price,
$a$ can be significantly larger than $d^2$. Then 
$\sqrt{-D}\sim \sqrt{a}$,
and the {\it quasi-period\,} 
for the {\it logarithmic time}  $\log(t)$ is about
 $2\pi/\sqrt{a},$
which clarifies the role of $a.$
%\vskip 0.2cm

Let $d=1/2$ for pure $a$-investing (when $b=0$). This gives  
$c=1$, i.e. the initial news-function $N(t)$ grows
linearly. Then the $p$-function behaves as a sum of  
random independent jumps of the share-price
by $\sigma$ or $-\sigma$ for proper $\si$,  "heads or tails",  
distributed uniformly. So our equations have some
statistical meaning; cf. \cite{Gu}.

For the pure $b$-investing with $b>0$:  
$p(t)=C_1t^{c-b}+C_2$ and its 
leading term is  $C_1t^{c-b}$, since $a=0$ and $c>b$.  
By the way, $p(0)$ may be non-zero here; for instance, we can 
set  $p(t)=(P_t-P_{t_0})/P_{t_0}$ for
any point $t_0$ in equation (\ref{2}).
\vskip 0.2cm

Now let us assume that $c>2b$, and let 
$\widehat b=0, \widehat a=b(c-b)$. Then the
corresponding $\widehat{D}$ is $(c/2-b)^2$, 
$\widehat{r}_{1,2}= c/2\pm (c/2-b)=\{c-b, b\},$ and  
$\widehat p(t)=\hat{C}_1t^{c-b}+\hat{C}_2t^b$.
Since $c-b>b,$ the leading term here coincides with that from
the previous formula. We conclude that for  $c>2b$, pure 
$b$-investing gives essentially the same as pure $a$-investing 
for proper $a$. This happens for sufficiently large $t$:
then $b$ eventually tends to zero.   

The difference between these two regimes
becomes  significant only when $b<c<2b,$ 
i.e. during the middle stage of the ``impact period''.
Indeed, the exponent  $r_1$ cannot be made smaller than $c/2$ for 
$a$-investing. As to $b$-investing, $r_1=c-b<b$ 
approaches zero when the news "fades" and the contribution
of $du/dt$ to $p$ can be disregarded.
\vskip 0.1cm
\vfil

{\sf Discussion.}
The leading exponent  $r$ in $t^r$, which is
$r_1\,$ in (\ref{6}) and $d\,$ in  (\ref{7}),
satisfies $d\le r\le 2d\,$ for $d=(c-b)/2$. Here $r\sim 2d$ 
occurs when 
$b$-investing dominates. The lower bound
 $r\sim d$ can be reached  only when $a$-investing
is strongly present. If $b=0$, then $r= c/2$ for
sufficiently large $a$. Recall that the news-reduction 
coefficient $c$
is generally from $0$ to $1$; it is close to $1$ 
when the initial news-functions $N(t)$ grows linearly.
Practically, the values $r< 0.5$ indicate short-term positions.
If $c=1$, then $r\sim 1$ only if $a\sim 0\sim b$. 
Each type of investing has its own natural
time-intervals, {\it prime time-units}, and its own
typical average durations of positions. The time-unit can
be from hours (or smaller) to months,\; it was  mostly $2h$ in
our trading system. Let us refer to \cite{Bo} on power-laws
for price functions, though are approach is
 different (we study short-term news impacts).
%We note that the relation
%$r=\log(P_3-P_0)-\log(P_1-P_0)$ makes some sense, where $P_i$
%is as above the share-price at the $i$-th moment, but this 
%is too approximate. 

\vskip 0.2cm
The $C$-constants above 
are essentially proportional to the value of the news and
are related to the company {\it momentum volatility,} which
depends on the investment horizons (reflected in the tables below).
%It generally depends on  the stream of company-related 
%news, on the investment or forecast horizons, and of course
%on the general market conditions. 
See e.g. \cite{EN,ABL,FPSS}. The dependence of the 
volatility on the horizon is reflected in our tables below;
it is connected with the exponent $r$. 
The $t$-periodicity due to {\it profit-taking\,}
is an important factor here. Then the {\it stochastic
volatility} can be generally modeled via {\it Bessel processes},
similarly to the usage of {\it fBM\,} discussed a bit  below.

\comment{
\vskip 0.1cm
These constants $C$  can vary a lot.  
The coefficient $r$ is significantly more stable. It
is related to the kind of trading the company is involved in
(day-trading  etc.). I.e. it is different for different
{\it trading  horizons\,} and reflects the 
preferences of investors trading in this range, which are 
sufficiently stable. 
In terms of $r$, the
parameter  $\rho=-\log_2 r$ 
gives some measure of
short-term  volatility theoretically and practically.
Indeed, the value $r=1$ corresponding to $\rho=0$ means
that the  short-term volatility is modest. When
$\rho=1$ ($r=1/2$), the volatility becomes significant, and
it grows quickly when $\rho\to \infty.$
The latter typically occurs when the durations of 
positions are in hours (as in day-trading).
}

\vskip 0.3cm
{\sf Connection to 
statistical framework}. The leading term $t^r$ of our $p(t)$  
is the  square root of the {\it variance\,}
$Var(B^H)$ of the {\it fractional Brownian motion\,}
 $B_H(t)$  ({\it fBM} for short)
for the {\it Hurst exponent\,} $H=r$, where $r$ is as above.
%:$r=r_{1,2}$ or $r=d$ for (\ref{6}) or (\ref{7}).
It also appears in the self-similarity property 
of {\it fBM}:  $B_r(ts)\sim t^r B_H(s)$. 
%As above, $r=r_{1,2}$ or $r=d$ for (\ref{6}) or
%(\ref{7}).
%Accordingly,
%for $r<1/2$ (the case of short-term investing), the 
%the {\it increments} of this process are negatively correlated. 
One can try to introduce generalized
{\it fBM}  for the {\it full\,} solutions from
(\ref{7}) or even for those from Section \ref{sec:Bessel} below
in terms of Bessel functions. A more systematic way to link our 
ODE to SDE is via the Kolmogorov-type
equations for the {\it transition probability 
density\,}; see e.g. equation (1.7) from \cite{Kat} for
Bessel processes.
\vskip 0.1cm    
 
We refer to \cite{Che, GJR, GNR} for the basic properties of 
$B_H(t)$
and their applications in financial mathematics; {\it fBM} is
an important tool for modeling 
{\it volatility} of stock markets.
A {\it qualitative\,} reason for the connection with our approach 
is that the expected (percent) growth of the share-price is 
essentially proportional to the 
standard deviation of the corresponding stochastic process. 
Another (essentially equivalent) connection goes via
expected values of {\it options}. We will not discuss
the passage to SDE any further in this paper; at least,
it explains that $r$ is closely  correlated with the market
volatility.

\vskip 0.2cm
\subsection{\bf Two events, comments}
The impact of two events at $-\tau<0$ and $0$ on the
share-price $p(t)$
can be naturally described by the system
\begin{align}\label{10}
& \frac{du(t)}{ dt}\ =\ \frac{c_0}{ t} u(t) 
+\frac{c_\tau}{ t+\tau} u(t)- \frac{1}{ \si (t+\tau)} p(t),\\
\label{11}
&\frac{dp(t)}{ \si dt}\ =\ \frac{a}{ t} u(t)+
b \frac{du(t)}{ dt} \for c\equal c_0+c_\tau.
\end{align}
When $c_\tau=0$, it describes the case when there is no
news at $-\tau$, but this moment is taken as the  
{\it support\,} for the price-target; generally, 
price-targets do depend on historical levels. 
Let $b=0$ here and below. We obtain:
$t(t+\tau)d^2p/dt^2+((1-c)t+(1-c_0)\tau)dp/dt+ap(t)=0$,
%$t(t+\tau)d^2p/dt^2+(1-c)(t+\tau)dp/dt+ap(t)=0$, 
which 
can be integrated in terms of hypergeometric functions.
Namely, 
$p(t)=F(\al,\be; \ga,-t/\tau)$
is a solution for $\ga=1\!-\!c_0,\, \al\!+\!\be=-c$,
$\al,\be=-c/2\pm \sqrt{c^2/4-a}$; see e.g.\cite{AS},Ch.15, 
or use Mathematica function
{\tt Hypergeometric2F1$[\al,\be,\ga,x]$}.
 One can also take here
$p_1(t)=t^{-\be} F(\be,-\al-c_{\tau}, 1+\be-\al,-\tau/t)$ 
and $p_2(t)$
upon $\al\leftrightarrow\be$ in $p_1(t)$. When $\tau/t\sim 0$, 
such $p_{1,2}$ with proper coefficients of  proportionality 
approach $t^{r_{1,2}}$ as $t>\!>0$ for $r_{1,2}$ from (\ref{5}) 
under  $b=0$.
%\begin{align}\label{5}
%&r_{1,2}=d\pm \sqrt{D},\ d=(c-b)/2,\  D=d^2-a.
%\end{align}

\vskip 0.1cm
{\sf Using deviations.}
{\it Hedging\,} vs. SPY or some index is an example;
see e.g. \cite{BLSZ,BSV}.
The assumption is that after the companies within the index reacted to
some {\it index news\,} at the moment $-\tau$, a
specific company's news arrives 
at $0$. So if a position in a stock is {\it hedged\,} 
by investing an equal amount in the corresponding index in the 
opposite direction, the return will be  
$p(t)-p_{ind}(t)$, where $p(t)$ is 
governed by (\ref{10}\&\ref{11}) an $p_{ind}$,
the index' rate of return, is in the form of 
(\ref{6}) or (\ref{7}) for proper
parameters. Practically, our trading system automatically
determines $\, r_{ind}, C_{ind}, r, C\,$ such that 
$p(t)\!-\!p_{ind}(t) \simeq C\, t^r\!-\!C_{ind}\,t^{r_{ind}}$.
However, more refined $p(t)$, solutions of
(\ref{10}\&\ref{11}) instead of
$C\,t^r$, are significant here, especially
for (relatively) small $t$.

We mention that it makes perfect sense to switch here to 
the corresponding
{\it difference\,} equations. See e.g. \cite{ChA}, Section 1, 
concerning the one-dimensional {\it global
hypergeometric function}.
\comment{
they are represented
by series convergent {\it everywhere} in contrast to the
$F$-functions 
above. Actually, $p(t)/t$ and $u(t)/t$ already "belong" to
difference theory.}
%\begin{align}\label{12}
%& \frac{du(t)}{ dt}\ =\ 
%\frac{c_1}{ (t-t_1)} u(t)+ \frac{c_2}{ (t-t_2)} u(t)-
%\frac{1}{ \si (t-t_1)} p(t),\\
%\label{13}
%&\frac{dp(t)}{ \si dt}\ =\ 
%\frac{a_1}{ (t-t_1)} u(t)+\frac{a_2}{ (t-t_2)} u(t)+
%b\frac{du(t)}{ dt}.
%\end{align}

\vskip 0.2cm
{\sf Practical matters.}
Such adjustments  are 
quite natural, but it appeared that the elementary solutions 
of system (\ref{6})\&(\ref{7}) already describe well 
the real market processes when we  
focus on the impact of a {\it single\,} event and when the
time interval is not too large. 
The following key features of these solutions can be
observed in stock markets: 
\vskip 0.2cm

\noindent
{\bf (i)} $t^r$-dependence of the envelope of the
price-function for $0<r\le 1,$ 

\noindent
{\bf (ii)} quasi-periodic oscillations of the 
price-function in terms of  $\log(t).$
\vfil
 
Here $t$ is the time from the event. In our trading system,
(i) is the key; the periodic oscillations are 
addressed  using different tools, not really connected with
solving differential equations. 

\vskip 0.2cm 
Quasi-periodic oscillations (our second observation)
are more difficult to observe and measure.
Mathematically, such
oscillations are typical for $a$-investing and do not
appear for pure $b$-investing. They are
"around"  the mean values, and generally
require involved statistical analysis; cf. \cite{FPSS}.  
They can be mostly seen only for relatively
big $t,$ so they can be "overwritten"  by other general
market and company news and trends.
As to (i), the market evidence is solid.    

\vskip 0.2cm
From the perspective of $a$-investing, the term $p(t)/t$
is some kind of {\it profit-taking\,}, though we will argue
below that taking $p(t)$ instead of $p(t)/t$
is more relevant for "pure" 
profit-taking. 
Practically, the events or commentaries
are sometimes used simply as triggers when profit-taking.
Under $a$-investing, such "overreacting" 
mathematically means  that the coefficient
$a$ becomes relatively large; see (\ref{7}).

\vfil
Generally, our model "predicts"
that in the absence of other major news, the intervals
between consecutive rounds of
$a$-type profit-taking tend to grow 
approximately as a geometric sequence, i.e. we arrive at 
some kind of Elliot waves (associated with Fibonacci 
numbers). Strictly speaking, the profit-taking is the effect of
second order, i.e. for the share-price minus its expected average.
Mathematically, the average satisfies (\ref{7}) in our model.
The oscillations of this  difference are actually  $t$-periodic, 
not just $\log t$-periodic as for $a$-investing, which  will be 
addressed below using Bessel functions.

\vskip 0.1cm
It is important that {\it long-term returns\,} of
different companies become comparable in spite of
very different trading patterns and volatility. 
I.e. they become closer to each
other  almost regardless of their short-term behavior.
There are of course winners and losers, but the long-term
rate of change is sufficiently uniform even for quite different
types of companies. Mathematically,
it means that the smaller $r$, the bigger the
constants $C$ in (\ref{6},\ref{7}). 
We will reflect this
in our $g$-functions (\ref{g1}-\ref{g7}) 
and tables, making "basic returns" comparable
after 3-4 months. This can be important for extending
our system to trading {\it options}.
\vskip 0.2cm

To finish this discussion, let us emphasize that the analysis
above is by no means restricted to stock markets. 
Market instruments and
tools have various counterparts 
beyond trading equities. For instance, 
short-trading, profit-taking, hedging, doing derivatives 
are quite common in some forms, though reach the most 
sophisticated  levels in stock markets. 
The discontinuous nature of market data is not unusual too;
it will be addressed "practically" in Section \ref{sec:3}. 
%\vskip 0.2cm

\subsection{\bf Profit-taking etc}\label{sec:Bessel}
The model above addresses well {\it quasi-periods}
under $a$-investing (or mixed investing). The periodicity
with respect to $\log(t)$ is some
kind of profit-taking, but the actual one is 
significantly more  momentum:
{\it sell when  $p(t)$ reaches some level\,}. This is a major
reason for  short-term "periodic" volatility, which is
an important feature of stock markets; see also \cite{ACH}. 
Its role is
crucial not only
for short-term trading; see \cite{FPSS}. 
Figures 3,8 there are the key for them
and for us. The short-term 
volatility is "around" the mean value   
$\overline{p}(t)=p_{avrg}(t)$. 
By the way, the periodicity of the volatility provides an
explanation of the profitability of 
counter-trend (contrarian) strategies. 

For "pure" profit-taking, $u(t)$ must be understood as some 
market "consensus" on
keeping a stock at its current price. So the "upgrade
function"  must react here to $p(t)$, not to $p(t)/t$ as 
above. This is relative to $\overline{p}$, 
an effect of "second order", so we
will need to switch to $\tilde{p}(t)= 
(p(t)-\overline{p}(t))/\overline{p}(t)$
and the corresponding $\tilde{u}(t)$.

The most natural assumption is the proportionality
of $d\tilde{p}(t)/dt$
to $\tilde{u}(t)$. Adding the
term $a\,\tilde{u}(t)/t\,$ 
to (\ref{15}) is possible too (see (\ref{17})), 
but the key change is the
replacement of $p(t)/t$ there by $p(t)$. One has:
\begin{align}\label{14}
& \frac{d\tilde{u}(t)}{ dt}\ =\ \frac{c}{t}\, \tilde{u}(t)-
\frac{1}{ \si}\, \tilde{p}(t),\\
\label{15}
&\frac{d\tilde{p}(t)}{ \si dt}\ =\ e\,\tilde{u}(t).
\end{align}
This is almost exactly (3.14) from \cite{CM}. Generally,
the {\it spinor Dunkl eigenvalue problem} is the differential 
equation for $\overline{v}=\{v_0(t),v_1(t)\}$:  
\begin{align}\label{spin-D}
& \frac{d\overline{v}(t)}{ dt}\equal 
\{\frac{d}{dt}v_1,\frac{d}{dt}v_0\}=
\{\frac{c}{t} v_1,0\}-
\{\la v_0, \la' v_1\},
\end{align}
%i.e. $\overline{v}$ is treated as a super-function.
See \cite{CM}, Sections 2,3, e.g. Lemma 3.4 
there, and (\ref{pauw3},\ref{pauw4}) below. 
\vskip 0.1cm
\noindent
%We arrive at (\ref{14})\&(\ref{15}), 
This is a spinor variant
of the equation $\frac{dv(t)}{ dt}=
\frac{c}{2t}(v(-t)-v) - \la v,$
 where we switch to
$v_0=\frac {v(t)+v(-t)}{2}$, $v_1=\frac {v(t)-v(-t)}{2}$,
considering them as independent functions.
I.e. we switch from $v$ to a 
{\it super-function\,} $\overline{v}$, where  $\la$ is  
extended to a pair $\{\la,\la'\}$ acting on $\overline{v}$\,
"diagonally".

To solve equations (\ref{14})\&(\ref{15}), we obtain:
\begin{align}\label{ptut}
&t^2\frac{d^2\tilde{p}}{dt^2}-c t \frac{d\tilde{p}}{dt}
+et^2p\ =\ 0\ =\  
 t^2\frac{d^2\tilde{u}}{dt^2}-c t \frac{d\tilde{u}}{dt}
+et^2u+cu,\\
&\tilde{p}\!=\!A_1\tilde{p}_1\!+\!A_2\tilde{p}_2,\  
\tilde{p}_{1,2}(t)\!=\!t^{|\al|} 
J_{\al_{1,2}}(\sqrt{e}t) \hbox{\, for\, }
 \al_{1,2}\!=\!\pm\frac{1\!+\!c}{2}.\label{ptut1} % \\
%&\tilde{u}\!=\!B_1\tilde{u}_1\!+\!B_2\tilde{u}_2,\  
%\tilde{u}_{1,2}(t)\!=\!x^{|\al|} 
%J_{\be_{1,2}}(\sqrt{e}t) \hbox{\, for\, }
% \be_{1,2}\!=\!\pm\frac{1\!-\!c}{2}.
\end{align} 
Here the parameters $a,c$ are assumed generic,
$A_{1,2}$ %, B_{1,2}$ 
are undermined constants, and we
use the {\it Bessel functions\,} of the first kind:
$$ J_\al(x)=\sum_{m=0}^\infty 
\frac{ (-1)^m (x/2)^{2m+\al}}{m! \Ga(m+\al+1)}.
$$
See \cite{Wa} (Ch.3, S 3.1). We will also need the
asymptotic formula from S 7.21
there:
$$ J_\al(x)\sim \sqrt{\frac{2}{\pi x}}\cos(x-\frac{\pi\al}{2}
-\frac{\pi}{4}) \for x>\!> \al^2-1/4. 
$$
The latter gives that $\tilde{p}_{1,2}(t)$ are approximately
$\tilde{\c}\, t^{c/2}\, \cos(\sqrt{e} t-\phi_{1,2})$ for
some constants $\tilde{\c},\phi_{1,2}$. Interestingly, the phases 
$\phi_{1,2}=\pm\frac{1+c}{2}\pi+\frac{\pi}{4}$ are
uniquely determined by $c$. We conclude that
for sufficiently big $t$, the function $p(t)$ under the
profit-taking as above is basically:

\begin{align}\label{pas}
\tilde{p}(t)\approx 
t^{c/2}\bigl( A\cos (\sqrt{e}t+\pi c/4)+B\sin(\sqrt{e}t-\pi c/4)
\bigr),
\end{align}
for some constants $A,B$; the $t$-period is
$\frac{2\pi}{\sqrt{e}}$. 

\comment{
More generally, $N(t)$ can satisfy the 
{\it Dunkl-type eigenvalue
problem} $\frac{dN(t)}{ dt}=\frac{c}{t}\,N(t)
\!+\!\la N(t)$,  which can be managed in terms of Bessel functions.
See \cite{CM}, Sections 2,3, e.g. Lemma 3.4 there,
and (\ref{pauw3},\ref{pauw4}) below. 
 Say, $\la$ can
be positive right after the event or negative due
to {\it profit-taking}. Also, we disregard that
$N(t)$ can be bounded. So adding the "saturation" can be needed,
which will be addressed later.
}

Let us now replace  $\tilde{p}/\si$ by 
$\hat{p}\, t^{\nu-1}/\si$ for $0<\nu\le 1$
and $d\tilde{u}/dt-e\tilde{u}(t)$ 
by $d\hat{u}/dt-e\hat{u}(t)/t$ in (\ref{14}\&\ref{15}). 
The system
becomes:

\begin{align}\label{16}
& \frac{d\hat{u}(t)}{ dt}\ =\ \frac{c}{t}\, \hat{u}(t)-
\frac{t^\nu}{ \si t }\, \hat{p}(t),\\
\label{17}
&\frac{d\hat{p}(t)}{ \si dt}\ =\ \frac{e}{t}\,\hat{u}(t).
\end{align}

It can be solved in terms of Bessel function too.
The corresponding fundamental solutions are\, 
$\hat{p}_{1,2}(t)=t^{c/2}
J_{\pm c/ \nu}\,(\,\frac{2\sqrt{e}}{\nu}\,t^{\nu/2}\,)$.
One has:
$$
\hat{p}_{1,2}(t)\, \approx\, \hat{\c}\ t^{\frac{c}{2}-\frac{\nu}{4}}
\,\cos(2\sqrt{e}t^{\nu/2}/\nu\!-\!\psi_{1,2}) \hbox{\ as\ } 
t>\!>0.
$$ 
I.e. $\hat{p}(t)$ it is slower than $\tilde{p}(t)\,$ 
from (\ref{pas})  and the periodicity 
is for $t^{\nu/2}$ in this case; it even 
tends to $0$ as $t\to\infty$ for $c<\nu/2$.
%and  $\hat{p}_{1,2}$ are $\sin(2\sqrt{x}-\psi_{1,2})$ 
%for $c=1/2, e=1$.  

%When $c=1/2, e=1$,
%using the formula 
%$J_{1/2}(x)=\sqrt{\frac{2x}{\pi}}\frac {\sin(x)}{x}$, we
%obtain $\hat{p}= \sin(2\sqrt{t})$.

%c=0, \al=1/2, p=sin x; c=2, \al=3/2, p=sin x -x cos x
%c=1, \al=-1, p=\sum_{m=0}^\infty \frac{(-1)^m m (x/2)^{2m}}
%{(m!)^2}  ; checked for x, x^3 in xp''-cp'+xp=0.  

%with au/t: for c=1/2, a=1, p=sin(2\sqrt{x}). It does satisfies
%x^2p''+(x/2)p' +xp=0. 

Finally, combining (\ref{pas}) with 
$p_{avrg}=\overline{p}\,$ taken from (\ref{7}),
$p(t)$  can be assumed a 
linear combination
of $\, t^r\! \cos(\rho\log(t)\!+\!\ze)
\bigl(1\!-\ep\cos(\varrho t\!+\!\xi)\bigr)\ $
for proper parameters $r,\rho,\varrho,\ze,\xi,\ep$.
This holds asymptotically,
%Systematically, the terms from (\ref{4}) with $a,b$
%must be added to (\ref{15}), but we will not do this. 
but seems basically sufficient for practical modeling 
momentum trading. We note here a connection with
\cite{Che}, where the sum of Brownian motion, {\it BM\,}, with 
{\it fBM} was considered; see also the end of Section
\ref{sec:regimes}.

The $t$-periodicity of profit-taking is directly
related to short-term volatility in stock markets.  
This  is generally
a stochastic phenomenon \cite{EN, FL, FPSS}. However, 
as we see, the volatility due to profit-taking
has solid "algebraic origins".
Namely, relatively simple algebraic-type formulas 
{\it with few parameters\,}, which
reflect investors' trading preferences,
can {\it look\,} quite chaotic.  
This was actually the key for us: {\it there are very many traders,
but possibly only very few trading patterns}.

Let us provide a numerical example of such "algebraic volatility".
Using the $g$-functions from (\ref{g1},\ref{g3}) with
$1\le t \le 150h$ (1 month), let :
\begin{align}\label{fakp}
&p(t)=
   0.4 (1\! -\! \sin(t)/3)  \cos( 2\pi\log(t)) g(t,1)  \\
   +& 0.5\, (1\! -\! \sin(t/5)/3) 
\sin( 2\pi\log(t)) g(t\! +\! 12,3). \notag
\end{align}

\begin{figure}[htbp]
%\begin{center}
\vskip -0.1in
\hskip -0.2in
\includegraphics[scale=0.5]{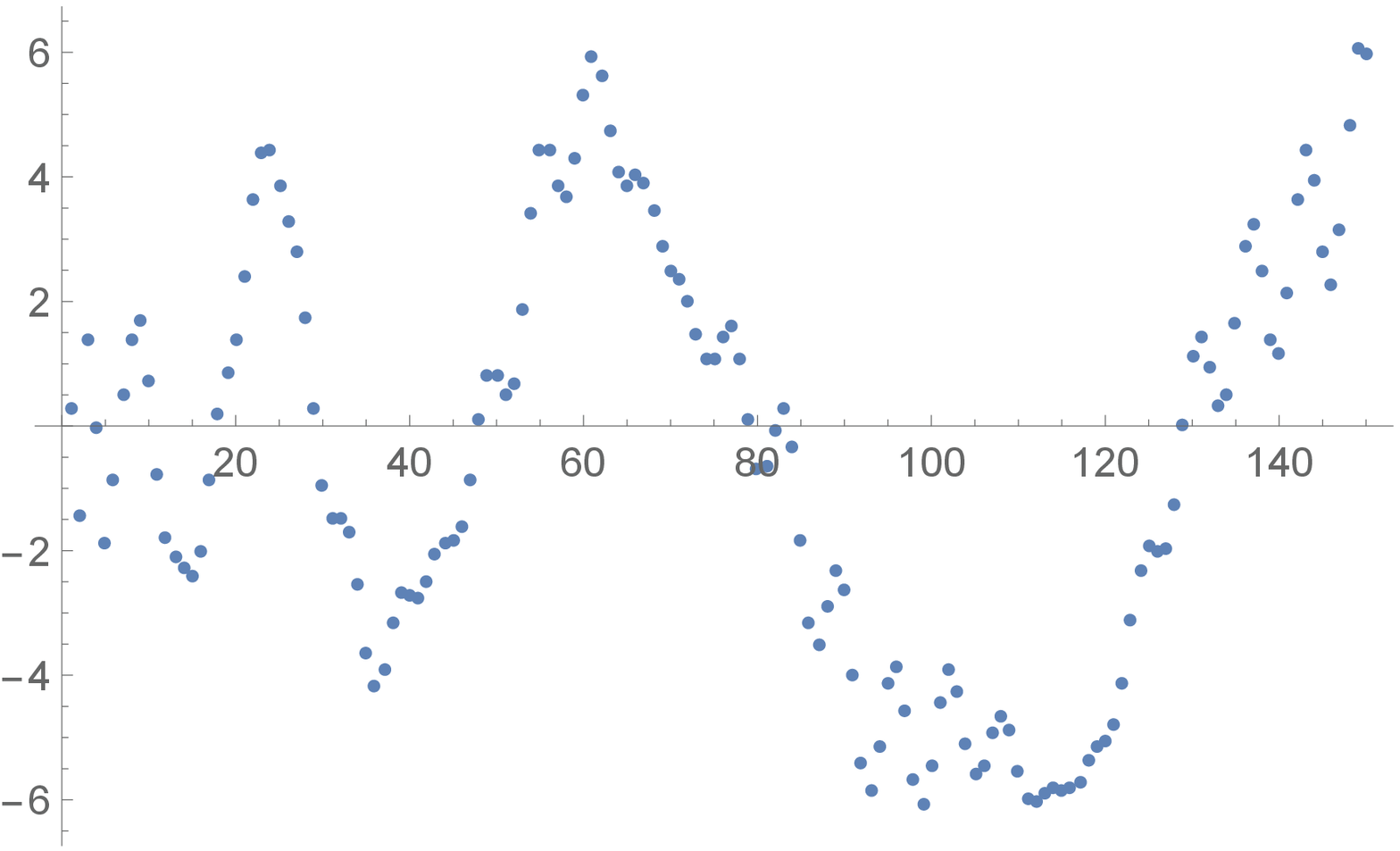}
\vskip -0.2in
%\vskip -1.0cm
\caption{Model chart: "algebraic volatility"}
\label{fig:fake}
\end{figure}

In spite of relatively simple formula, the {\it fake chart}
in Figure \ref{fig:fake}
exhibits  a lot of volatility, which is mathematically
hardly surprising for such trigonometric expressions.
Before managing
real charts, the system was "trained" to trade
profitably such {\it fake} ones. It is momentum;
catching the periods and quasi-periods was not an objective.
We do not have sufficient "stability theory" for the
periods. However the exponents $r$ can be reasonably 
found by the system (automatically) for fake and real charts.
In (\ref{fakp}), $r=0.137, \ 0.418$ for $g(1),g(3)$. 
  
%\vspace* {-6.5in}

%\comment{
\subsection{\bf Toward Macdonald processes}
A natural passage from our
differential equations to random processes is via 
considering  {\it ensembles\,} of news-functions:\,
$\{u^{(1)},u^{(2)},\ldots,u^{(n-1)}, u^{(n)}\}$, at the moments
$t_1>t_2>\cdots >t_{n-1}>t_n$. Let us assume that
they are governed by the 
same reduction coefficient $c$. Then a counterpart
of (\ref{0}) is the following system:
\begin{align}\label{KZ1} 
&\pa u^{(i)}/\pa t_j= c\frac{u^{(i)}-u^{(j)}}{t_j-t_i}
\for i\neq j,\\
&\pa u^{(i)}/\pa t_i= c\sum_{j\neq i} \frac{u^{(i)}-u^{(j)}}
{t_i-t_j},\label{KZ2}
\end{align}
which is a special variant of {\em rational\,} 
Knizhnik-Zamolodchikov equation of type $A$ in the standard 
$n$-dimensional representation of the symmetric group $\S_n$.
See e.g. \cite{ChB}, Section 1.1. This means that news at $t_i$
is positively influenced by that at $j$ if $u^{(j)}> u^{(i)}$ 
or by that at $j>i$ if $u^{(j)}<u{(i)}$; the impact
is negative otherwise. Qualitatively, a stronger news 
{\it before\,} $t_i$
increases the impact of $u^{(i})$ and diminishes
its impact if it arrives {\it after\,} $t_i$ (and vice versa).
Quantitatively, we divide here by $(t_i-t_j)$.
Note that $u^{(1)}+\cdots
u^{(n)}$ does not depend on $t_j$, so this is a process of
"news-redistribution".  
In the case of $n=2$, we obtain:
\begin{align*} 
&\pa u^{(1)}/\pa t_1= c\frac{u^{(1)}-u^{(2)}}{t_1-t_2},\ 
\pa u^{(1)}/\pa t_2= c\frac{u^{(1)}-u^{(2)}}{t_2-t_1},\\
&\pa u^{(2)}/\pa t_2= c\frac{u^{(2)}-u^{(1)}}{t_2-t_1},\,\,
\pa u^{(2)}/\pa t_1= c\frac{u^{(2)}-u^{(1)}}{t_1-t_2}.
\end{align*}

\comment{
One can add price-functions $p$ and 
profit-taking as follows:
\begin{align}\label{KZ3} 
&\pa u^{(1)}/\pa t_1= c\sum_{j=2}^n
\frac{u^{(1)}-u^{(j)}}{t_1-t_j}-ap,\ \pa p/\pa t_1=u^{(1)}
\end{align}
for some constant $a$. I.e. we assume here that the change of
$p$ depends only on the latest 
news $u^{(1)}$ and also that $p(t_1=t_2)=0$.

In the case of $n=2$, let us assume that $p$ depends only
on $y=(t_1-t_2)/2$.
\begin{align*} 
&\pa u^{(1)}/\pa t_1= 2c\frac{u^{(1)}-u^{(2)}}{t_1-t_2},\ 
\pa u^{(1)}/\pa t_2= c\frac{u^{(1)}-u^{(2)}}{t_2-t_1},\\
&\pa u^{(2)}/\pa t_2= c\frac{u^{(2)}-u^{(1)}}{t_2-t_1},
\pa u^{(2)}/\pa t_1= c\frac{u^{(2)}-u^{(1)}}{t_1-t_2}.
\end{align*}
\begin{align*} 
&\pa u^{(1)}/\pa y= c\frac{u^{(1)}-u^{(2)}}{t_1-t_2}-a p,\ 
\pa u^{(2)}/\pa y= c\frac{u^{(2)}-u^{(1)}}{t_1-t_2}.
\end{align*}
}

%\comment{
{\sf Spinor Dunkl problem.}
This consideration is a special case of
the following more general approach.  
Let $\S_n$ be the symmetric
group, 
% of permutations 
%$w:(12\ldots n)\mapsto w(1)w(2)\ldots w(n))$,
 $s_{ij}$ the transpositions. To link our systems
to the {\it Macdonald processes\,} from
\cite{BC} (in the nonsymmetric setting), we 
consider an {\it ensemble\,} of scalar "news-functions"
$\{u^{(w)}, w\in \S_n\}$ 
depending on $\{t_i, 1\le i\le n\}$ from the {\it sector\,}
$t_1>t_2>\cdots >t_{n-1}>t_n$. We set $w(f)\equal
%f(w^{-1}(t_1,\ldots,t_n))=
f(t_{w(1)},\ldots,t_{w(n)})$ for $w=(w(1),\ldots,w(n))$.
A counterpart of (\ref{0}) is the following system:

\begin{align}\label{pauw}
\pa u^{(w)}/\pa t_i= c\sum_{j\neq i}
\frac{u^{(w)}-u^{(ws_{ij})}}{t_{i}-t_{j}} +\la_{w(i)} u^{(w)}
\for 1\le i \le n.
\end{align}
It is equivariant under the action $u^{(w)}\mapsto u^{(vw)}$,
$\la_i\mapsto \la_{v(i)}$ 
of $v\in \S_n$ (without touching $\{t_i\}$). This is 
the {\it spinor Dunkl eigenvalue problem\,} of type $A$ in
the terminology of %\cite{ChO} and 
\cite{ChA}. 
It can be integrated in terms of {\it multi-dimensional
Bessel functions\,}; the theory of the latter is relatively
new even in (classical) symmetric setting \cite{Op}.
The original {\it Dunkl eigenvalue problem\,} is
a system of differential-difference equation for
a single $u=u^{(id)}$ above defined for {\it all\,} sufficiently
general $\{t_i\}\in \R^n$, not only in the {\it sector\,} above, 
where we plug in: $u^{(w)}=w(u)$.
This one is equivariant
with respect to the action of $v\in \S_n$ on 
$\{t_i\}$ and $\{\la_i\}$  and $u$ by permutations.

The $\la$-parameters are adjusted to the original
{\it non-spinor\,} Dunkl eigenvalue problem. One can 
consider more general linear combinations in 
(\ref{pauw}) (in the
spinor case). Note that the price functions are not involved 
here; the usage of $\la$  serves as some substitute.
 \vskip 0.2cm

Let us consider the case of $n=2$, setting
$u^{(0)}=u^{(id)}, u^{(1)}=u^{(s_{12})}$:
\begin{align}\label{pauw1}
&\frac{\pa u^{(0)}}{\pa t_{1}}= c\frac{u^{(0)}-u^{(1)}}{t_1-t_2} +
\la_1 u^{(0)},\ 
\frac{\pa u^{(0)}}{\pa t_{2}}= c\frac{u^{(0)}-u^{(1)}}{t_2-t_1} 
+\la_2 u^{(0)},\\
&\frac{\pa u^{(1)}}{\pa t_{1}}= c\frac{u^{(1)}-u^{(0)}}{t_1-t_2} +
\la_2 u^{(1)},\ 
\frac{\pa u^{(1)}}{\pa t_{2}}= c\frac{u^{(1)}-u^{(0)}}{t_2-t_1} 
+\la_1 u^{(1)}. \label{pauw2}
\end{align}

Switching to $\tilde{u}^{(i)}=
\exp(-(\la_2+\la_1)(t_1+t_2)/2)u^{(i)},$ $\la=(\la_1-\la_2)/2$,
and $x_{1,2}=t_1\pm t_2,\, t_{1,2}=(x_1\pm x_2)/2$, 
we obtain that $\pa \tilde{u}^{(i)}/\pa x_1=0$.
Setting $\,y=x_2, \ f^{0,1}=\tilde{u}^{(0)}+ \tilde{u}^{(1)}\,$, 
$f^{1}=\tilde{u}^{(0)}- \tilde{u}^{(1)}\,$,
we arrive at:

\begin{align}\label{pauw3}
&\frac{\pa \tilde{u}^{(0)}}{\pa y}= 
c\frac{\tilde{u}^{(0)}-\tilde{u}^{(1)}}{y} +
\la \tilde{u}^{(0)},\ 
\frac{\pa \tilde{u}^{(1)}}{\pa y}= 
c\frac{\tilde{u}^{(1)}-\tilde{u}^{(0)}}{y} -
\la \tilde{u}^{(1)},\\ 
&\frac{\pa f^{0}}{\pa y} = \la f^{1},\ 
\frac{\pa f^{1}}{\pa y} = 2c\, \frac{f^{1}}{y}+\la f^{0},\ \,
\frac{\pa^2 f^{0}}{\pa y^2}=
\frac{2c}{y}\, \frac{\pa f^{0}}{\pa y}+ \la^2 f^{0}.
%
%\frac{\pa^2 f^{1}}{\pa y^2}=
%c \frac{\pa}{\pa y}\bigl(\frac{f^{1}}{y}\bigr)+ \la^2 f^{1}.
 \label{pauw4}
\end{align}
The last equation can be integrated in terms of Bessel 
functions; cf. (3.14) from \cite{CM}, equation
(\ref{ptut}) above, and also \cite{ChB}, Section 2.4.
However in contrast to the solution
from (\ref{ptut1}), the function $f^0$ will not oscillate
due to the positivity of $\la^2$; we assume that $\la\in \R$.

Similar to (\ref{ptut1}), adding the profit-taking term 
results in $\la^2\mapsto \la^2-a$ for some constant
$a>\la^2$. This will provide the desired
oscillations; we omit details and generalizations.
%However in contrast to the solution
%from (\ref{ptut1}), the function $f^0$ will not oscillate
%due to the positivity of $\la^2$; we assume that $\la\in \R$.

%}    %end comment 

\setcounter{equation}{0}
\section{\sc Market implementation}\label{sec:3}
\subsection{\bf Major challenges}
{\it The first major challenge\, }  with the mathematical 
analysis of stock charts and other
market information is that the corresponding 
functions are of discontinuous nature. 
Automated high-frequency trading adds a lot of volatility too
\cite{CJP}.  
This makes the separation of the signals  
from noise and trading involved.

%\vfill\eject %!!!

{\it The second 
challenge\,} is that even if the news and 
has clear meaning, the corresponding trading decisions
can depend on many factors. For instance, it can be simply too
late to invest in this particular news. Executing large orders 
can be with significant losses right after the news, and
so on. By the way, 
the {\it counter-trend\,} (contrarian) variants of our 
trading system, i.e. those selling when the share-price goes 
up and so on, can outperform the {\it pro-trend\,} variants. 

{\it The third challenge\,} is picking right moments
for closing positions.  We use the {\it termination
curves\,} discussed below and the "signals" 
opposite to the direction (long or short) of the  position taken,
determined automatically.  
Obviously, the {\it bid-ask spread\,} reduces 
the profitability; see \cite{KS}. This is one of the reasons
why we optimize {\it returns per positions\,};  the positions
generally last from 5 to 10 days.

{\it The fourth challenge\,} is that a {\it significant\,}
variety of (profitable) strategies is needed to address market
volatility. In our system,
using counter-trend and pro-trend variants
simultaneously, employing different 
{\it opti-parameters\,}, and varying the
moments when the system receives quotes provide reasonable 
stability.
The number of different profitable variants of the system
is practically unlimited: 12 "production lines"
were used in real-time experiments. 

{\it The fifth challenge\,} is using weights, which
for us are mainly those  based on the results of the
prior optimization. We obviously 
rely mostly on the equities the most suitable for our 
system, i.e. those performed the  best during the
optimization process. However, the {\it opti-parameters\,}
and weights based on the past performance can fail
in future.

{\it The sixth\,} is simply due to the novelty of our
approach. The usage of our {\it 2-bid tables\,} for creating 
momentum trading systems, trading options and
{\it technical analysis\,} of stocks requires experience. 
The pont-tables from 
Section \ref{sect:PP} can help to get used to our {\it 2-bids\,}.
Also, we provide various performance results of our own system,
which can be used as "benchmarks" for those who follow our
approach.

%\vskip 0.1cm
\subsection{\bf Forecasting} 
The work of our system is based on the {\it forecasting curves\,},
automatically produced 
time-predictions for share-prices.
The {\it termination curves\,} are their shifts 
up or down with some coefficients of proportionality
providing some room before their intersection with 
the actual share-price graphs. 
These intersections trigger the 
terminations of the taken positions (if any). 
This is similar to trading US-style options, when 
the termination curves are horizontal lines shifted up or down
for calls or puts. The curves we use are essentially \ 
$b$\,(time)${}^r$ \ for "bids" $b$ and exponents $r$, assigned to 
the 7 "categories" discussed below (main 4 and their 3 
consecutive averages).

The bids 
are discrete and must be large enough (at least $1$)
to form an {\it admissible 2-bid\,}, 
which is a pair $\{ b=bid, c=category\}$.
The 2-bids are ranked {\it lexicographically}, first
with respect to $b$ (the bigger the better)
and then, if the bids coincide,
with respect to $c$: the smaller $c$ and its prime time-interval
the better. The "winner" is the top bid. {\it Bids below the 
{\it threshold} in their categories are ignored as noise.} 
The thresholds  for prime-intervals are $1,2,3.5,7$  
for $c=1,3,5,7$
times some common rescaling coefficient $\be$;  see the tables below.

\comment { 
The  "bid-thresholds", smallest admissible bids,
the shifts of termination curves  and other parameters 
are determined on the basis of prior performance 
via machine learning.} 
\vskip 0.1cm

The basic functions  we use are as follows:
 \begin{align}
g(t,1)=& 0.5\cdot\hbox{Floor}[1548\,(\,0.26t+0.74)^x-1548]/100+1,
\ x=0.137,\notag\label{g1}\\
&\hbox { in the case of  the  {\it super} category ($c=1$),}\\
g(t,3)=& 2\cdot\hbox{Floor}[10\,(\,2(t/d)-1)^x]/10 
\for  x=0.418,\notag\label{g3}\\ 
%\log(10)/\log(250)
&\hbox { in the case of  the {\it ultra} category ($c=3$),}\\
g(t,5)=& 0.1\cdot\hbox{Floor}[22.875\,(\,2.024\, \frac{t}{5d}\,
-1.024)^x+12.125],\ 
x=0.5678,\notag\label{g5}\\ 
&\hbox { in the case of the  {\it extra} category ($c=5$),}\\
%&g(t)= \hbox{Floor}[10t^x]/10 \hbox{\ for\ } x=0.634
%\hbox {\ in\ premium\ category,}\\
g(t,7)=& 3.5\cdot(\hbox{Floor}[10.25\,(\,t/(22d))]/10+1),
\hbox{i.e.\ here\ } x=1,\notag\label{g7}\\
&\hbox { which serves  the   {\it regular\,}\ category ($c=7$), }
%&g(t)= (2\cdot\hbox{Floor}[10t]/10+2)/4\ (\hbox{i.e.\ } x=1)
%\hbox {\ in\ regular\ category,\ and} \\
\end{align}
where $t$ is 
measured in hours; $1h$ is the  prime time-interval in
the super case, $d=6.5h$, the duration of 
one Wall Street business day, is that 
in the ultra-category. Accordingly, the prime
time-intervals are 1 week $= 5d$ in the extra category
and 1 basic month $=22d$ in the regular category. Here
$x\approx 0.137+\log(u)/6$ for 
$u=\{1,\, 1d,\, 2.5d,\, 22d\}$, where $2.5d$ (instead of $5d$ for
$c=5$) is due to some practical reasons. Qualitatively, $x\,$
is supposed to depend linearly on the logarithm of the corresponding
prime time-interval, but this can vary.  
\vskip 0.1cm

Here Floor$[z]$ means the maximal integer no greater than $z$.
For $0<t<t^\bullet$, where $t^\bullet=1, 1d, 5d, 22d$ 
correspondingly ($t^\bullet_i$ will be used here for $i=1,3,5,7$),
we extend the functions above by a uniform
linear formula: $g(t^\bullet)(2t+t^\bullet)/(3t^\bullet)$. Also, we 
define $g$-functions for even categories $c=2i$,
where $2i=2,4,6$, as the averages
of the neighboring $g$, i.e.  
$g(t,2i)=(g(t,2i-1)+g(t,2i+1))/2$; the prime time-intervals
are $t^\bullet_{2i}=2t^\bullet_{2i-1}$ (not the 
corresponding averages).

Finally, the basic functions will be  $\,b\,g(t,c)\,$, 
where $b$ is the bid (an integer), $c$  the category. 
The trading 
system automatically determines the bids {\it backward\,}
as  price-changes in percent   
divided by the corresponding $g$. This is performed at every moment 
when the system obtain quotes in all 7 categories, and with
some {\it depth\,}, the number $m$ of steps back.  
I.e. it constantly calculates for the {\it rescaling coefficient\,}
$\be:$
\begin{align}\label{bim}
b_i(m) = \hbox{Floor}\bigl[\,100\,\be\,
\frac{\mid\, p_{t}-p_{t-mt_i^\bullet}\,\mid}
{g(mt_i^\bullet,i)p_{t-mt_i^\bullet}}\,\bigr], 
\ 1\le i\le 7, \ \,\be\ge 1,
\end{align}
for the
corresponding $t_i^\bullet$ and a 
sequence $m=1,2,3,\ldots$ (mostly, 1 month back); 
here $p_t$ is the share-price at  $t$, $|\cdot|$ the absolute value. 

Then the highest 2-bid $b_{i_\circ}(m_\circ)$
among all $i$ and $m$ becomes the {\it top 2-bid}; 
if two 2-bids coincide, the smaller $m_\circ$ the better.
The corresponding 
$b_{i_\circ} g(t-t_\circ)$ for  
$t_\circ=t-m_\circ t_{i_\circ}^\bullet$, 
shifted and with some proportionality
coefficient, becomes the {\it termination curve\,},
which can be changed if a higher {\it top 2-bid} arrives. 
To improve the performance,
the top 2-bids are renewed only when 
$\pm p(t)$ {\it decelerates\,} with some threshold
(subject to optimization);
$\pm$ for long/short or $\mp$ for the "counter-trend".
Also, the system constantly produces 
{\it top start 2-bids}, changed when  $\pm p(t)$ 
{\it accelerates\,} (with their threshold). 
 They are used for opening positions, forecasting and
terminations of the trades in the opposite mode.

Finally, the {\it trading signals} are the 
increases of the {\it top 2-bids\,} or 
{\it top start 2-bids\,} 
and the intersections with 
the termination curves.  

Consecutive increases of top bids for the same equity
in the same direction are used 
to open multiple positions: of level 1 on the first bid,
of level 2 for the first increase and so on. The trades
based on level 2,3 bids 
mostly outperform those of level 1. However omitting
level 1 bids significantly reduces the total amount 
that can be invested; in professional
trading, the greater the better.
\vskip 0.3cm

\subsection{\bf Tables of two-bids}\label{sect:2bids}
Recall that there are four categories 
{\it super (1), ultra (3), 
extra (5), and regular (7)}, and also intermediate
even categories.
They are governed by different bid-tables, where 
{\it 2-bids\,} are  pairs
$(b,c).$ Usually  $b$ are integers from 1 to 5. Practically,
2-3 categories are mostly
used for individual companies,
 though the system
becomes less stable with 2 categories. This can be
greater than 3 when trading indices, but 3 seems 
reasonably optimal.  
The average durations of
positions are mostly in the range from 3-15 days for us, so the
regular category rarely occurs in our simulations and 
real-time runs.
\vskip 0.2cm

The termination can be {\it only\,} due
to the signals, unless for clear "hangs", which requires
a special consideration; see e.g.
\cite{BDM}.  The signals here are intersections with termination
curves or {\it start bids\,} in the opposite direction.
So the average durations can
be adjusted only by choosing proper combinations of 
categories and initial parameters;  all parameters
are subject to machine optimization.  The
system finds many "profitable" and stable combinations of 
parameters, which can be used to obtain
desired durations of positions and for other adjustments. 
New positions are mostly open due to the new {\it start bids}.

\vfil

Using different initial values of parameters,
pro-trend and counter-trend (contrarian) modes, weights and so on
results in many different variants. Also, much depends on
the moment the system enters the market, obtain quotes
and the prior history.
 {\it The system was proved to be able to produce
a lot of profitable 
trading lines\,}, which resembles very much human
decision-making. Even with playing simple games, there are 
almost always various ways to win; so one can choose.
 
\vfil 

\vskip 0.3cm
{\sf Super table ($c=1$):}
\vskip 0.1cm

\halign{\ \hfil #\hfil\       &  \  #\hfil$|$\     
& \ \hfil #\hfil\     &  \ \hfil #\hfil\    
&  \ \hfil #\hfil\   
 &  \ \hfil #\hfil\     & \ \hfil #\hfil\     
&  \ \hfil #\hfil\    \cr
&$\underline{b\setminus h}$ &  1 h &  2 h &  1 d &  5 d &  1 m 
&  3 m  \cr
& 1             &  1   &  1.5 &  3   & 6.5  &  11  &  15   \cr
& 2             &  2   &  3   &  6   &  13  &  22  &  30  \cr           
& 3             &  3   &  4.5 &  9   & 19.5 &  33  &  45  \cr
& 4             &  4   &  6   &  12  &  26  &  44  &  60  \cr
& 5             &  5   &  7.5 &  15  & 32.5 &  55  &  75  \cr
& 6             &  6   &  9   &  18  &  39  &  66  &  90.  \cr
}
\vskip 0.3cm
Here and below $1d$ equals 6.5 hours,
$1 m$ means 22$\cdot$6.5h, $3 m =$65$\cdot$ 6.5h
(working days only). The (expected) return at $t$ for a bid $b$ 
and category $i$ is simply $b g(t,i)$, assuming that the initial
moment is $t=0$. 

\vskip 0.2cm
{\sf Ultra category ($c=3$):}
\vskip 0.1cm

\halign{\ \hfil #\hfil\       &  \  #\hfil$|$\     
& \ \hfil #\hfil\     &  \ \hfil #\hfil\    &  \ \hfil #\hfil\   
 &  \ \hfil #\hfil\     & \ \hfil #\hfil\     &  \ \hfil #\hfil\cr
&$\underline{b\setminus d}$ &  1 d &  2 d &  5 d &  15 d &  
45 d &  6 m  \cr
& 1            &  2  &  3  &  5  &  8   &  13  &  20  \cr
& 2            &  4  &  6  &  10 &  16  &  26  &  40  \cr           
& 3            &  6  &  9  &  15 &  24  &  39  &  60  \cr
& 4            &  8  &  12 & 20  &  32  &  52  &  80  \cr
& 5           &  10 &  15  & 25 &  40  &  65  &  100  \cr
& 6           &  12 &  18  & 30 &  48  &  78  &  120.  \cr
}
\vskip 0.1cm

Here, additionally, $6 m$ means 6 months, which is
126d, 2 months
are (approximately) 45d;  $d$ always means 6.5h.
Only working days are counted.

\vskip 0.2cm
{\sf Extra category ($c=5$):}
\vskip 0.1cm

\halign{\ \hfil #\hfil\       &  \ #\hfil$|$\     
& \ \hfil #\hfil\     &  \ \hfil #\hfil\    &  \ \hfil #\hfil\   
 &  \ \hfil #\hfil\     & \ \hfil #\hfil     \cr
&$\underline{b\setminus w}$&  1 w & 2 w &  1 m  &  3 m  &   9 m \cr
& 1            &  3.5  &  5.5  &  8.5 & 15.5  &   28 \cr
& 2            &  7   &  11   &  17   &  31   &   56 \cr
& 3            & 10.5 &  16.5 &  25.5 & 46.5  &   84 \cr
& 4            &  14  &  22   &  34   &  62   &  112 \cr
& 5            & 17.5 &  27.5 & 42.5  & 77.5  &  140, \cr 
}
\vskip 0.1cm

\noindent
where, as above, 1 week = 5 days, 1 months  = 22 days, 
3 months= 65 days, 9 months = 191 day, 12 months= 252 days
(to be used next).

\vskip 0.2cm
{\sf Regular category ($c=7$):}
\vskip 0.1cm

\halign{\ \hfil #\hfil\       &  \ #\hfil$|$\     
& \ \hfil #\hfil\     &  \ \hfil #\hfil\    &  \ \hfil #\hfil\   
 &  \ \hfil #\hfil\  &\ \ \ \ \ \ \ \hfil #\hfil   \cr
&  $\underline{b\setminus m}$&  1 m  &  2 m &  4 m  &  12 m  \cr 
& 1            &  7  &  10.5   &   17.5 &  44.5 \cr       
& 2            &  14  &  21   &  35 &  89 \cr            
& 3            &  21 &  31.5  &   52.5 &  133.5  \cr    
& 4            &  28 &  42  &   70 &  178.  \cr  
}
\vskip 0.3cm
\vfil

{\sf Comparing  the categories.}
Let us compare the minimal admissible bids (basic returns) 
in the different
categories for the 13 basic durations, mostly taken from the
tables above. 
Those from the tables above  are in bold;
the others are calculated using the corresponding 
$g$-functions:
\vskip 0.3cm

\halign{\hfil #\hfil\ \ & \hfil #\hfil\ & \hfil #\hfil\ 
& \hfil #\hfil\ 
& \hfil #\hfil\ & \hfil #\hfil\ & \hfil #\hfil\ & \hfil #\hfil\ 
& \hfil #\hfil\ & \hfil #\hfil\ & \hfil #\hfil\ & \hfil #\hfil\ 
& \hfil #\hfil\ & \hfil #\hfil\ & \hfil #\hfil\  \cr
&cat& 1h    & 2h     & 1d     & 2d     & 1w     & 2w      & 3w     
& 1m     & 2m    & 3m     & 4m     & 6m     & 9m        \cr
& 7& ---   & ---    & ---    & ---    & ---    & ---     & ---    
&{\bf 7} &{\bf 10.5}& 14   &{\bf 17.5}& 23.8   & 34.3     \cr
& 5& ---   & ---    & ---    & ---    &{\bf 3.5} &{\bf 5.5}  & 6.9    
& {\bf 8.5}& 12.7  &{\bf 15.5}& 18.0  & 22.2  &{\bf 28} \cr 
& 3& ---   & ---    &{\bf 2} &{\bf 3} &{\bf 5} & 6.8    &{\bf 8}     
& 9.6   &{\bf 13}& 15.2  & 17.0   &{\bf 20}& 23.8      \cr
& 1&{\bf 1} &{\bf 1.5}&{\bf 3}& 4.3 &{\bf 6.5}& 8.5    & 9.7   
&{\bf 11}& 13.6   &{\bf 15} & 16.1   & 17.8   & 19.7   \cr
}
\vskip 0.3cm

\noindent
Recall that we set as above:
\begin{align}
& 1d=6.5h,\  1w\ =\ 5d,\ \  1m\ =\ 22d,\  2m\ =\ 45d,\notag\\
& 3m\ =\ 65d,\ 4m=86d,  \  6m=126d,\  9m=191d. \label{categ}
\end{align}
Also, recall that 2-bids are ranked naturally: 
first $b$, the bigger the better, 
then $c$ (when $b$ 
coincide) with the priority to smaller $c$, the shorter
the durations of positions the better. 

Note that for $b=1$, which is the smallest bid, the
returns after 3 or 4 months are approximately comparable
for all 4 categories. This is by design. Also, the expected
return at $2t_i^\bullet$ is $1.5$ greater than that at 
$t_i^\bullet$, which
is the prime time-interval for the corresponding 
category ($i=1,3,5,7$), with
a minor deviation for $i=5$ (the extra category).  
Also, the curves we use for prediction (and termination)
heavily depend on the category, but 
they produce reasonably
comparable returns after 3-4 months;
we aim at using and trading options here. 

Any bid is automatically considered
in all "higher" categories. For instance, 
the smallest possible bid, which is the return
of $1\%$  next hour, in super category,
is "equivalent to"  $3\%$ next day, 
so it "beats" the smallest ultra-bit, which is $2\%$
a day. Then it is supposed to generate  $6.5\%$ next week 
(vs. minimal 3.5\% in the extra category), and $11\%$
next month (vs. 7\% in the regular category). To make
this table working, 2 times every bid
in the same column from the comparison table (with the
same durations) is supposed to be
greater than any bid there, which holds. This matches well
bidding in contract card games: the greatest bid wins
regardless of the suit. 

The functions we used above are designed to provide such
natural logical inter-relations when comparing bids from 
different categories. Also, an integrality of some (not all)
bids is a consideration. This can help to
use these tables manually without computers, though the
mathematical discretization is the main point here.

To avoid any misunderstanding, the bids above begin with $1$
($1\%$ per hour in the
super category) mostly for the sake of 
readability. The trading system
divide these tables (all of them) by  the {\it common\,} 
rescaling coefficient $\be$.
For instance, the
division of all bids by $2$ makes 
sense: $0.5\%$ per hour is  more realistic
than $1\%$. Such rescaling significantly increases the 
number of "admissible 2-bids", which is generally needed for the
trading system to be stable and react promptly to 
the changes of share-prices. This coefficient $\be$
is  subject to machine optimization, as well 
as all other parameters. 
\vskip 0.2cm

Finally, let us provide the table where we compare in the
same way the minimal bids in all 7 categories:
\vskip 0.1cm

\halign{\hfil #\hfil\ \ & \hfil #\hfil\ & \hfil #\hfil\ 
& \hfil #\hfil\ 
& \hfil #\hfil\ & \hfil #\hfil\ & \hfil #\hfil\ & \hfil #\hfil\ 
& \hfil #\hfil\ & \hfil #\hfil\ & \hfil #\hfil\ & \hfil #\hfil\ 
& \hfil #\hfil\ & \hfil #\hfil\  \cr
&cat& 1h    & 2h     & 3h    & 1d      & 2d     & 4d      & 1w   
& 2w     & 1m    & 2m     & 3m     & 4m   \cr
&1& 1.     & 1.49 & 2.27 & 3.   & 4.31 & 5.92 & 6.49 
& 8.44 & 10.99& 13.57& 15.01& 16.16\cr 
&2& ---    & 1.28 & 1.87 & 2.5  & 3.65 & 5.16 & 5.74 
& 7.62 & 10.29& 13.28& 15.1 & 16.57\cr 
&3& ---    & ---  & ---  & 2.   & 3.   & 4.4  & 5.   
& 6.8  & 9.6  & 13.  & 15.2 & 17.  \cr 
&4& ---    & ---  & ---  & ---  & 2.54 & 3.71 & 4.25 
& 6.15 & 9.05 & 12.85& 15.35& 17.5 \cr 
&5& ---    & ---  & ---  & ---  & ---  & ---  & 3.5  
& 5.5  & 8.5  & 12.7 & 15.5 & 18.  \cr 
&6& ---    & ---  & ---  & ---  & ---  & ---  & ---  
& 4.97 & 7.75 & 11.6 & 14.75& 17.75\cr 
&7& ---    & ---  & ---  & ---  & ---  & ---  & ---  
& ---  & 7.   & 10.5 & 14.  & 17.5. \cr 
}

\vskip 0.2cm
\subsection{\bf Basic system operations} 
{\ }\vskip 0.1cm

{\sf SIGNALS}.
Producing {\it buy signals\,} and {\it sell signals\,}
is the main purpose of our (any) trading system.
When trading, our system generally 
processes the quotes for the periods about 
one month backward, employing the parameters 
obtained during the prior optimization and 
the weights based on the optimization too.   

There can be multiple signals in the same direction, the
first, the second and so on. The consecutive number of
a signal is called the {\it level of the signal}. 
Using such levels is a special feature of our system. 
Generally the signals of levels 2-3
are better ``protected" than those of level 1, the first signals
in a certain direction; only the signals of level 1,2,3,4 were
used in real-time runs.

Statistically, the number of signals of level 1, $NL1$, 
matches that for 2+3+4: $NL1\sim NL2+NL3+NL4$. 
Then $NL2 \sim NL3+NL4$,  and so on. The combination of signals 
of level 2 and 3 gave better performance than the usage of all 
(statistically, about 20\% better than that for level 1),  
but the signals of level 1 are also of good quality.
\vskip 0.1cm

The signals are mostly treated as  
{\it orders\,}. For instance,
one sells short on a {\it sell 
signal\,} and then buys to cover upon the first 
{\it buy signal\,}.  This is the other way
around for the counter-trend trading. 
The signals can be due to sufficiently big bids or
intersections with termination curves.
The positions can be opened on the first, 
second signal or the signals of
higher {\it levels\,}.  The positions of all levels 
are terminated altogether after  the 
{\it first\,} signal comes in the opposite direction.

\vskip 0.1cm
Practically, up to 4 simultaneous 
positions can be open with an equity if the signals
of all 4 levels were present.
All of them will be closed at once  upon the
{\it first\,} signal in the opposite direction. We
suggested some ways to split the termination of
big positions into several steps, say, involving 
``neighboring lines", however this was not tested.
Executing large orders is a 
well-known market concern \cite{MCL,GRS,CJP}. 

Using {\it levels} resembles using {\it leverage\,},
but the system does it in its own ways.
Also, we note that the signals are produced independently 
for different equities, although the system can work in more 
sophisticated regimes, including different variants of hedging.
\vskip 0.2cm
%\vfil

{\sf RETURNS.}
The {\it return per one position\,} is the main 
quantity the system optimizes. Here the ask-bid spreads, 
the slippage with execution of the orders,
and the broker commission  must be subtracted from the 
{\it returns\,}, 
practically, about 0.15-0.25\% per one position for 
``professional trading". {\it We always calculate pure returns, 
without taking the spread and similar losses
into consideration.} The returns we provide below are mostly
pure returns {\it per position}, but we always
calculate the usual (pure) returns during the periods under 
consideration too.

Pure returns like 0.4\% per position 
are, generally, sufficient for the profitability;
the system can do 
better than this in spite of relying on quotes only,
as the source of market information, 
various delays and charges.
The actual durations 
of the positions 
the system created were mostly  in the range of 5-10 days.

\vfil
{\sf OPTIMIZATION.}
The optimization procedures can be
for trading {\it Longs Only\,},  {\it Shorts Only\,}, or 
(mostly) for trading both, 
{\it L \&\,S\,}.

The optimization (``education") periods 
are of obvious importance. Our system
does not have any prior information about the market
and equities 
beyond the information that it can extract from the data provided 
during the optimization periods. They can be historical or based 
on prior trades by the system.
Generally, the optimization periods have to be 
1 year or longer.  Ideally, they must be {\it diverse\,}, 
i.e., must contain sufficiently long periods when the 
stock goes up and when it goes down. The
more "difficult" the optimization period, the better and
more stable the out-of-sample returns. 

These factors are of importance for choosing 
the optimization periods, and creating real 
"trading lines". 
However after this, the real-time
adjustment of parameters becomes  entirely automated. 
Mostly, the "real-time optimization" is 
for 6-months periods backward.

Generally, the durations from 1 to 2 years of the 
optimization periods  are  statistically reasonably to
react properly to different types of volatility 
and various market trends. However 6 month periods
and a simplified optimization
are good enough to keep "lines" running, until they
are redesigned on the basis of more systematic  optimization.

\vskip 0.1cm
{\sf DURATIONS.}
The end-user can request the 
{\it desired average durations\,} of 
positions. For our system, the range from 5 to 10 days
was considered reasonable.   
However, if the categories, trading modes and the companies 
to trade are prescribed, it is for the system to determine the 
most optimal "lengths" of positions. 
The positions are opened and closed 
entirely on the basis of the {\it signals\,}, so the 
{\it desired duration\,} is not imposed in any form during 
trading and tests. Generally,
if the actual duration ({\it length})
of positions during the control (out-of-sample) period
appears sufficiently close to the {\it desired duration\,}, 
then this is just a confirmation that the optimization 
was relevant. Stable {\it rhythm} is an important 
indicator of stability of the system.
%\smallskip

\subsection{\bf Testing the system}
Multiple experiments were conducted using the historical
and real-time data. Special attention was paid to trading  
liquid companies and \,SPY, the
trust that owns stocks in the same proportion as that 
represented by the SP500 stock index.
\vskip 0.3cm

{\sf CONTROL PERIODS.}
The most systematic  historical testing was for the period
2006/01/01-2007/04/13. More exactly, five 4 month's 
{\it control periods\,} (out-of-sample!) were taken:\\
Period 1: 2006/01/01-2006/04/30,\ \ Period 2: 2006/04/01-2006/07/30,\\
Period 3: 2006/07/01-2006/10/30,\ \ Period 4: 2006/10/01-2007/01/30,\\
Period 5: 2007/01/01-2007/04/13.
The last period was a little shorter. 
  
The historical testing consisted of\\
 (i) optimization during the 12 month's {\it optimization period\,} 
taken backward from the beginning of the {\it control period}, \\
 (ii) "trading from scratch" during the next 4 month's 
{\it control period\,} with closing all positions 
at the end of the period.

\vskip 0.2cm
Note that the control periods overlap (1 month), 
to simulate continuous trading, without closing 
all open positions at the ends of periods;
this is how the system really works. 
The optimization periods and the {\it corresponding} 
control periods do not overlap of course. The system
was used in the {\it pro-trend\,} variant in this test.
\vskip 0.2cm

We evaluate the  {\tt AVERAGE 4 MONTH RETURN}
for five 4 month's {\it control
periods\,} by the formula:
$$
\hbox{\tt AVRG RETURN\ }= 
88*(\sum_{i=1}^5 \hbox{RET}_i*\hbox{NUM}_i)/
(\sum_{i=1}^5 \hbox{LNGTH}_i*\hbox{NUM}_i),
$$
\noindent
where 88 is the average number of business days during
4 months, and $\hbox{RET}_i,\hbox{NUM}_i,
\hbox{LNGTH}_i$ are the corresponding
{\tt RET, NUM, LNGTH}, the average return per
position,  
the number of positions and the average 
length (duration in business days) of one position 
during the corresponding 4 month's period. 
\vskip 0.3cm

{\sf TRADING SPY (LONG ONLY).}
Let us provide the results of control "trading" \,SPY\,,
without short positions and in the pro-trend regime.
%Omitting short trading can be imposed by the end-users of 
%the system. 
Generally, trading SPY is quite a challenge;
see e.g. \cite{FPSS} concerning some aspects 
of its fluctuations. Mathematically, long and short trading,
are on equal grounds; addressing possible negative developments
is part of any risk-managements, which is quite universal.
\vskip 0.2cm

The results for the signals of 4 levels are
presented separately. By
{\tt num, ret, lngth\,} we denote the number of
(long only) positions, the returns per position, and
their durations  for each level.
The number in $(\cdot )$ is the corresponding 
{\it standard deviation\,}. The averages for all 5
periods,  
{\tt RETURN}, {\tt LNGTH}, and {\tt AVR CHANGE}
are provided. We mention that {\tt RETURN} becomes
$15.3\%$ in the (well-tested) variant with {\tt LNGTH}$=5.53d$, 
instead of $3.0d$, which can be more suitable for end-users; the
duration can be made even longer, but this can reduce profitability. 

\begin{verbatim}
TRADING SPY (LONG ONLY)
AVERAGE POSITION LNGTH: 3.0 d;
AVERAGE 4 MONTH RETURN: 14.9%;
AVR SPY 4 MONTH CHANGE: 4.80%.
\end{verbatim}
\vskip 0.1cm
\begin{verbatim}
PERIOD: 20060101-20060430, SPY CHANGE=4.6%
NUM=18      RET=0.72(0.37)    LNGTH=3.0d    ALL
num=10      ret=0.58(0.38)    lngth=3.1d   lev=1
num=4       ret=0.87(0.23)    lngth=4.0d   lev=2
num=2       ret=0.79(0.19)    lngth=3.1d   lev=3
num=2       ret=1.1(0.15)     lngth=0.5d   lev=4
PERIOD: 20060401-20060730, SPY CHANGE=-1.0%
NUM=13      RET=0.45(1.26)    LNGTH=5.2d    ALL
num=4       ret=-0.23(1.15)   lngth=7.0d   lev=1
num=3       ret=0.17(1.05)    lngth=6.3d   lev=2
num=3       ret=0.97(1.12)    lngth=3.7d   lev=3
num=3       ret=1.11(1.19)    lngth=3.3d   lev=4
PERIOD: 20060701-20061030, SPY CHANGE=9.0%
NUM=23      RET=0.56(0.43)    LNGTH=2.2d    ALL
num=13      ret=0.44(0.42)    lngth=2.1d   lev=1
num=5       ret=0.44(0.26)    lngth=2.2d   lev=2
num=3       ret=0.8(0.15)     lngth=2.9d   lev=3
num=2       ret=1.28(0.22)    lngth=2.0d   lev=4
PERIOD: 20061001-20070130, SPY CHANGE=8.5%
NUM=12      RET=0.59(0.35)    LNGTH=2.2d    ALL
num=8       ret=0.46(0.33)    lngth=2.4d   lev=1
num=3       ret=0.89(0.12)    lngth=2.3d   lev=2
num=1       ret=0.8(0.09)     lngth=0.8d   lev=3
PERIOD: 20070101-20070413, SPY CHANGE=2.0%
NUM=17      RET=0.1(1.47)     LNGTH=2.4d    ALL
num=8       ret=0.08(1.58)    lngth=2.4d   lev=1
num=5       ret=0.22(1.7)     lngth=2.2d   lev=2
num=3       ret=0.31(0.52)    lngth=2.2d   lev=3
num=1       ret=-0.94(0.02)   lngth=3.1d   lev=4.

\end{verbatim}

Short trading with a market that essentially
goes up is quite a challenge for any trading system.
Short trading here provides some "insurance" for
the periods when \,SPY\, goes down. Some losses can be 
acceptable  when it goes up, but the system actually
remains profitable.
Let us demonstrate this for the same periods and data. 
As we wrote, the bid-ask spread is not counted, not
too high for liquid assets.

 \smallskip

\begin{verbatim}
TRADING SPY (SHRT ONLY)
AVERAGE POSITION LNGTH: 3.2 d;
AVERAGE 4 MONTH RETURN: 3.15%;
AVR SPY 4 MONTH CHANGE: 4.80%.

PERIOD: 20060101-20060430, SPY CHANGE=4.6%
NUM=33      RET=0.02(0.72)    LNGTH=3.7d    ALL
num=14      ret=-0.06(0.81)   lngth=3.5d   lev=1
num=10      ret=0.19(0.69)    lngth=3.2d   lev=2
num=5       ret=-0.11(0.51)   lngth=4.6d   lev=3
num=4       ret=0.(0.62)      lngth=4.5d   lev=4
PERIOD: 20060401-20060730, SPY CHANGE=-1.0%
NUM=46      RET=0.5(0.61)     LNGTH=2.7d    ALL
num=18      ret=0.31(0.65)    lngth=2.8d   lev=1
num=13      ret=0.6(0.58)     lngth=2.8d   lev=2
num=8       ret=0.65(0.49)    lngth=2.7d   lev=3
num=7       ret=0.64(0.53)    lngth=2.0d   lev=4
PERIOD: 20060701-20071030, SPY CHANGE=9.0%
NUM=66      RET=0.04(0.77)    LNGTH=2.9d    ALL
num=24      ret=0.01(0.83)    lngth=2.7d   lev=1
num=15      ret=0.03(0.75)    lngth=3.4d   lev=2
num=14      ret=0.04(0.75)    lngth=3.1d   lev=3
num=13      ret=0.09(0.65)    lngth=2.6d   lev=4
PERIOD: 20061001-20070130, SPY CHANGE=8.5%
NUM=42      RET=0.05(0.64)    LNGTH=4.4d    ALL
num=14      ret=-0.18(0.7)    lngth=4.5d   lev=1
num=12      ret=0.11(0.56)    lngth=4.4d   lev=2
num=10      ret=0.21(0.62)    lngth=4.0d   lev=3
num=6       ret=0.18(0.49)    lngth=4.8d   lev=4
PERIOD: 20070101-20070413, SPY CHANGE=2.0%
NUM=68      RET=0.(0.93)      LNGTH=2.5d    ALL
num=31      ret=0.09(0.96)    lngth=2.0d   lev=1
num=17      ret=0.06(1.08)    lngth=2.6d   lev=2
num=11      ret=-0.17(0.68)   lngth=2.8d   lev=3
num=9       ret=-0.22(0.7)    lngth=3.2d   lev=4.

\end{verbatim}
\smallskip

\vskip 0.2cm
{\sf TRADING LIQUID COMPANIES.}
For the same periods,
let us present data for "trading" 
of 165 stocks, mostly liquid. 
It is for {\it longs and shorts\,}
and pro-trend, i.e. essentially under  
the {\it mean reversion trading}.
The
{\tt AVERAGE LNGTH}$=5$ and {\tt RETURN}$=9.56\%$ are
the averages over all 5 periods;
{\tt NUM} and {\tt num} are the numbers of
positions. 

\begin{verbatim}

AVERAGE POSITION LNGTH: 5.0 d;
AVERAGE 4 MONTH RETURN: 9.56%;
AVR SPY 4 MONTH CHANGE: 4.80%.

PERIOD: 20060101-20060430, SPY CHANGE=4.6%
NUM=2236    RET=0.64(3.4)     LNGTH=5.2d    ALL
num=1105    ret=0.55(3.57)    lngth=5.4d   lev=1
num=602     ret=0.68(3.25)    lngth=5.2d   lev=2
num=344     ret=0.81(3.31)    lngth=5.1d   lev=3
num=185     ret=0.79(2.89)    lngth=4.7d   lev=4
PERIOD: 20060401-20060730, SPY CHANGE=-1.0%
NUM=2433    RET=0.14(4.08)    LNGTH=5.4d    ALL
num=1169    ret=0.13(4.19)    lngth=5.3d   lev=1
num=628     ret=0.16(4.12)    lngth=5.6d   lev=2
num=394     ret=0.09(3.89)    lngth=5.4d   lev=3
num=242     ret=0.25(3.78)    lngth=4.9d   lev=4
PERIOD: 20060701-20071030, SPY CHANGE=9.0%
NUM=2401    RET=0.66(3.93)    LNGTH=4.5d    ALL
num=1248    ret=0.64(3.92)    lngth=4.4d   lev=1
num=619     ret=0.74(3.91)    lngth=4.5d   lev=2
num=344     ret=0.53(3.98)    lngth=4.5d   lev=3
num=190     ret=0.7(3.99)     lngth=4.2d   lev=4
PERIOD: 20061001-20070130, SPY CHANGE=8.5%
NUM=2174    RET=0.71(3.67)    LNGTH=5.2d    ALL
num=1101    ret=0.67(3.73)    lngth=5.2d   lev=1
num=566     ret=0.77(3.66)    lngth=5.2d   lev=2
num=324     ret=0.74(3.54)    lngth=5.d    lev=3
num=183     ret=0.73(3.62)    lngth=5.2d   lev=4
PERIOD: 20070101-20070413, SPY CHANGE=2.0%
NUM=1812    RET=0.65(3.05)    LNGTH=5.d     ALL
num=934     ret=0.56(3.1)     lngth=5.1d   lev=1
num=476     ret=0.79(3.05)    lngth=5.d    lev=2
num=257     ret=0.71(3.06)    lngth=4.9d   lev=3
num=145     ret=0.62(2.63)    lngth=4.9d   lev=4.
\end{verbatim}

\vskip 0.1cm
\noindent
The list of stock symbols of these companies is as follows:
\vskip 0.1cm

%\smallskip
\begin{tiny}
\begin{verbatim}
"AA",  "AAP",  "AAPL",  "ABC",  "ABT", "ACAS", "ADBE", "ADM", "ADP", "ADSK", 
"AIG", "AIV",   "ALL", "AMAT", "AMGN", "AMTD", "AMZN", "ANF",  "ANN", "APA", 
"APC", "ATI",  "AVP",  "AXP",  "BA",  "BAC",  "BBBY",  "BBY", "BEAS", "BEN", 
"BHI", "BJS", "BMET",  "BMY", "BNI", "BP", "BRCM", "BSC", "C", "CAL", "CAT", 
"CCU", "CELG", "CEPH", "CFC", "CHK", "CHRW", "CHS", "CMCSA", "CMCSK", "CMI", 
"COF", "COP",  "COST", "CSCO", "CTSH",  "CVS",  "CVX",   "D",  "DE", "DELL", 
"DO", "DVN",  "EBAY",  "EK",  "EOG", "EQR",  "ERTS",  "ESRX",  "FD",  "FDO", 
"FDX", "FNM", "FPL", "FRE", "GE", "GENZ", "GG",  "GILD", "GLW", "GM", "GPS", 
"GRMN", "GS", "GSF", "HD",  "HON", "HPQ", "IBM", "INTC", "IP", "ITG", "ITW", 
"JCP", "JNJ", "JPM", "JWN", "KLAC", "KO", "KR", "KSS", "LEH",  "LLY", "LMT", 
"LNCR",  "LOW",  "LRCX",  "MCD",  "MER",  "MET", "MIL", "MMM",  "MO", "MON", 
"MOT",  "MRO",  "MRVL",  "MSFT",  "MXIM", "NBR", "NE",  "NEM", "NKE", "NOV", 
"NSC",  "NUE",  "ORCL",  "OXY",   "PEP",  "PFE", "PG", "POT", "PRU", "QCOM", 
"RIG", "ROK",  "SBUX",  "SLB",  "SNDK",  "SPG", "STN", "SU", "SUN",  "SUNW", 
"SYMC", "TEVA", "TGT",  "TWX",  "TXN",   "UNH", "UNP", "UTX", "VLO",  "VNO", 
"VZ", "WAG", "WB", "WFMI", "WMT", "WYE", "X", "XLNX",  "XOM", "XTO", "YHOO". 
\end{verbatim}
\end{tiny}
\vskip 0.2cm

Let us combine all 5 control intervals in one period (avoiding
terminations of the ends of the intervals) and 
show all levels and the corresponding numbers
of positions taken, {\tt NUM} for all and {\tt num} for
levels; the lengths are the average durations
of the positions. One has:

\begin{verbatim}
Period: FROM  1/1/2006  TO  4/13/2007

NUM=9332     RET=0.6      LNGTH=5.5d      ALL
num=4143     ret=0.52     lngth=5.6      lev=1
num=2228     ret=0.67     lngth=5.4      lev=2
num=1285     ret=0.63     lngth=5.3      lev=3
num=735      ret=0.69     lngth=5.1      lev=4
num=416      ret=0.76     lngth=5.3      lev=5
num=237      ret=0.6      lngth=5.6      lev=6
num=131      ret=0.55     lngth=5.7      lev=7
num=76       ret=0.57     lngth=5.5      lev=8
num=54       ret=0.99     lngth=4.9      lev=9
num=27       ret=0.52     lngth=5.       lev=10.
\end{verbatim}
%\noindent

\vfil
A simplified optimization was performed here,
with only 2 {\it fixed\,} categories ($c=2,4$)
and reduced number of iterations. For this period,
24 stocks (from 165) performed negatively, including
 INTC, DELL, EBAY. Trading such 
"heavy-weighters"  generally requires
full optimization and at least 3 categories.
However here we made the optimization fully uniform for 
all companies and fast, aiming at thousands of companies. 
The optimization 
for INTC or similar, if this is the objective, must be 
done more thoroughly. The following
24 companies had negative returns:
\vfil

\begin{verbatim}
ADBE   num=   90  ret=-0.29%      lngth=3.9
AMGN   num=   49  ret=-0.48%      lngth=9.1
APA    num=   66  ret=-0.25%      lngth=5.6
BJS    num=   68  ret=-0.88%      lngth=6.9
CHK    num=   58  ret=-0.7%       lngth=8.1
CHS    num=   74  ret=-0.4%       lngth=6.1
COF    num=   49  ret=-0.05%      lngth=6.4
COP    num=   45  ret=-0.51%      lngth=9.2
DELL   num=   88  ret=-0.32%      lngth=4.8
EBAY   num=  101  ret=-0.51%      lngth=4.3
EOG    num=   82  ret=-0.64%      lngth=5.4
HD     num=   47  ret=-0.05%      lngth=8.5
INTC   num=   89  ret=-0.53%      lngth=7.1
JNJ    num=   26  ret=-0.94%      lngth=11.6
MMM    num=   50  ret=-0.53%      lngth=7.2
MOT    num=   67  ret=-0.86%      lngth=5.3
NBR    num=   80  ret=-0.99%      lngth=5.3
NOV    num=   87  ret=-0.66%      lngth=5.1
SNDK   num=   90  ret=-0.73%      lngth=2.8
SUN    num=   79  ret=-1.04%      lngth=3.8
SYMC   num=   83  ret=-0.69%      lngth=4.2
TEVA   num=   80  ret=-0.15%      lngth=4.2
TWX    num=   46  ret=-0.27%      lngth=10.4
XLNX   num=   82  ret=-0.16%      lngth=5.5.
\end{verbatim}

Here and above
only signals of levels no greater than 4 were used for trading.
We invested symbolic \$100 in every position,
so multiple signals in one direction increased this amount
up to \$400, which resembles trading on margin. The first 
signal in the
opposite direction (for this stock) results in the termination
of all positions. This regime can significantly improve
profitability. Higher levels are more frequent
for actively traded companies, so this is some kind 
of leverage.

%\vfil
We do not use weights here.
Let us just mention  that investing only in  
100 companies from 165 above with
the best optimization results
constantly improves the
performance of the systems; which is a variant
of using weights. 
However, some companies with solid
optimization returns, i.e. suitable for our 
system, performed just so-so during
the control periods. This is
the nature of stock markets, discussed well in the
literature; see e.g. \cite{YZ}.
%\vfil

Let us now provide some auto-generated results 
of {\it real-time\,} trading 
simulation with 170 companies, similar to
those listed above, under
{\it long\,\&\,short\,} with 4 levels 
(L1,L2,L3,L4),
and for 3 "production lines" (A,B,C). The lines were 
with different "opti-parameters" and/or different entry points;
"B" was counter-trend. 
The first half, "no weights",
describes the uniform trading of all companies, 
the second half is for the 100 companies 
with the best returns during the optimization:
\vskip 0.1cm

\begin{verbatim}
TRADING FROM 2007, 2, 20 TO 2007, 6, 4; ALL,  NO WEIGHTS:

RET AVR A: RETL1=0.68  RETL2=0.76  RETL3=0.89  RETL4=1.04  
RET AVR B: RETL1=0.67  RETL2=0.7   RETL3=0.86  RETL4=0.84  
RET AVR C: RETL1=0.61  RETL2=0.7   RETL3=0.75  RETL4=0.75  

TRADING FROM 2007, 2, 20 TO 2007, 6, 4; FOR 100 FROM 170:

RET AVR A: RETL1=0.57  RETL2=0.79  RETL3=1.16  RETL4=1.4   
RET AVR B: RETL1=0.96  RETL2=1.04  RETL3=1.23  RETL4=1.23  
RET AVR C: RETL1=1.08  RETL2=1.11  RETL3=1.19  RETL4=1.17.  
\end{verbatim}
\vskip 0.1cm

The returns here are per position; the average position lasted
about $5$ days; SPY increased 
5.5\% during 2007/02/20-2007/06/04.
Actually about 1000 companies were traded for this period
combined in
groups based on trading volumes, with
about 170 in each.
Every company was traded in 12 different "lines", so the
total was 72 lines. The average return was
about 0.7\% per position; the average position was
about 5 days. The results above are for 3 lines only.
%\vfil

The optimization procedure is
based on the gradient method and is actually not
far from the methods 
used in {\it networks}; see \cite{BBO,HG}. It
was almost always with solid returns
for any equities and "learning periods" in spite
of using very few parameters. This alone is some
discovery. However predicting the future is of course 
much more subtle and much less certain, in spite of
the fact that risk-taking  preferences of investors are
quite conservative. In our approach, we only try 
to predict the ways investors react to news, 
but not the news itself! See here e.g. \cite{CT} for various
algorithms used in financial mathematics.

\vskip 0.5cm
\subsection{\bf Some charts}
To clarify the logic of the decision-making inside
the system we will provide the performance graphs
describing in detail pro-trend, long\&short
"trading" \, SPY\, and\, XAU\, (Gold \& Silver) using
the historical stock quotes {\it once a day\,}. All signals,
trades, positions and returns  can be
seen under sufficiently high magnification. These
charts are upon the optimization, so we provide them 
mostly to clarify the "logic"
of the system. Generally, using day-quotes only is 
a serious demerit; the system works reasonably,
but the performance is worse than trading SPY above
with 3 quotes a day. 

\begin{figure*}[htbp]
%\begin{center}
%\vskip -0.3in
\hskip -0.2in
\includegraphics[scale=0.6]{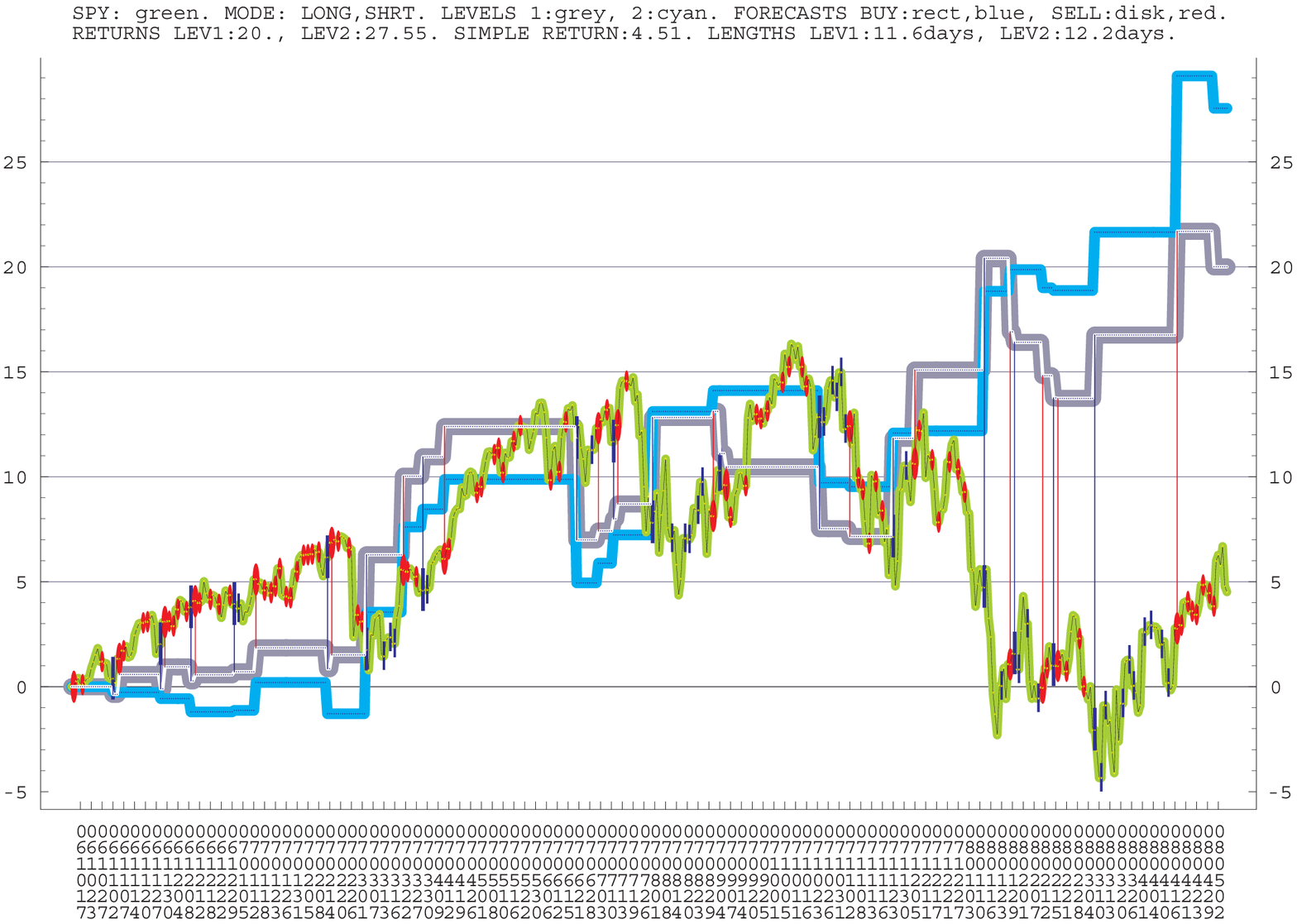}
\vskip -0.7in
\vskip -2.5cm
\caption{SPY, Long-Short, Daily Historical Quotes}
\label{fig:spy2008}
\end{figure*}

\begin{figure*}[htbp]
%\begin{center}
\vskip -1.2in
\vskip -0.4cm
\hskip -0.2in
\includegraphics[scale=0.6]{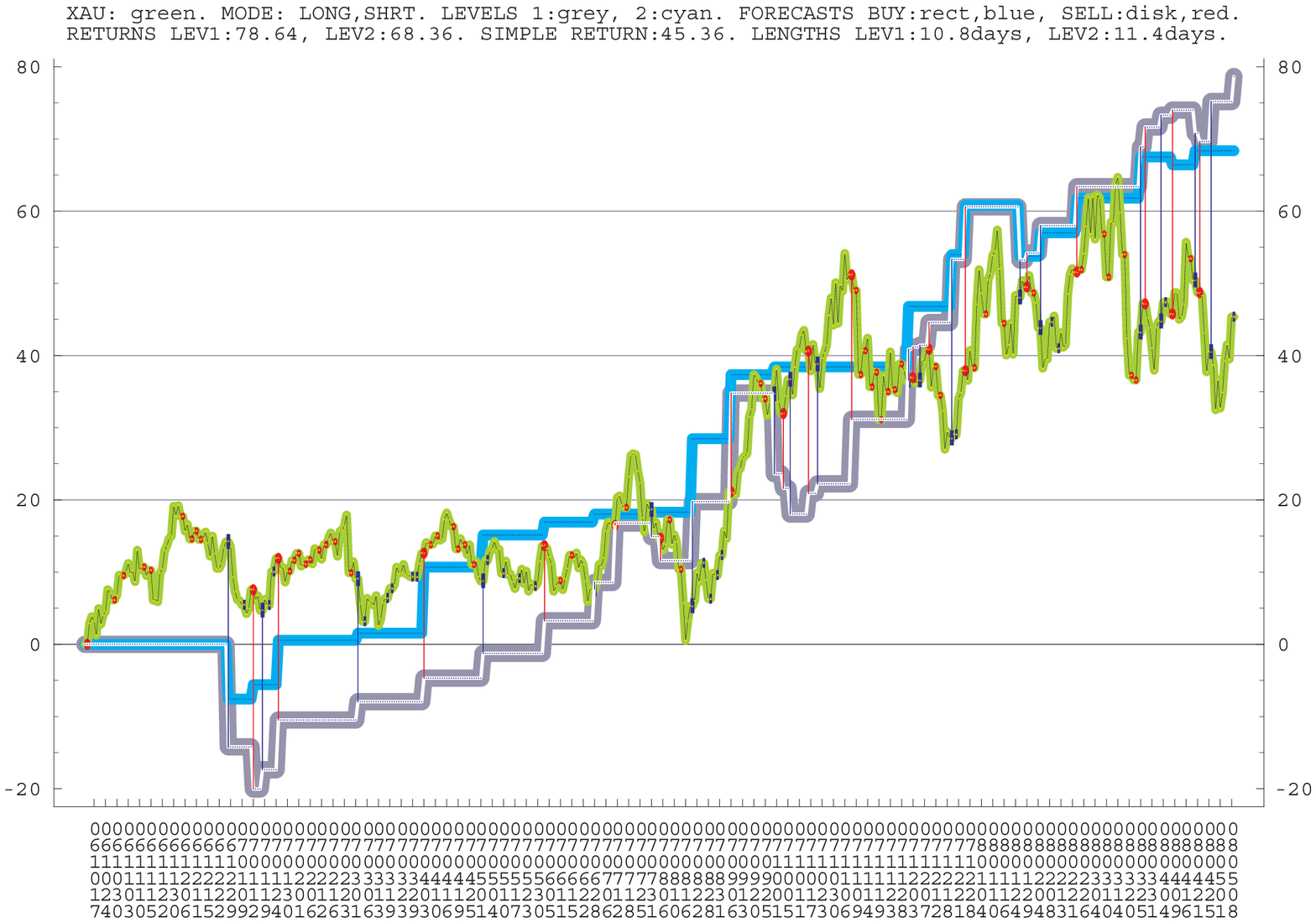}
\vskip -1.1in
\vskip -1.8cm
\caption{XAU, Long-Short, Daily Historical Quotes}
\label{fig:xau2008}
\end{figure*}

We use
green, grey and cyan correspondingly for the price-change,
the returns based on level 1 signals, and those based on
level 2 signals. Correspondingly, buy-sell signals are marked 
by blue-red rectangles-ovals; large ones mark trades for 
level 1 signals. See 
Figures \ref{fig:spy2008},\ref{fig:xau2008}. 

\vfil
\vspace* {-1.4in}

\vspace* {-0.5in}

\vskip 3.5in
%\vfil

%\vspace* {-6.5in}
\begin{figure*}[htbp]
%\begin{center}
\vskip 0.1in
\hskip -0.2in
\includegraphics[scale=0.5]{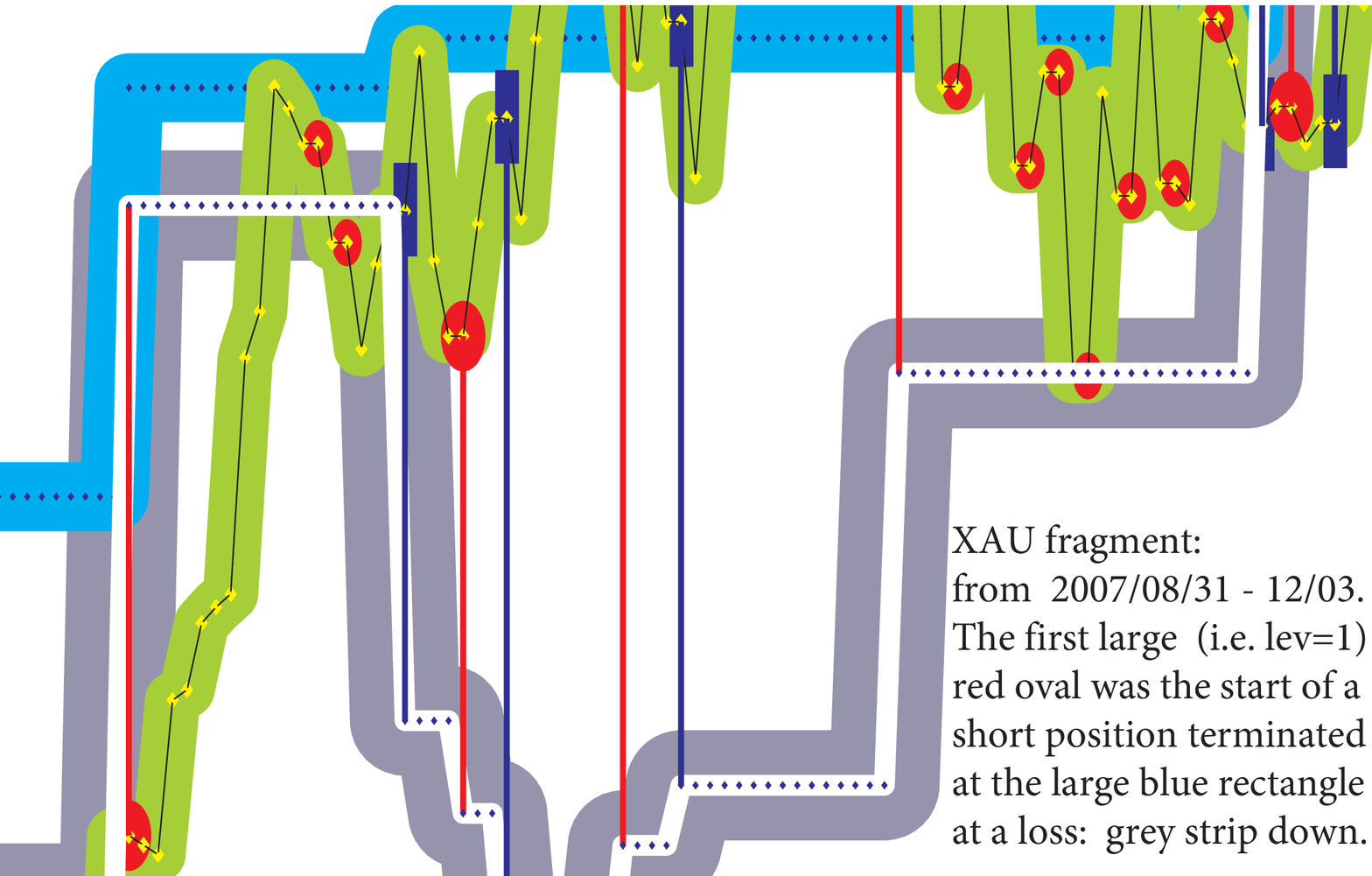}
\vskip 0.2in
%\vskip -1.0cm
%\caption{Fragment XAU, Long-Short, Daily Historical Quotes}
%\label{fig:xau-frag2008}
\end{figure*}

\vfil
Trading indices and commodities generally requires
special app\~roaches; see e.g. \cite{FPSS,GTW}.
Our system manages them reasonably, but it appeared necessary
to increase the number of used categories to 4, especially 
for \, SPY\,, versus
our usual 2-3 for individual companies. This is natural,
since indices and some
commodities are subject to many kinds of investing and hedging.
This generally  creates a lot of "noise" and makes it
difficult to catch timely their "response to news".  Their charts,
especially short-term, are of very stochastic nature. 
Nevertheless our (automated) discretization procedures and other 
parts of our algorithms proved to be efficient.
\vskip 0.2cm

The moments of buy signals (all of them, of all levels) are marked
by blue rectangles; the large ones correspond to level 1
signals. Accordingly, the sell signals are marked by red ovals;
large for level 1. The  blue and red vertical lines
connect the level 1 execution  points in the middle of 
the {\it grey} graph with those of the green equity chart.
{\it The
returns graphs are changed only upon the terminations}.  
The cyan graph is for the trades based on level 2 signals;
here vertical lines are not used. The returns are in percent
from the beginning of the graph.

\vskip 0.2cm 
To help the readers, we provide a fragment of the XAU 
Figure \ref{fig:xau2008}. For example, here the first level 1
trade, marked by a large red oval (the first such),
lasted till the first large 
blue rectangle and was executed at a loss:
a vertical drop of the grey strip after the termination;
XAU went up significantly and "unexpectedly" here.
However the next trade,
which was short on the sell signal of level 2,
shown by the next (small) red oval, 
appeared successful: a small increase of the cyan strip.

\vskip 0.2cm
{\it These two
charts are  upon the optimization\,}, so they only evaluate
the quality of the optimization, i.e. what our automated
optimization procedure produced for this period.
Only control periods (out-of-sample!)
can be used to estimate real profitability. However these
charts clarify the "logic" of the system. By the way, its unstable 
performance in the beginning can be expected; 
the system needs sufficient "history". 

\vfil   
By {\it simple returns\,} here, we mean
the total returns of \, SPY\, and \, XAU\, during the 
considered period (green
curves). Only signals of  levels 1,2 were used for
"trading" (grey and cyan). 
%\vfill
%\eject

\vskip 0.2cm
{\sf USING WEIGHTS.}
Let us provide the performance results for the following
2 periods: 3/21/2001/3/21 (9:30) - 6/14/2001/6/14 (13:30),
2000/10/24 (9:30) - 2001/6/10 (13:30), with correspondingly 
60 and 113 days  The graphs below are in terms of
"trading points", when the system "visits the market" (receives
quotes), 
here 3 times a day.
So the number of points is approximately 180 and 339 for
these control periods.  We
focus on using {\it weights\,} based on the prior 
optimization returns. Namely,  the better optimization
returns, the greater amounts to invest in this stock.
Picking the companies with optimization returns greater than 
some limit is a variant of using such weights.
The 75 companies were traded, long \& short, pro-trend
(i.e. essentially under mean reversion trading); they 
were mainly taken from the list of the most liquid ones. 
Sharpe Ratio (SR) is Mean Standard Deviation. 

\vskip 0.0cm
By "straight", we mean that symbolic \$100 were invested per
any position (long or short) for the companies with the
optimization (prior!) returns
$>0\%$ and $>20\%$. The latter bound was adjusted to reduce
the number of traded companies approximately by $50\%$. 
Generally, using the weights (or using "$>20\%$") improves the
performance, but not always significantly 
vs. "$>0\%$", depending on the market types. "Red" is used for
simple (actual) portfolio returns 
(based on the changes of share-prices), "blue" for
the returns the system achieves.

%\vspace* {-6.5in}
\begin{figure*}[htbp]
%\begin{center}
\vskip -0.0in
\hskip -0.2in
\includegraphics[scale=0.6]{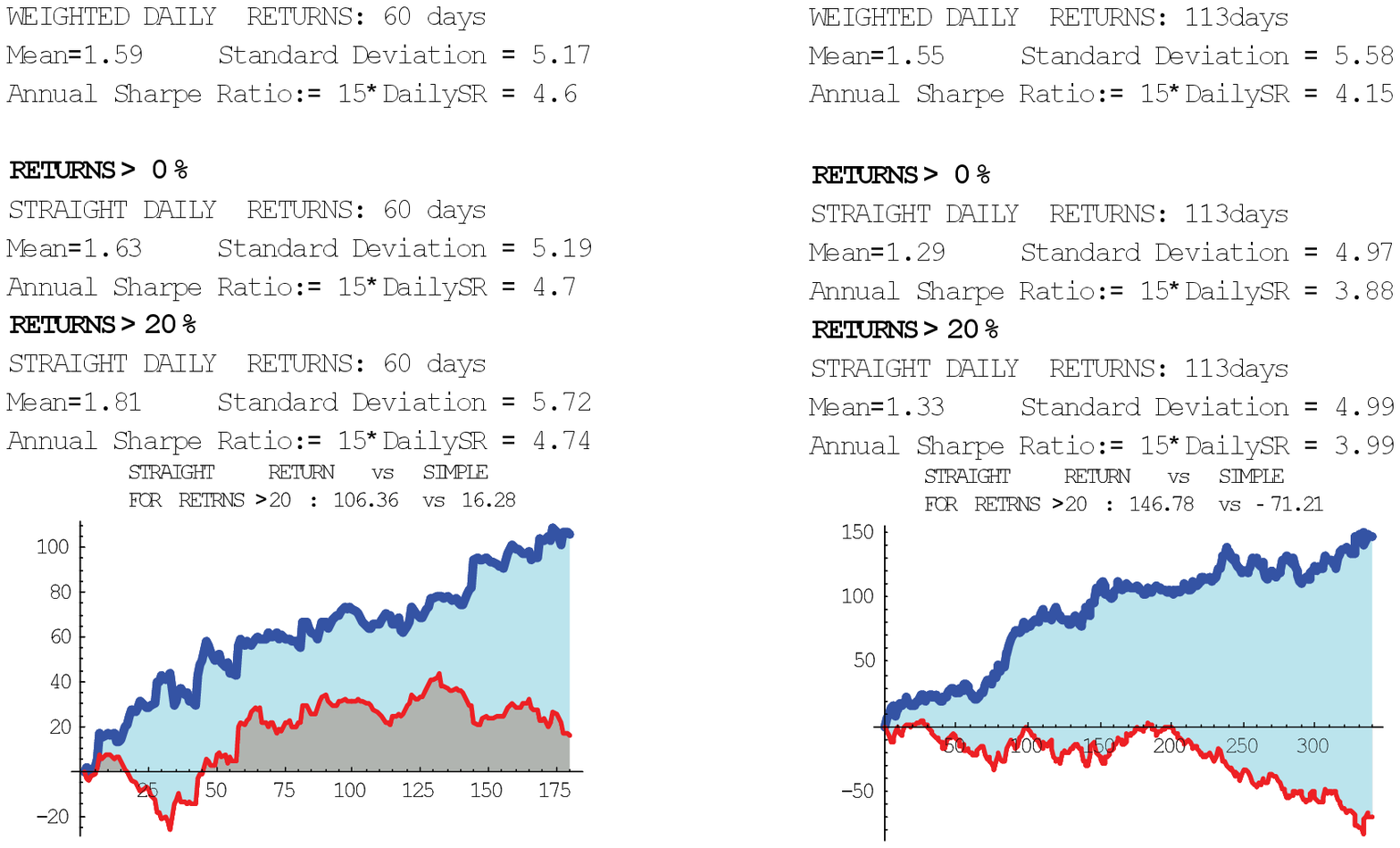}
\vskip -3.1in
\vskip -1.5cm
\caption{75 Companies, L\&S, 3 Quotes a day}
\label{fig:75}
\end{figure*}

\vskip 1.0cm
%\vfill
\setcounter{equation}{0}
\section{\sc Pont, a card model} \label{sec:4}

\subsection {\bf General design}

This game is a combination of {\it bridge\,}
and {\it Russian preference\,} with
poker-style auction. The name "bridge" was
derived from earlier  "biritch", so we make it
further from the origin (and shorter).  
It utilizes a standard  deck
of 52 cards or a smaller one of 36 cards.
The auction is quite different from that of bridge and
involves more risks. See here and below \cite{Pa}.
The bidding 
%{uses the relative number of tricks to be taken and} 
does not use the denomination
of suits. The player who starts the auction
has no advantage. The cards may be updated while bidding, 
which resembles draw poker.
% Poker lovers will enjoy it.
The winner of the auction, the declarer, determines the final
number of cards per hand as part of the  declaration of the contract:
the trump and the minimal number of tricks to be taken. 

\vskip 0.2cm
Following suit and the use of trump cards is similar to 
bridge-type games. The scoring is simpler than that of bridge.
The declarer's award is based on 
the value of the contract depending upon whether or not it  was made.
The game can be for 2, 3, 4 players, 2 partnerships, or 
for 1 versus computer.
There is also a poker-like version.
All variants are almost equally dynamic and playable.
% {and exciting}
%suitable for players of all ages.
%The game, especially the auction, is a certain model of short term 
%investing {with  the bullish strategy prevailing over the bearish one}.

\vskip 0.2cm
{\sf Stock market connections.}
The game, especially the auction, can be considered as a
simple  model of playing the market, especially
under momentum "investing on news". The bids then
are some counterparts 
of the forecasts of share-prices. 
The play checks  the quality of the bid, 
but this is not related to real trading, where this quality
is the return upon the termination of the position taken. 
 
The number of cards per hand 
and the number of taken tricks reflect respectively
the duration of the investment  and the return.
The downplay and {\it mis\`ere} resemble a bit selling short,
but this is superficial. 
This is a game: just a model.  

The suits are substitutes for the time-horizons of investments or
the companies considered for investing. They are on 
equal grounds in {\it pont\,} in contrast to other bridge-type games. 
Given a suit, the better cards the more reasons to make it a trump.
In our trading system, the category of the top bid
determines the time-horizon of the investment, though
the categories are ranked in contrast to suits in {\it pont}. 

The players compete to become  {\it the declarer}, 
which is somewhat similar  to winning  the
"right" to invest. The upgrades and increases  are designed 
to reflect real-time actions. The bids are actually {\it 2-bids\,},  
which adds some "timing"; they depend on the size of hands
(from 6 to 9), which has no 
counterparts in other bridge-type games. 

The {\it play\,} itself 
has little to do with real
playing the stock markets.
For instance, the use of trump cards and 
positions of players around the table have no 
market analogues. The role of such special
elements of card games  is diminished in {\it pont\,}.
However they are inevitable; the game must be 
not too primitive. Also, more playable games have 
stronger roots in our psychology. Making {\it pont\,}
playable was a challenge since it uses unusual
{\it fractional bids\,}, related to 
our approach to risk-taking. This was a 
test of the principles of our trading system.
we think that playing {\it pont\,} can help
to get used to our {\it 2-bids\,} and in real
playing the stock markets, possibly  better than 
playing poker or bridge.

%\vfill

\vskip 0.3cm
\vfil
\subsection{\bf Description}

%\vskip 0.3cm
The game uses a standard card  deck of 52 cards for 4 players 
or a smaller, four-suit deck of 36 cards (from the ace
down to the 6), when there are two or three players. 
In the case of 4 players, they may  divide themselves into
two partnerships; here the whole deck is used too. 
The dealers are changed clockwise after each game. 
The cards are dealt singly in the clockwise order
and face down, giving each player six cards.
After the players pick up their hands,
the dealer starts the auction by making the bid or passing.
\vskip 0.3cm

\vfil
{\sf Auction.} A {\it bid} is  a fraction N/D 
with the denominator D is from  6 to  8 and the numerator  N is
no larger than  D.
Generally speaking,
the bid is the expected
number of tricks to be taken (N) divided
by the final 
number of cards per hand (D). The latter may be from 6 to 9.
The fraction must be no smaller than 3/6 for 3 or 4 individual 
players,
and no smaller than 4/6 for 2 players or partnerships.
The fractions 4/8, 7/7, 8/8 are excluded.
The bids 3/6, 4/7, and 5/8 are not allowed for 2 players, 
but are accepted  for 3 or 4 players.

The auction proceeds clockwise with each player either making a bid
that is not lower then the previous ones of other players;
for instance, 4/6, 5/7, 6/8, 5/6, 6/7, 7/8,
6/6  may be claimed after 4/6. Otherwise say "pass".
Bidding is forbidden after the first bid was made
if a player has already passed.  
Passing is allowed after bidding only if there are other players who
did not pass; also, 
the last remained (survived) player may not pass.
The round of bidding continues until the {\it last bid}, when 
a player (who then becomes the {\it closer}) 
repeats his/her previous bid for the first time, or simply says 
``close''. If the others (two opponents for the team variant)
passed after this, the closer becomes  the {\it declarer}. 
Otherwise there is no declarer.  

More rounds are necessary if  all players passed or at
least two of them claim the same bid.  
To start the next round, the dealer  {\it upgrades} the cards,
giving out a card per hand face down. Then
each player picks up the card and after this removes
one card from the hand by laying it face down. I.e.
the hands must be 6 again. Then the closer (or the dealer 
if all passed)
claims first, repeating or enlarging his/her last bid,
and the auction continues following the same rules
until the first repetition.
Those who passed during the 
previous rounds do not bid, unless all passed.
The cards may be upgraded only twice.
\vskip 0.3cm

\vfil
{\sf Taking no tricks.} If all passed after the last (the second)
upgrade, the dealer leads to start
the {\it downplay} notrump,  where the players are trying to win
the smallest number of tricks. Also, the closer starts the
downplay if  two or more players (or both teams, 
if applicable) 
do not pass after the second upgrade, which is the last, but
claim {\it coinciding\,}
 bids, i.e, neither of them is the winner.   

At the end of the game, the numbers of taken tricks
will be  diminished by the minimum number, which is 
to make it zero at least for one player, 
and subtracted from the corresponding scores. 
In the case of 2 players, this diminished  number
must be divided by two before subtracting; e.g.
the player who took  4 tricks will  loose 1 point, 
which is 
4 minus 2, the number of tricks of the opponent, divided
by 2.

A player may claim  {\it mis\`ere},
which means that no tricks will be taken.
This may be done only before the first upgrade, and
bitten by 6/7 or higher for 2 players (teams), 
by 5/6 or higher for 3 or 4 players. 
{ \it Mis\`ere} is played notrump.
The declarer makes the opening lead by placing the card on the
table face up. 
If there are 3 or 4 individual players 
all cards are placed face up on the 
table after this. 
It is the same for partnerships, but the 
partner  does not participate  laying his/her
cards  face down.
The mis\`ere contract is defeated if either of the
opponents finds the way
where the declarer takes at least one trick.

\vskip 0.3cm
\vfil
{\sf The play.} After the auction, the declarer may {\it increase},
asking the dealer to deal out 1 card per player face down.
The procedure can be repeated several times,
but the maximal number of cards per hand must  be no 
greater than 9.
The declarer picks up the cards every time. The others will do
this only after the declaration of the contract.
Then the declarer declares the {\it contract},
choosing the trump suit or notrump, which is allowed,
and stating the minimal number of tricks to be taken
(including the partner's tricks for the partnerships).
The denominator "D" equals the number of cards per hand after the
last increase.

{\it The number of tricks to win
cannot be smaller than the final number of cards per player
(after the last increase)
times the fraction from the last declarer's bid}. 
The bid "{\it mis\`ere}" can be changed by the declarer
by the contracts 6/7, 7/8, 8/9, 6/6, 7/7, 8/8, 9/9.
It is the same for 2,3,4 players, and the partnerships.
Also, the last bid 6/8  can be changed by {\it mis\`ere}
if there were no upgrades and increases.
The partner's hand is discarded {\it face down\,} when playing 
{\it mis\`ere}
in the team variant.

%\vfil

The declarer starts the {\it play} trying to take enough 
tricks to fulfill 
the contract or take no tricks for {\it mis\`ere}.
For partnerships, anytime during the play 
the declarer may ask the partner to place
all his cards face up on the table and then he/she
starts playing both of the 
partnership hands (unless in {\it mis\`ere}).
All players have to follow suit if they can.
Otherwise they must trump. Only the declarer may lead a trump.
Other players may do this only if they have no other suits left.
The play lasts until
the declarer (together with the partner
if applicable) takes the necessary number of
tricks or the contract is defeated. 

\vskip 0.3cm
\vfil
{\sf Score system.} At the end of the play, the declarer's 
{\it score} goes up
by the {\it value} of the contract, the number of tricks
from the contract minus 3 for 2 players (or partnerships) and 
minus 2 for 3-4  individual players, if
the contract was made.
Otherwise  this value  is subtracted from the score.
If the last bid {\it before the first upgrading\,}
was more or equal than 5/6, then
this value goes up by one, called {\it premium}
(when adding or subtracting). 
The same {\it premium} is added to 
{\it mis\`ere}, treated as 5/6 when calculating the score
(3 points for 2 players/teams and 4 points for 3,4 
individual players).
A fulfilled  contract  of fraction $=N/D=1$,
gives  1 {\it bonus point\,} for 2 players (partnerships)
and 2 bonus points for 3-4 individual players. 
For 3 or 4 individual players, successful contracts
5/6, 6/7, 7/8, 8/9, 9/10, and  mis\`ere 
add 1 bonus point to the declarer's score. 
In contrast to the premium,
the bonus is not subtracted from the score if the contract 
fails. 

\vskip 0.2cm
There is another, somewhat more involved,
variant of the {\it pont score system\,} with more "punishment"
for defeated contracts. The play goes till the end.
If the number of taken tricks is less than it was declared
than the score of the declarer is diminished by the 
value of the contract multiplied by the number of missed
tricks. Say, if the declare took the necessary tricks but one,
the score becomes smaller by the value of the contract. This
score system is for experienced players.

Finally, the {\it rewards} will be proportional to the scores 
of the players diminished by
their arithmetic mean, that is the total
of all scores divided
by the number of players.
The partners may redistribute
the total partnership reward (the sum of their rewards).
The standard recommended way is as follows. If both rewards 
are positive
or negative then it is the same as for individuals.
If the first reward is positive, the second is 
negative, and the total is
negative, then the first partner doesn't pay.
If the total is positive here,
then the second partner receives nothing (and pays nothing).

%\vfil
\vskip 0.3cm
\vfil
{\sf Bidding table.} The following table is the list of  bids in the
increasing order and the corresponding minimum
contracts for different numbers of cards per hand.
The stars (adding  1 point each to the score) show
the premium {\bf p}
for declaring during the {\it first} round of the auction 
and the bonus {\bf b} for making the contract. 
\vskip 0.2cm
\halign{\hfil\ \ #\quad\hfil&
\ \ \ \ \quad\hfil#\ \ \ \ \ \hfil&\hfil\quad#\ \ \hfil\cr
  3--4 individual players    
& contracts 
& 2 players(partnerships)\cr
}
\halign{\hfil\ \ \it#&\quad\hfil\bf#&\quad\hfil#\hfil&\quad\bf#\hfil
&\quad\it#\hfil\cr
  names
& b\ p\ \ bids:
& tricks / cards 
& :bids\ \ p \ b 
& names \cr 
  \quad 1
& 3/6:
& 3/6, 4/7, 4/8, 5/9
& : --- 
& ---\cr
  1+1
& 4/7:
& 4/6, 4/7, 5/8, 6/9 
& : ---
& --- \cr
  1+2
& 5/8: 
& 4/6, 5/7, 5/8, 6/9
& : ---
& --- \cr
  \quad 2
& 4/6:
& 4/6, 5/7, 6/8, 6/9 
& :4/6
& 1  \cr
  2+1
& 5/7:
& 5/6, 5/7, 6/8, 7/9
& :5/7
& 1+1 \cr
  2+2
& 6/8:
& 5/6, 6/7, 6/8, 7/9
& :6/8
& 1+2 \cr
  m
& *\ *\ \  m/6:
& ...., 6/7, 7/8, 8/9
& :5/6\ \ \ *\ \ 
& 2\  \cr
\quad 3
& *\ *\ \ \ 5/6:
& ...., 6/7, 7/8, 8/9
& :m/6\ \ *\ \ 
& m \cr
  3+1
& *\ *\ \ \ 6/7:
& 6/6, 6/7, 7/8, 8/9
& :6/7\ \ \ *\ \ 
& 2+1 \cr
  3+2
& *\ *\ \ \ 7/8:
& 6/6, 7/7, 7/8, 8/9  
& :7/8\ \ \ *\ \ 
& 2+2 \cr
  \quad 4
& **\  *\ \ \ 6/6:
& 6/6 ,7/7, 8/8, 9/9
& :6/6\ \ \ *\ *
& 3.\cr
}
\vskip 0.2cm
\noindent Here {\it mis\`ere} ({\it m}\ =\ {\bf m/6})
has the same list of admissible contracts
as {\bf 5/6} but is ranked  higher for 2 players (partnerships) and
lower for 3 or 4 individuals. 
Recall that the {\it mis\`ere} contract 
may be played after the last bid {\bf 6/8}
or smaller; {\bf m/6} is omitted in the
column of contracts.

The names  of the bids are convenient when
bidding. The name gives the number of additional
card (after +) and the value of the 
(lowest)  contract coinciding with the bid,
calculated without the premium and bonus. 
For instance, the value of {\it 2+2}\ =\ {\bf 6/8} for 3,4 players 
equals 2+2=4. For 2 players, the contract {\it 1+2}\ =\ 
{\bf 6/8} gives 3 points.

\vskip 0.3cm

\subsection{\bf Variants}\label{sect:PP}
{\ }
\vskip 0.2cm

%\vfil
{\sf Basic-Pont (BP).} The simplest version of the game is
the {\it basic pont}, which is  played without {\it mis\`ere},
and "premium".
The table is  also simplified by dropping the bids of
denominator 8 (the {\it +2}-bids):
\vskip .2cm
\halign{\hfil\ \ #\quad\hfil&
\ \ \ \ \quad\hfil#\ \ \ \ \ \hfil&\hfil\quad#\ \ \hfil\cr
\ \ \ \ \  3--4 individuals     
&\ \ \  contracts 
&\ \ \ \ 2 players(teams)\cr
}
\halign{\hfil\ \ \it#&\quad\hfil\bf#&\quad\hfil#\hfil&\quad\bf#\hfil
&\quad\it#\hfil\cr
  names
& b\ \ \ \ bids:
& tricks / cards 
& :bids\ \ \ \ b 
& names \cr 
  \quad 1
& 3/6:
& 3/6, 4/7, 4/8, 5/9
& : ---
& ---\cr
  1+1
& 4/7:
& 4/6, 4/7, 5/8, 6/9 
& : ---
& --- \cr
  \quad 2
& 4/6:
& 4/6, 5/7, 6/8, 6/9 
& :4/6
& 1  \cr
  2+1
& 5/7:
& 5/6, 5/7, 6/8, 7/9
& :5/7
& 1+1 \cr
  \quad 3
& *\ \ \ \ \ 5/6:
& 5/6, 6/7, 7/8, 8/9
& :5/6
& 2 \cr
  3+1
& *\ \ \ \ \ 6/7:
& 6/6, 6/7, 7/8, 8/9
& :6/7
& 2+1 \cr
  \quad 4
& **\ \ \ \ \ 6/6:
& 6/6, 7/7, 8/8, 9/9
& :6/6\ \ \ \ \ *
& 3.\cr
}
\vskip 0.3cm

\vfil

{\sf Poker-Pont (PP).} Another variant is {\it poker pont\,} for
2, 3, 4 individual players. It follows the table of the {\it basic 
pont\,},
without bonuses.

\vskip .2cm
\halign{\hfil\quad\bf#\hfil&\quad #\hfil&
\hfil\quad\bf#\hfil&\quad #\hfil
\cr
  bids:
& contracts 
     & 5/7:
     & 5/6, 5/7, 6/8, 7/9
\cr  
  3/6:
& 3/6, 4/7, 4/8, 5/9
     & 5/6:
     & 5/6, 6/7, 7/8, 8/9
\cr
  4/7:
& 4/6, 4/7, 5/8, 6/9
     & 6/7:
     & 6/6, 6/7, 7/8, 8/9
\cr
  4/6:
& 4/6, 5/7, 6/8, 6/9 
     & 6/6:
     & 6/6, 7/7, 8/8, 9/9.
\cr
}
\vskip .2cm

%\vfil
{\it PP-betting.} As in  poker, each player puts up  {\it ante}
(one chip or more) to form  a {\it pool}, which
consists of the {\it pot} and the sectors, one for a player.
A player always puts chips in the corresponding sector.
The dealer starts {\it betting}, adding chips to the pool
or putting nothing, {\it passing}.
The player on the dealer's left may pass,  
{\it call} by putting the same,
or {\it raise} by adding extra chips of his/her own. Other players
continue clockwise until all  have finally called any raises.
A player may 
raise after passing if the latter was before the first raise.
Passing is allowed after raising or calling
if there are other players who did not pass.

The first player who calls without adding  is the {\it closer}.
If all other players passed, the closer is the {\it declarer}.
If there is no declarer,  the dealer upgrades cards and 
the closer starts another round
of betting by raising or doing nothing. Those who 
passed before if at least one player raised  may not bid.
There are 
{\it optional  upgrades} in {\it poker pont\,}; they  may be omitted.
Then the card will be dealt to the next player.
A player who upgrades  puts a chip to his/her sector (per each 
new card). 
The number of upgrades is no more than 3 (for 4 rounds of betting). 
If still all pass after the last upgrade then 
the {\it ante} goes to the pot,
the dealer is changed clockwise, and a new game starts.

%\vskip 0.3cm
%\vfil
{\it PP-play.} If there are two or more players who put the same 
number of chips
(regardless of the extra chips for upgrades which may be 
different), then
the closer begins one round of bidding among those players
only.  It is as in the {\it basic pont\,};
the declarer is a player claiming the highest bid.
If still there is no declarer, there will be no more upgrades.
Then the dealer moves all chips but {\it ante} from  the sectors 
to the pot, and the next dealer starts a new game.

The declarer may increase several times (no more than 3), adding
a chip per increase to the pool. 
After the declaration of
the contract all opponents  pick the  cards and
{\it respond}  or pass clockwise
starting with the first on the declarer's left.
One must add one chip per each increase (totally, the current
number of cards per hand minus 6) to the pool
to respond  and become an {\it active opponent}. 
Other opponents are {\it passive}.
However all participate in the play, which follows
the standard rules.

The declarer leads and wins the pool 
(including the pot) when making the contract.
If the latter is defeated then all the opponents, active and passive,
take  chips back from their sectors and the active opponents
divide the declarer's chips and the pot
among themselves proportionally to the number
of taken tricks. The fractions are ignored and the remaining chips
(if any) go to the pot.
If there are no active opponents, the declarer takes his/her chips
back even in the case of the failure (but not the pot).
The contract has to be the minimal possible for the current number
of cards per hand. Namely, 3/6, 4/7 for 3-4 players,
4/6, 5/7 for 2 players, and  6/8, 6/9 for either. It 
may not be  lower than the last bid if there has been 
a round of bidding to determine the declarer.

%\vfil
%\vskip 0.3cm
\subsection{\bf Comments}
{\ }\vskip 0.3cm

%\vfil
{\sf Additional rules.}
Extra penalties can be added for breaking the
rules. The opponents may decide to diminish the declarer's or 
partnership's score by  the value of the contract if the declarer 
(partnership) 
made a mistake against the rules when playing. 
Vise versa, in the case of opponent's mistake,
the declarer has the right to consider the
contract to be fulfilled and the other player(s) 
may decide to subtract its value (or its doubled value) 
from the score of the 
opponent whose fault is it. In the {\it poker pont\,}, the contract
is considered to be defeated in the case of a declarer's mistake.
If it is an opponent's mistake,
the chips from the pool are distributed as if the contract were
defeated, and the
opponent who made a mistake gives this very number of chips
to the declarer. These are basics, to be developed by players.
%\vfill

\vskip 0.1cm
The following regulation could improve the
coordination of the opponents (for 3 or 4 individual players)
and may be added to the rules. 
The opponents  has to   play
the lowest  card higher than the card of the  declarer\& partner
to win the trick if they can.  
However the card must be  the lowest possible
to leave the trick to an
opponent whose card already beats the cards
of the declarer\&partner.
As to the partnerships, a general regulation is to  
at least repeat the bid of your partner if
you have 2 sure tricks or more,
i.e. could win two tricks for any trump.
For instance, it may be either "A A", or "A K" in the same suit, 
or "A" in one suit and "K Q" in another. When you pass, 
but the opponents don't, it can stimulate your partner
to pass or  claim {\it mis\`ere}. To avoid this,
it makes sense to bid if you can count on 3 (or more) tricks upon 
declaring your trump, especially if you have honors and the hand 
seems good for the increases. Just to give an example 
of such {\it coordination}. 

\vskip 0.2cm
\vfil
{\sf A computer version.}
The computer realization of the variant for two players
is based on the following principles. The computer is programmed
to selects the best one
considering several random choices of the hand of the player
(taking into account all information about
the cards of the player appearing during the play). 
It does the same when bidding and declaring, but
diminishes the most likely bid and contract by one level.
The simplest {\it one-way version} is when the computer never bids
(and has no score),
and the player either determines the contract (without upgrades)
and then plays  following
the standard rules or passes subtracting
2 points from his/her  score.
It follows {\it basic pont}. 
However the bidding scale and the admissible contracts starts
from 4/7 considered as {\it 0+1} and giving
1 point. More generally, {\it 0+k}, which means (3+k)/(6+k),
 is counted 
as k points for k=1,...,6.

The computer basic strategy
is to win  the trick leaded by the player with the smallest
possible card and to play the lowest
card otherwise. If it has no proper suit and no trump left,  the
card can be the lowest (from the shortest suit,
if there are several cards of the same 
rank). However the suits where the player has no cards 
according to the information during the process of play 
are considered  the best.
When the computer leads, then
the suit where the player has no cards 
is the first choice too.
If the  highest card (one of them if there are
several of the same rank) 
has the {\it adjacent\,} one in the same suit 
(say, the pairs "A K" or  "K Q"
are adjacent),
or the next card in the suit is lower by 4 or more 
(say, "A 10"), then this is the second choice for leading.
Otherwise the suit must be the
longest and the card the highest among the longest suits
However the  longest suit where the two highest cards  are
adjacent is considered first. 
If still there are several
choices the computer decides randomly. These are of course
very basic considerations; the actual computer
program can be significantly more developed. 
%The player is trying to make the score the best. 

\vskip 0.1cm
\subsection{\bf Concluding remarks} Let us stress that
our trading system is not a {\it black box\,}; the logic of 
its decisions concerning trading stocks
(any instruments) can be fully reconstructed  and
understood; cf. \cite{HG}.  We found not many situations
where its decisions could be questioned on the basis of the 
usual {\it technical analysis\,}, though the system uses
the stock-charts and its own prior decisions in novel ways.
{\it Pont\,} clarifies some principles
of our approach and test them "psychologically".
We also hope that playing {\it pont\,} can help to get used to
our {\it 2-bids}.
%; it seems closer to momentum investing
%than poker and bridge.
%\vfil

The bidding
table of {\it pont\,} and the one  used for system's 2-bids
$(b,c)$ are similar, and this is not just an analogy!
The auction and bidding seem fundamental for any 
intelligence. This  can be 
within some expert system, inside our brain or AI. 
Poker and contract card 
games serve well the humankind as a
risk-taking playground: they obviously capture
something important about human cognition. See \cite{Pa}.
    
\vfil
Obviously, using computers makes bidding 
formal and not "immediately understandable". 
The automated optimization and {\it deep learning} are  
even more difficult to interpret, even if every optimization step 
can be seen in detail, as in  our programs. Generally, machine 
learning
is fully "trustworthy", only if the results can be clearly 
interpreted "humanly".    In our trading system,
the optimization is mostly of this kind
due to a small number of parameters our system deals with.
The main are  the categories, the modes (long/short, 
pro/counter), key thresholds, and some
derived parameters like the average duration of positions;
all are meaningful to investors. Our usage of power functions
in the tables of 2-bids has solid grounds too, as we tried
to demonstrate.
 
\vfil
The discretization, which is necessary to separate noise
from signals, is not really "intuitive", but the "action 
potentials"  are always necessary and
any usage of computers requires discretization. In our 
trading system we made the discretization as "human" as possible.
The author of the paper is a specialist in discrete theories
(mostly "integrable"), but the market reality 
resulted in non-standard auction-style 
{\it stratified discretization}.
It is new, though
using the data stratification and sample curves is common in 
{\it neural networking}. It is likely that our approach
reflects the risk-taking processes in our
brain; its successful market
implementation can be regarded as some confirmation.

%\vfil
The importance of finding optimal relation between
the decisions and sampling frequencies is well recognized. 
Let us quote \cite{Si}:

{\it Though available data are sampled at discrete intervals of 
time - daily, weekly, and so on - it need not be the case that 
economic 
agents make their decisions at the same sampling frequency. 
Yet it is not uncommon for
the available data, including their sampling frequency, 
to dictate a modeler's
assumption about the decision interval of the economic agents in the
model. Almost exclusively, two cases are considered: discrete-time 
models typically
match the sampling and decision intervals - monthly sampled data
mean monthly decision intervals, and so on - whereas 
continuous-time models
assume that agents make decisions continuously in time and then 
implications
are derived for discretely sampled data. There is often no sound
economic justification for either the coincidence of timing in 
discrete-time
models, or the convenience of continuous decision making in 
continuous-time models.} 
\vskip 0.1cm

This is actually the key problem we address in our trading
system and this paper: how to coordinate different 
"decision intervals"
and what is optimal decision-making based on a simultaneous
analysis of several "frequencies". This is a must
for AI systems focused on trading and of obvious 
importance well beyond stock markets.
\vskip 0.1cm

Timing the market is and always was a great challenge, but
now we have a new chapter: a systematic 
AI-based {\it research} and optimization of the process of
investing. The usage of AI is a must here because 
the only reliable way to test performance of any trading system
is {\sf (a)} when it is fully(!) automated (machine learning 
included), \ {\sf (b)} when someone else (not the creator)
runs it, \ {\sf (c)} the design and analysis of the experiments
is as rigorous as possible, and \ {\sf (d)} all findings are
confirmed by real-time trading, which obviously requires 
full automatization. 
\vfil

We provide a sufficiently complete description of the basic principles 
of our trading system and the ways it was tested.
Not all aspects of our approach were addressed here.
The system consists of a lot of programs;
many are used for technical processing data, including
but not limited to managing historic and real-time quotes, 
practical matters like
splits-dividends, and so on. Quite a few serve the 
optimization, historical and real-time. The real-time
optimization uses the system own 
history of trades, upgrading the parameters is 
"while trading" (normally during weekends). 
Historic simulations require a lot of special
software too. This is on top of {\it actual trading programs\,} and
those monitoring the performance. 
The coordination of such
a ramified combination of service, optimization and action 
programs  is quite a test for any system; this is no different
from the ways our brain works.

%Obviously quite a few
%traders have their own systems and programs.

%We hope that the presentation of the main principles of  our approach
%and some of our findings can stimulate 
%research in the direction of {\it AI-based general 
%risk-management}.
\vskip 0.1cm

{\sf Beyond stock markets.}
 The stock markets can be considered as a great model
for many aspects of
decision-making. In our approach, the
impact of "events" is measured {\it indirectly}, via the
responses of the "agents", which is quite standard  in 
sociology and statistical physics. Many of
our findings seem of universal nature.
%
% In sociology, this is similar to
%theories o {\it mass behaviour\,}, not only for humans of course.
For instance, our equations connecting
{\it price-functions\,} with the {\it news-functions\,}
can be equally used to model 
the relation between the {\it expected\,} resources for a task and 
those actually used, presumably including our brain. 

"Investing" has its special features.
Under {\it momentum risk-taking\,}, 
the agents seek to optimize
their actions: {\it (a)}\ entering the "game" quickly  when
a clear signal is detected, and {\it (b)}\ exiting when some 
price-function reaches expected levels. Practically, we use
here {\it tables of  2-bids\,} and 
{\it forecasting-termination curves}. Theoretically, 
power functions and their generalizations,
Bessel and hypergeometric functions, are of significance here,
as it was demonstrated in Section
\ref{sec:2}.

Importantly, we focus on the time-intervals when
the {\it news impact} remains growing. The main reason is
obvious: an objective of any trading
is to capture local maxima of price-functions. Also, it is
quite likely that {\it other\,} events and profit-taking will
occur before the "natural decay" of the news impact.
This is common in mathematical finance and physics
to analyze the asymptotic behavior of correlation
functions, which generally vanish at infinity; we 
model only short-term impacts.
\vskip 0.1cm

Our brain can be considered as a kind
of social system, though with  huge number of neurons and 
very complex interactions. Assuming this,
"events" reveal themselves via some waves of
"mass behavior". Accordingly, such waves are likely 
to be the main information available to individual neurons
and the key source of their "decisions",  
governed by {\it action potentials\,} and similar mechanisms.
Eventually our brain creates some "images" of the 
underlying events. We are even able to 
form  this way {\it abstract concepts\,}, 
such as {\it space-time}. 
Philosophically, let us at least mention here
Kant; see e.g. \cite{Ja}. 

Our analysis of stock markets,
especially the simplicity of the basic differential 
equations we propose, can be an indication that the 
{\it power-laws}\, for the impacts of "events", auction-type
procedures, and
certain {\it price-functions} are present 
in the biology of the brain at the neural level. 
This is  related to {\it neural networking}.
The {\it price-function} generally measures the current 
{\it importance\,} of the event and the corresponding {\it expected\,}
resources needed for its analysis. 
Our brain will try to diminish the neural activity 
when some "price-levels" are reached, though {\it the price\,}
(the importance) varies depending on the intensity of the
triggered brain activity. This can even result in {\it periodic\,} 
"waves of interest"
in an event: an auto-mechanism for its abiding analysis, 
which we mathematically associate with Bessel functions. 
%We hope that finding such mechanisms can be within the reach of modern 
%biology. 
There is of course some "macro-management" too,
say timely corrections of the failed decisions. The
mechanisms of such conscious or unconscious  "re-visiting" 
the analysis of past events are obviously complicated.

\vskip -0.2cm

\subsection{\bf MRT: main findings} 
{\ }

{\sf (1) Cognitive science.}
The origin of our approach in cognitive science is
the concept of {\it momentum risk-taking}, {\it MRT}, which can be
defined as short-term decision-making and forecasting based 
on the real-time monitoring the actions of other agents.
%Accordingly, the information on the actual events is indirect
%and mostly incomplete. 
Poker and our {\it pont} are good
examples of games with similar data, but stock markets are of 
course the main source of this concept. In contrast to 
{\it thinking-fast\,} and {\it thinking-slow\,}
from \cite{Ka}, when the "agents" can generally choose between
two modes of thinking (unless in specially crafted 
experiments), there is no such choice here and high 
uncertainty is generally involved. 
Investors are assumed to decide "optimally" on the basis of
the current news impact.
 Our restriction to short-term
decisions and forecasts makes it possible to 
propose a mathematical, {\it quantititative\,} 
model of {\it MRT}, in contrast
to {\it thinking-fast}, which is generally {\it qualitative\,}.

\vfil
\vskip 0.1cm
{\sf (2) Toward general-purpose AI.}
 The restriction to {\it MRT\,} seems a realistic 
approach to {\it general
purpose\,} AI systems. {\it MRT} is obviously one of the key parts
of any intelligence, not only with humans. There is
an astonishing universality of momentum risk-taking; those who master
it in one field, can generally use their expertise in other
fields upon proper (sometimes little) training. We think that 
almost the same {\it risk-taking curves\,} 
(we call them {\it forecasting or termination curves}) govern
quite a spectrum  of short-term risk-management tasks
and that the corresponding "learning" is quite uniform almost 
regardless of the concrete tasks. The neural 
{\it action potentials\,} provide some discretization and 
timing, but there must be other mechanisms in the 
biology of the brain serving {\it MRT}. Some are beyond
the immediate purpose of {\it MRT}, for instance the
analysis of prior decisions. Expecting errors and correcting 
them is what intelligence is about.

\vfil
\vskip 0.1cm
{\sf (3) Modeling MRT.}
 Importantly, {\it MRT} can be modeled mathematically, which
we perform thanks to our focus on {\it short-term\,} management. 
The power growth of our forecasting curves holds 
only for relatively short periods "after the event". The
corresponding differential equations modeling news impact 
seem sufficiently reliable to us. The
trading system described in Section \ref{sec:3} is an experimental
confirmation: it is based on the "power-law" for price-functions
with exponents depending on the corresponding {\it investment horizons}.
Mathematically, an argument in favor of our 
approach is a model of
{\it profit-taking} in terms of Bessel functions. This relates the
periodicity of profit-taking to the asymptotic periodicity
of Bessel functions: a new approach to the
{\it market volatility\,}, one of the key subjects
in quantitative finance.

\vfil
\vskip 0.1cm
{\sf (4) Market volatility}. The closest approach to our one
we found in the vast literature on volatility in stock 
markets is based on the {\it fractional Brownian motion\,}, 
{\it fBM\,} 
with {\it Hurst exponents\,} reflecting the investment horizons.
For instance, the usage of
{\it fBM\,} explains theoretically why the volatility is extreme for
day-trading (with low Hurst exponents).
%which is however 
%insufficient for the "practical" implications (say, in trading systems). 
Some statistical variant of our approach is a consideration of 
a linear combination of 2-3 {\it fBM\,} corresponding to 
"heterogeneous time scales". Let us refer here at least
to  \cite{Che,DB}; see Section \ref{sec:2} above.
The approach via {\it fBM} does not separate the 
{\it profit-taking\,} from  the "stochastic"
volatility of stock markets, which is of key importance
for practical trading (and our system).
Our theoretical analysis indicates that 
{\it Bessel processes}, generalizing {\it fBM}, are likely
to emerge here. 

\vskip 0.1cm
{\sf (5) Some perspectives.}
As it was quoted in the Introduction, we 
are decades away from {\it general purpose} AI 
(USA National Science \& Technology Council). 
%in spite of the obvious progress with {\it narrow AI}. 
However, one can hope that some 
"prototypes", can be designed faster than this. Even limited 
"deep learning" we (mostly) use in our 
experiments described in Section \ref{sec:3}, provided 
efficient  "human-like" behavior of our trading system.
It was entirely focused on investing, but 
designing this kind of {\it MRT} for
various tasks (with uniform and sufficiently fast
machine learning) seems quite doable. It will require\,
(i)  further developing the mathematical model of 
{\it MRT} we suggested,\, (ii) finding its roots in the biology of
the brain and psychology, \,
(iii) improving the learning and risk-taking algorithms and 
making them really universal,\,  (iv) experiments, and more 
experiments.

%Even for the main creator of this system, the author
%of the present paper, it is not that simple to navigate
%at this software sea.    
%Each and every segment here, including data processing,
%%required a lot of special inventions, but this is no different
%from our brain!

\smallskip
%\vfil

\vskip -2cm
%\medskip
\bibliographystyle{unsrt}

\end{document}